\documentclass[fleqn,usenatbib]{mnras}

\usepackage{newtxtext,newtxmath}
\usepackage[T1]{fontenc}
\usepackage{placeins}
\usepackage{hhline}
\usepackage{makecell}
\usepackage{cancel}
\usepackage{booktabs}
\usepackage{multirow}
\usepackage{amsmath} 
\usepackage{bm}
\usepackage{physics}
\usepackage{natbib}
\usepackage[dvipsnames]{xcolor}
\usepackage{graphicx}
\usepackage{xurl}
\usepackage{hyperref}
\usepackage{enumitem}

\usepackage{footnote}
\makesavenoteenv{figure}
\makesavenoteenv{figure*}

\usepackage{array}

\usepackage{eso-pic}

\DeclareRobustCommand{\VAN}[3]{#2}
\let\VANthebibliography\thebibliography
\def\thebibliography{\DeclareRobustCommand{\VAN}[3]{##3}\VANthebibliography}

\title[Implications from Redshift Duality]{Redshift Duality with Pantheon+SH0ES in a Planck-anchored Flat $\Lambda$CDM Framework: Implications for Hubble Tension and Observational Inference}

\author[T.\,K.\ Lee]{
Tae-Kyoung Lee$^{1}$\thanks{E-mail: tklee0301@gmail.com; affiliation updated for this author version.}\\
$^{1}$Department of Social Welfare, Graduate School of Public Policy, Sejong University, 209 Neungdong-ro, Seoul 05006, Republic of Korea, Earth
}

\date{Accepted 2026 April 15. R2 received 2026 April 12; R1 received 2026 March 10; in original form 2025 August 11}

\pubyear{2026}

\makeatletter

\gdef\@journal{%
  \makebox[\textwidth][s]{%
    \mbox{\normalfont\footnotesize Author's version compiled in the MNRAS \LaTeX\ style.}%
    \hfill
    \mbox{\normalfont\footnotesize Published in MNRAS; \href{https://doi.org/10.1093/mnras/stag726}{doi.org/10.1093/mnras/stag726}.}%
  }%
}

\def\@printed{}

\def\ps@headings{%
  \let\@mkboth\markboth
  \def\@oddhead{\Large\hfill{\it\@shorttitle}\hspace{1.5em}\rm\@ddell\thepage}%
  \def\@evenhead{\Large\@ddell\thepage\hspace{1.5em}\it\@shortauthor\hfill}%
  \def\@oddfoot{\footnotesize\copyright\ \@pubyear\ The Author. \,CC BY.\hfill}%
  \def\@evenfoot{\hfill\footnotesize\copyright\ \@pubyear\ The Author. \,CC BY.}%
  \def\sectionmark##1{\markboth{##1}{}}%
  \def\subsectionmark##1{\markright{##1}}%
}

\def\ps@titlepage{%
  \let\@mkboth\@gobbletwo
  \def\@oddhead{\@journal}%
  \def\@evenhead{\@journal}%
  \def\@oddfoot{\footnotesize\copyright\ \@pubyear\ The Author. \,CC BY.\hfill}%
  \def\@evenfoot{\hfill\footnotesize\copyright\ \@pubyear\ The Author. \,CC BY.}%
  \def\sectionmark##1{}%
  \def\subsectionmark##1{}%
}

\makeatother

\AtBeginDocument{\pagestyle{headings}}
\defcitealias{Planck2020}{P20}
\defcitealias{Tristram2024}{P24}
\defcitealias{Riess2022}{R22}
\defcitealias{Riess2024}{R24}
\defcitealias{Brout2022}{B22}
\defcitealias{Scolnic2022}{S22}
\defcitealias{Scolnic2025}{S25}
\defcitealias{Freedman2001}{F01}
\defcitealias{Freedman2012}{F12}
\defcitealias{Freedman2021}{F21}
\defcitealias{Freedman2025}{F25}
\defcitealias{Vogl2025}{V25}
\defcitealias{Gupta2023}{G23}

\begin{document}
\label{firstpage}
\pagerange{\pageref{firstpage}--\pageref{lastpage}}
\maketitle

\begin{abstract}

We test an operationally defined redshift duality in which the observed redshift comprises the standard metric-expansion component together with an additional line-of-sight quantum contribution arising from the cumulative conversion of photon energy into effective mass as a function of path length and frequency. Fitting this hybrid model to the Pantheon+SH0ES compilation, we find that the metric-expansion Hubble constant, $H_\Lambda$, is recovered to a value consistent with the Planck baseline of $67.4~\mathrm{km\,s^{-1}\,Mpc^{-1}}$ within $\lesssim 0.33\sigma$. Redshift-binned analyses show that while the flat Lambda cold dark matter ($\Lambda$CDM) model produces an apparent drift in the inferred Hubble parameter across the Hubble flow, the hybrid model restores the constancy of $H_\Lambda$ across redshift bins. The correctional trends of cosmological physical quantities re-inferred under this framework further indicate the potential to alleviate anomalies associated with high-redshift galaxies. These results suggest that redshift duality warrants further consideration in observational processing and inference, while preserving consistency with a Planck-anchored flat $\Lambda$CDM baseline.

\end{abstract}


% 1
\section{Introduction}
\label{1}

\subsection{Hubble Tension from Local Measurements} 
\label{1.2}

The $\Lambda$CDM framework provides a baseline description of the cosmic thermal and expansion history, constrained by measurements of cosmic microwave background (CMB) anisotropies, yielding $H_0 = 67.4 \pm 0.5 \mathrm{\ km\ s^{-1}\ Mpc^{-1}}$~\citep{Planck2020}.

\vspace{0.8mm}

Against this baseline, the \textit{Hubble tension} reflects a dichotomy between the Hubble constant inferred indirectly from the CMB and that inferred from local-Universe measurements. The SH0ES collaboration, using a three-rung distance ladder based on Hubble Space Telescope (HST) observations that calibrates SNe~Ia via Cepheid variables anchored by geometric distances, infers $H_0=73.04\pm1.04~\mathrm{km\,s^{-1}\,Mpc^{-1}}$ \citep{Riess2022}

\vspace{0.8mm}

However, alternative distance indicators yield conflicting results. \citet{Freedman2019} reported a lower value of $H_0 = 69.8\pm0.8~(\mathrm{stat})\pm1.7~(\mathrm{sys})~\mathrm{km\,s^{-1}\,Mpc^{-1}}$ using the Tip of the Red Giant Branch (TRGB). \citet[hereafter F21]{Freedman2021} reinforced this, emphasizing that TRGB stars originate from old, metal-poor halo populations, thereby minimizing the systematic uncertainties associated with extinction and crowding that plague Cepheid measurements in the disk. This perspective was further supported by \citet[hereafter F25]{Freedman2025}, where the use of James Webb Space Telescope (JWST) data for TRGB and J-region Asymptotic Giant Branch (JAGB) indicators yielded $H_0$ values of $68.8$ and $67.8~\mathrm{km\,s^{-1}\,Mpc^{-1}}$, respectively, suggesting that the tension might be alleviated through methodologically cleaner tracers.

%\vspace{3mm}

Conversely, \citet[hereafter R24]{Riess2024} argued that reducing subsample bias requires a large, distance-limited sample. By analyzing a comprehensive suite of Cepheids, TRGB, and JAGB hosts within a consistent framework using JWST, they reported consistent $H_0$ values of $73.4\pm2.1$ (Cepheids), $72.1\pm2.2$ (TRGB), and $72.2\pm2.2~\mathrm{km\,s^{-1}\,Mpc^{-1}}$ (JAGB), contending that the lower values found by other groups result from proximity-driven sample selection rather than intrinsic physical differences.

%\vspace{2mm}

Extensions using the DESI fundamental-plane relation of early-type galaxies, anchored by SNe~Ia in the Coma cluster, imply even higher values, reaching $H_0=76.5\pm2.2~\mathrm{km\,s^{-1}\,Mpc^{-1}}$ \citep[hereafter S25]{Scolnic2025}. This suggests that the discrepancy is already ``baked-in'' to the local distance scale (out to $D\sim100~\mathrm{Mpc}$ or $z\sim 0.023$), rather than arising solely from high-redshift extrapolation.

%\vspace{2mm}

Furthermore, a diverse suite of probes independent of the Cepheid--SNe~Ia ladder consistently favors higher values. Infrared surface brightness fluctuations (SBF), calibrated via JWST TRGB distances and thus independent of Cepheids and SNe~Ia, yield $H_0=73.8\pm2.4~\mathrm{km\,s^{-1}\,Mpc^{-1}}$ \citep{Jensen2025}, reinforcing earlier HST-based SBF results ($H_0=73.3\pm2.5$; \citealt{Blakeslee2021}). Similarly, the baryonic Tully--Fisher relation reports $H_0=75.1\pm2.3~\mathrm{km\,s^{-1}\,Mpc^{-1}}$ \citep{Schombert2020}. Geometric methods such as strong-lensing time-delay cosmography infer $H_0=73.3^{+1.7}_{-1.8}~\mathrm{km\,s^{-1}\,Mpc^{-1}}$ \citep{Wong2020}, with individual systems like DES~J0408$-$5354 yielding $H_0=74.2^{+2.7}_{-3.0}$ \citep{Shajib2020}. Finally, Type~II supernovae, analyzed through both the distance ladder and the physics-based expanding-photosphere method (EPM), consistently align with the high-$H_0$ regime: $H_0=75.8^{+5.2}_{-4.9}~\mathrm{km\,s^{-1}\,Mpc^{-1}}$ \citep{deJaeger2020} and $H_0=74.9\pm1.9~\mathrm{km\,s^{-1}\,Mpc^{-1}}$ \citep[hereafter V25]{Vogl2025}.

\vspace{2mm}

Although \citetalias{Freedman2025} noted that their investigation into the Cepheid-anchored value remains ongoing due to significant photometric differences between analysis software packages (e.g., \textsc{daophot} vs.\ \textsc{dolphot}) in crowded disk environments, it is notable that historically, when the same lead author utilized Cepheids as the primary anchor, the inferred Hubble constants were $H_0 = 71\pm6~\mathrm{km\,s^{-1}\,Mpc^{-1}}$ \citep[hereafter F01]{Freedman2001} and $H_0 = 74.3\pm2.1~\mathrm{km\,s^{-1}\,Mpc^{-1}}$ \citep[hereafter F12]{Freedman2012}. These values align with the results of \citet{Riess2016}, \citet[hereafter R22]{Riess2022}, and \citetalias{Riess2024} within $1\sigma$, challenging the view that the tension is merely an artifact of specific calibration methodologies.

% 1-3
\subsection{Existing Approaches: Dynamical Dark Energy}
\label{1.3}

If the tension is interpreted as a late-time expansion mismatch, a conventional approach is to relax $w=-1$ via constant-$w$ extensions \citep{Caldwell1998} or the CPL form \citep{Chevallier2001, Linder2003}; in that context, phantom-like values ($w_0<-1$) have been explored phenomenologically to raise the CMB-inferred $H_0$, and \citet{DiValentino2021} argued that such models can shift the inferred value upward, though this resolution degrades when combined with late-time probes (BAO, SNe Ia).   

DESI Year-1 baryon acoustic oscillation (BAO) constraints \citep{Adame2025} (over $6$ million extragalactic sources, $0.1<z<4.2$) and the Union3 SN analysis with the UNITY1.5 Bayesian framework \citep{Rubin2025} both favor dynamical dark energy with $w_0>-1$ and $w_a<0$, with best fits $(w_0,w_a)\approx(-0.83,-0.75)$ for DESI (+CMB+Pantheon+) and $(-0.74,-0.79)$ for Union3 (+BAO+CMB) (yielding $\approx(-0.64,-1.27)$ for the DESI+CMB+Union3 combination); in the CPL interpretation this implies a ``phantom crossing'', with $w(z)$ transitioning from phantom-like at earlier times to quintessence-like today.

In contrast, \citet{Abbott2024} analyze the 5-year Dark Energy Survey (DES)---a uniform sample of $\sim1600$ machine-learning-classified SNe~Ia---reporting constraints consistent with a cosmological constant within $\sim2\sigma$ (SNe-only: $w=-0.80^{+0.14}_{-0.16}$; combined: $w=-0.941\pm0.026$). While noting a mild preference for dynamical dark energy, they caution this may arise from unresolved systematics at redshift extremes (low-$z$ vs. high-$z$). Consequently, this large-scale reference illustrates that evidence for time-varying dark energy remains highly method- and data-dependent across recent surveys.
 
In addition to the lack of consensus on dark-energy evolution, attempts to alleviate the Hubble tension by invoking a temporal evolution of dark energy shift the explanatory burden, introducing stringent requirements: (i) preserving consistency with the Planck-anchored baseline; (ii) identifying a fundamental physical principle for such evolution; and (iii) harmonizing these phenomenological modifications with the history of cosmic evolution.

%1.4
\subsection{Motivation}
\label{1.4}

The essence of the Hubble tension lies in the discrepancy between the Hubble value inferred from the cosmic background radiation based on the standard theory and the value derived from observing optical radiation that has reached us through its journey along the line of sight. The distinction between ``early'' and ``late'' is  a criterion based on secondary inferential interpretations. The existence of a tension implies that, while one of the two values is treated as the fiducial reference~(\citealp[hereafter P20]{Planck2020}; \citealp[hereafter P24]{Tristram2024}), the other encodes \textit{distinct physical information}~(\citetalias{Riess2022}; \citetalias{Scolnic2025}; \citealp{Wong2020}; \citetalias{Vogl2025}). 

Specifically within the framework of supernova cosmology, this suggests that radiation propagating along the line of sight may acquire an additional imprint over its cosmic journey, which manifests observationally as a deviation from the modulus--redshift relation. On the Hubble diagram, with redshift on the horizontal axis and distance modulus on the vertical axis, the tension is therefore expressed as a mismatch between the observed modulus and the modulus predicted at that redshift under the CMB $H_0$ baseline. At least within the \textit{Cepheid-calibrated} SN~Ia local distance ladder~\citepalias{Freedman2001, Freedman2012, Riess2022, Riess2024}, such a trend is empirically suggested.

Yet, because the physical flux is the product of photon number and photon energy, this would seem to imply an observational paradox under the conventional interpretation: either the magnitude is preserved while the photon energy decreases (thereby increasing the inferred redshift), or the photon energy remains unchanged while the photon number effectively increases (thereby making the source brighter).

It is essential to recognize that Hubble tension is an issue of \emph{observation and inference}. Observables are constructed through calibration pipelines that apply multiple correction factors, including photometric cross-calibration, light-curve shape standardization, color--luminosity corrections, host-galaxy mass-step adjustments, and selection-bias corrections, which collectively define the operational mapping from raw photometry to standardized distance moduli~
(\citealp[hereafter B22]{Brout2022}; \citealp[hereafter S22]{Scolnic2022}; \citetalias{Riess2022}).

If there exists a mechanism by which a portion of photon energy is converted into effective mass in a wavelength-dependent manner along the propagation pathway---acting more strongly at shorter wavelengths and thereby imprinting an additional redshift component on spectral features while also altering the continuum color---then this imprint can become operationally degenerate with dust reddening (Appendix~\ref{AppC}). In that case, the pipeline would attribute the color change to dust and apply a brightening correction to the magnitude, while the additional redshift component remains encoded in the spectral features. The net outcome is that an ``excess'' redshift would be observed relative to the corrected, inadvertently restored magnitude; we refer to this inadvertent recovery of the intrinsic magnitude scale as Dust-mimicking Magnitude Compensation (hereafter DMC). Consequently, regressing such data under a standard framework can bias the fit toward a higher effective Hubble parameter.

Moreover, if the imprint accumulates with path length, light from larger distances would be processed as having ``too much'' redshift for its distance modulus (equivalently, a modulus that is ``too small'' for its redshift). This induces an apparent drift in which the inferred effective Hubble parameter, $H_{\rm eff}$, increases systematically with distance (higher $z$). When the CPL form is fit to data containing this pattern, it can be misread as ``phantom crossing'' ($w_0>-1,\; w_a<0$), i.e., an apparent late-time evolution of the equation of state. In this sense, the Hubble-diagram anomaly offers a direct observational pathway to the evolving dark-energy trends reported in large-scale analyses such as DESI \citep{Adame2025} and Union3 \citep{Rubin2025}.

A methodological reference point for such an additional, non-expansion (or non-thermal, or non-time-dilating) redshift is the CCC+TL (Conformal Cyclic Cosmology + tired light) hybrid scenario introduced by \citet[hereafter G23]{Gupta2023}, which treats the observed redshift as a composite observable comprising both metric and non-metric contributions. Irrespective of the specific physical validity of the CCC framework, its operational capability to fit Type~Ia supernovae illustrates that replacing the single-origin assumption with an explicit two-component hypothesis provides a computationally viable pathway that is directly testable against data.

To motivate a plausible route to non-metric redshift contributions, we consider settings in which radiative energy can be localized\footnote{Conceptually, this is analogous to the \textit{Ouroboros} biting its own tail: a radiative state following a circular path or a topological knot (an infinite loop). While the wave propagates indefinitely within this localized confinement, the propagation is effectively halted on scales external to it, thereby appearing as a stationary entity.} or effectively sequestered into weakly radiating degrees of freedom. First, Bound States in the Continuum, originally proposed by \citet{vonNeumann1929} and reviewed by \citet{Hsu2016}, show that wave energy can remain spatially confined despite lying within a scattering continuum, enabled by destructive interference and symmetry-induced decoupling from radiation channels; this mechanism has been demonstrated experimentally in photonic systems \citep{Plotnik2011}. Second, topological solitons such as Skyrmions illustrate how non-linear field configurations can form localized, finite-energy states with particle-like stability protected within a topological sector \citep{Skyrme1961, Manton2004}. Finally, \citet{Wheeler1955} introduced the concept of geons: localized bundles of electromagnetic energy that can be self-confined by their own gravitational field, effectively behaving as massive objects. 

Taken together, these examples provide precedents for how propagating wave/field energy may be trapped into localized, effectively mass-like degrees of freedom. Physically, for a localized configuration one may define a rest frame in which the net momentum vanishes; in that frame, the trapped energy can be assigned an effective rest mass via mass--energy equivalence, $m_{\rm effective}=E_{\rm localized}/c^{2}$. Additionally, if a small fraction of radiation propagating along the line of sight is diverted into such weakly radiating or non-radiative states, it would appear observationally as flux attenuation.

In this spirit, we adopt Wave-to-Mass Conversion (hereafter WMC) as an operational label for the possibility that a small fraction of photon-field energy is transferred into localized, effectively massive degrees of freedom along cosmological sightlines, even if such a process occurs irreversibly. This usage is intentionally agnostic about the microphysics, but it is conceptually analogous to established field-to-matter conversion channels, such as pair creation in the Breit--Wheeler process and the Schwinger mechanism \citep{Breit1934, Schwinger1951, Golub2021}.

\begin{table}
\setlength{\tabcolsep}{5.5pt}
\centering
\caption{Summary of local $H_0$ measurements categorized by the observing facility, the geometric anchor (luminosity calibrator), and the Hubble-flow tracer. Notably, despite varying facilities and analysis pipelines, the Cepheid-based determinations by \citetalias{Freedman2001, Freedman2012} and \citetalias{Riess2022, Riess2024} show remarkable consistency, agreeing within $1\sigma$. The SN~II entry \citepalias{Vogl2025} represents a distance-ladder-free determination using ground-based facilities, where the physics of the expanding photosphere acts as the intrinsic anchor without external calibration. SST denotes the \textit{Spitzer Space Telescope}.}
\label{tab0}
{\fontsize{7.4pt}{12.3pt}\selectfont
\begin{tabular}{c c c c c}
\Xhline{0.9pt}
Instrument & Anchor Type & Hubble-flow & ${H_0}$ [km\,s$^{-1}$\,Mpc$^{-1}$] & Citation \\
\Xhline{0.9pt}
JWST & JAGB & SN~Ia & $67.80 \pm 2.17$ & \citetalias{Freedman2025} \\
JWST & TRGB & SN~Ia & $68.81 \pm 1.79$ & \citetalias{Freedman2025} \\
JWST & JAGB & SN~Ia & $72.2 \pm 2.2$ & \citetalias{Riess2024} \\
JWST & TRGB & SN~Ia & $72.1 \pm 2.2$ & \citetalias{Riess2024} \\
\hline
HST & Cepheid & SN Ia & $73.04 \pm 1.04$ & \citetalias{Riess2022}\\ 
JWST & Cepheid & SN~Ia & $73.4 \pm 2.1$ & \citetalias{Riess2024} \\
HST & Cepheid & SN~Ia & $71 \pm 6$ & \citetalias{Freedman2001} \\
SST & Cepheid & SN~Ia & $74.3 \pm 2.1$ & \citetalias{Freedman2012} \\
\hline
(Ground) & (SN~II) & SN~II & $74.9 \pm 1.9$ & \citetalias{Vogl2025} \\
\Xhline{0.9pt}
\end{tabular}
}
\end{table}

%1.5
\subsection{Our Approach and Scope}
\label{1.5}

\subsubsection{Operational Definition of WMC}
\label{1.5.1}

The objective of this work is to evaluate whether the Hubble tension and the reported late-time dark-energy evolution can be resolved by introducing, in addition to the standard cosmological redshift induced by cosmic expansion, a wavelength-dependent, non-metric redshift component that manifests cumulatively in proportion to the photon path length along the line of sight. We operationalize the WMC hypothesis: photon number within a ray bundle is conserved, while a fraction of photon energy is sequestered into effectively massive degrees of freedom during propagation, thereby inducing flux attenuation.

If the conversion rate is wavelength-dependent, this attenuation is not gray but chromatic, and it is accompanied by an apparent color change and an effective spectroscopic shift. To be operationally effective, we posit that this wavelength-dependency must become negligible in the microwave regime, thereby ensuring that the standard understanding of the cosmic thermal history~\citepalias{Planck2020} and the near-perfect blackbody spectrum~\citep{Fixsen1996} remains preserved.

\subsubsection{Operational DMC Pathway  in SN~Ia Standardization}
\label{1.5.2}

Guided by this operational definition, we explore whether the wavelength dependence of WMC can be empirically constrained to mimic dust extinction, thereby satisfying the condition where the color--luminosity coefficient $\beta$ (associated with the color parameter $c$) in the Tripp estimator~\citep{Tripp1998, Guy2007, Betoule2014} approximates the empirically derived values---ranging from $\beta \approx 2.9$~\citepalias{Freedman2025} to $\beta \approx 3.1$~\citepalias{Riess2022, Scolnic2022}. We test whether a wavelength-dependent WMC attenuation exists that successfully mimics dust in the optical regime---effectively utilizing the $\beta c$ term as an operational pathway to restore luminosity—while simultaneously rendering the WMC attenuation rate negligible in the microwave domain to preserve the CMB blackbody spectrum.

\subsubsection{The Hybrid Redshift Model}
\label{1.5.3}

We define a Hybrid Model in which the observed cosmological redshift is modeled as the combined effect of metric expansion and an additional non-metric WMC contribution. The metric component is governed by a spatially flat, Planck-anchored baseline with $(H_0,\Omega_m,\Omega_k,w)\simeq (67.4~\mathrm{km\,s^{-1}\,Mpc^{-1}},\,0.315,\,0,\,-1)$~\citepalias{Planck2020, Tristram2024}.

We do not explicitly model the consequent underestimation of intrinsic luminosity induced by WMC. This choice follows from our operational definition: to the extent that the WMC-induced color variation mimics dust extinction, the intrinsic luminosity scale is effectively restored by the color-correction term in the standard Tripp estimator~\citep{Tripp1998, Guy2007, Betoule2014}.

\subsubsection{Data and Regression Strategy}
\label{1.5.4}

To determine whether the Planck-anchored Hubble constant can be recovered from the elevated late-time determination within the hybrid framework, we utilize the Pantheon+SH0ES SN~Ia compilation~\citepalias{Scolnic2022}. As summarized in Table~\ref{tab0}, Cepheid-based determinations of $H_0$ show consistency within $1\sigma$ despite methodological differences, motivating the use of this anchor-dependent dataset as a controlled testbed. 

We perform global and tomographic (redshift-binned) regressions to verify the following validation criteria:
\begin{quote}
    \begin{description}
        \item \textbf{I. Planck Parameter Recovery:} We test whether the hybrid framework recovers the Planck-anchored baseline ($H_0 \simeq 67.4~\mathrm{km\,s^{-1}\,Mpc^{-1}}$) both globally and consistently across individual redshift bins. We verify whether this recovery is accompanied by a nuisance magnitude offset of $M \approx 0$. This condition is required because, under our operational definition, the distance modulus (and thus the intrinsic luminosity) is effectively restored by the standardisation pipeline.
        
\item \textbf{II. Benchmark Tension and Drift:} Standard metric-only models (flat $\Lambda$CDM, $w$CDM, and CPL) are employed as benchmarks to evaluate the validity of the hybrid framework. We first investigate whether these models, when regressed against the Pantheon+SH0ES\footnote{\label{fn:2}\href{https://github.com/PantheonPlusSH0ES/DataRelease/blob/main/Pantheon\%2B_Data/4_DISTANCES_AND_COVAR/Pantheon\%2BSH0ES.dat}{Pantheon+SH0ES Data Release}. We utilize the Pantheon+SH0ES dataset in its released form, comprising a subset of 1701 light curves that passed the standard ``Cosmology'' quality cuts established by the Pantheon+ pipeline (\citetalias{Scolnic2022}; Table 2): e.g., $|x_1| < 3$, $|c| < 0.3$, $\sigma(x_1) < 1.5$, $\sigma_{(\rm pkmjd)} < 2$ days, and $E(B-V)_{\text{MW}} < 0.20$ mag. These cuts isolate standardizable SNe~Ia to minimize intrinsic scatter, introducing no model-dependent bias toward either the $\Lambda$CDM or hybrid frameworks.} dataset, consistently yield a tension-level Hubble constant. Furthermore, in tomographic regressions using the flat $\Lambda$CDM baseline, we specifically check if a systematic ``drift'' manifests in high-redshift bins---where the inferred $H_0$ increases with $z$---thereby mimicking the ``phantom crossing'' behavior observed in recent dark-energy studies~\citep{Rubin2025, Adame2025}.%($w_0>-1,\; w_a<0$)
        
        \item \textbf{III. Statistical Preference:} We perform a comparative analysis using the Bayesian Information Criterion (BIC)~\citep{Schwarz1978} to determine whether the hybrid framework is statistically preferred over the metric-only benchmarks. We assess whether the improvement in fit provided by the non-metric extension justifies its added complexity compared to the standard $\Lambda$CDM, $w$CDM, and CPL models.
    \end{description}
\end{quote}

\subsubsection{Concurrent Alleviation of Independent Anomalies}
\label{1.5.5}

We examine the model's capacity to alleviate anomalies reported independently of the Cepheid-calibrated SN~Ia framework by defining and exploring various cosmological parameters derived under a specific inferential bias: a scenario where the distance modulus is effectively restored to its real value via the standardization pathway (DMC), yet the excess non-metric redshift component induced by WMC is misinterpreted as standard expansion redshift. This investigation is extended to the context of the ``impossible early galaxy'' problem~\citep{Labbe2023, BoylanKolchin2023, Adams2023, Atek2023, Donnan2023, Glazebrook2024, Nanayakkara2024} reported by JWST.

% 2
\section{METHODOLOGY}
\label{2}

\subsection{Planck-Anchored Baseline Cosmology}
\label{1.1}

The \citetalias{Planck2020} marked a high-precision benchmark: assuming spatial flatness ($\Omega_k=0$), it reported Hubble constant $H_0 = 67.4\pm0.5~\mathrm{km\,s^{-1}\,Mpc^{-1}}$ and $\Omega_m=0.315\pm0.007$. Combining the CMB, which constrains the expansion rate and matter content through early-Universe physics, with baryon acoustic oscillations (BAO) and Type~Ia supernovae (SNe~Ia) to break geometric degeneracies, \citetalias{Planck2020} further constrained the dark-energy equation of state to $w=-1.03\pm0.03$, consistent with a cosmological constant.

\citetalias{Planck2020} showed a mild internal preference for $\Omega_k<0$ ($\sim3.4\sigma$ in some likelihood combinations), but the \citetalias{Tristram2024} re-analysis resolves this: using improved polarization handling to mitigate instrumental systematics, \citetalias{Tristram2024} finds $\Omega_k=-0.012\pm0.010$, consistent with a flat geometry. \citetalias{Tristram2024} also reproduces a expansion-rate inference, $H_0=67.66\pm0.53~\mathrm{km\,s^{-1}\,Mpc^{-1}}$.

We therefore adopt a Planck-anchored baseline defined by $(H_0,\Omega_m,\Omega_k,w)\approx(67.4~\mathrm{km\,s^{-1}\,Mpc^{-1}},\,0.315,\,0,\,-1)$. In this work, this choice provides a controlled reference point against which late-time inferences can be compared.

\subsection{\texorpdfstring{Metric-Expansion Component: Standard $\Lambda$CDM Relation}{Metric-Expansion Component: Standard LambdaCDM Relation}}
\label{2.1}

We adopt a spatially flat \(\Lambda\)CDM relation between the redshift $z_\Lambda$ induced by cosmic expansion and the corresponding line-of-sight comoving distance $X_{\Lambda}$ as our metric-expansion baseline:
\begin{equation}
X_{\Lambda}(z_{\Lambda}) = \frac{c_{\scriptscriptstyle l}}{H_{\Lambda}}
\int_{0}^{z_{\Lambda}} \frac{dz'}{\sqrt{\Omega_m (1+z')^{3} + \Omega_\Lambda}},
\quad \Omega_\Lambda \equiv 1-\Omega_m .
\label{eq1}
\end{equation}
In the Planck-anchored configuration, we impose priors on the parameters using the CMB best-fit values reported by \citetalias{Planck2020}, namely \(H_{0}^{\rm \scriptscriptstyle CMB}=67.4~{\rm km\,s^{-1}\,Mpc^{-1}}\), \(\Omega_m=0.315\), and \(\Omega_\Lambda=0.685\). In our regression analysis, the metric-expansion Hubble parameter \(H_{\Lambda}\) and \(\Omega_m\) are each either fixed to their Planck best-fit values or treated as free parameters, while spatial flatness is enforced via \(\Omega_\Lambda \equiv 1-\Omega_m\). The condition \(w=-1\) is already encoded in the relation itself. We fix the speed of light to \(c_{\scriptscriptstyle l} \equiv 299\,792.458~{\rm km\,s^{-1}}\).

% 2-2
\subsection{Non-metric Component: WMC Relation}
\label{2.2}

Under the operationally defined WMC framework, a fraction of isotropically emitted radiation is transferred during propagation into effectively non-radiative degrees of freedom, attenuating the radiative energy remaining in the line-of-sight beam with distance. By construction, the apparently ``missing'' energy is redistributed rather than destroyed, preserving global energy conservation while allowing the radiative component to carry an imprint consistent with an additional redshift. Although superficially reminiscent of Zwicky's tired-light idea \citep{Zwicky1929} in invoking distance-dependent attenuation, it differs in two fundamental respects: (i) the redshift-like imprint is tied to an explicit energy-redistribution channel rather than photon-energy loss, and (ii) it is formulated as an additional, subdominant non-metric component on top of, rather than in place of, a metric-expansion baseline.

We model the attenuation of the isotropically emitted specific flux, $\Delta F_{\nu}$, over a proper-distance increment $\Delta X$. We posit that this decrement is proportional to the band-limited WMC conversion coefficient per unit length $P(\nu,i)$, the local photon number flux density, and the photon energy $h\nu$. This relation is expressed as
\begin{equation}
\frac{\Delta F_{\nu}}{\Delta X} \;\equiv\; -\,P(\nu,i)\,\frac{n_{\nu}}{\Delta t\,\Delta A}\,h\nu,
\label{eq2}
\end{equation}
where $h$ is Planck's constant and $\nu$ is the photon frequency. The term $n_{\nu}/(\Delta t\,\Delta A)$ denotes the photon number flux density per unit frequency, defined as the number of photons of frequency $\nu$ crossing an area $\Delta A$ during a time interval $\Delta t$.

The coefficient $P(\nu, i)$ denotes a frequency-dependent conversion probability per unit length ($\mathrm{Mpc}^{-1}$) between radiative and effectively non-radiative degrees of freedom. We consider only the forward conversion channel, thus omitting directional subscripts (e.g., $P_{wm}$, $P_{mw}$). The variable $i$ serves as a proxy for line-of-sight environmental conditions that modulate the conversion efficacy. $i$ may be interpreted as an effective coupling parameter governing vacuum-assisted tunneling between the propagating radiative mode and the high-energy vacuum sectors (for general discussions on vacuum energy and high-energy contributions, see, e.g., \citealt{Weinberg1989, Martin2012, Burgess2013}), with laboratory evidence for macroscopic quantum tunneling phenomena \citep[e.g.,][]{Voss1981}.

\subsubsection{Operational Assumption: Photon Number Conservation}
\label{2.2.1}

Allowing photon-number non-conservation would generally increase the complexity of DMC-based magnitude restoration, introducing an additional, non-trivial coupling between excess redshift and intrinsic dimming. For regression clarity and model parsimony, we therefore impose photon-number conservation within the propagating ray bundle, so that the photon number flux density does not secularly evolve with proper distance:
\begin{equation}
\frac{d}{dX}\left(\frac{n_{\nu}}{\Delta t\,\Delta A}\right)\approx 0.
\label{eq3}
\end{equation}

\subsubsection{Operational Treatment: Homogeneous-Vacuum}
\label{2.2.2}

We adopt a spatially flat metric baseline ($\Omega_k=0$), and operationally treat the environmental parameter $i$---a proxy for the vacuum interaction efficacy---as approximately uniform within a homogeneous domain along the line of sight. Under this convention, the secular drift of the conversion coefficient is neglected:
\begin{subequations}
\begin{align}
&\frac{d}{dX}P\!\bigl(\nu,i(X)\bigr)=\frac{\partial P}{\partial i}\frac{di}{dX}\simeq 0, \label{eq4a}\\[2.0ex]
&P\!\bigl(\nu,i(X)\bigr)\simeq P\!\bigl(\nu,i_0\bigr)\equiv P(\nu). \label{eq4b}
\end{align}
\end{subequations}
Here $i_0$ denotes the representative (constant) value of $i$ within the homogeneous domain. In what follows, the conversion coefficient is therefore taken to depend only on frequency (or equivalently wavelength), and we write $P(\nu)$ by convention.

\subsubsection{Operational Treatment: Effective Frequency Dependence}
\label{2.2.3}

Astrophysical tracers emit broad spectral energy distributions (SEDs), so the frequency-dependent coefficient $P(\nu)$ is operationally constrained only through an SED-weighted effective coefficient for each tracer $\kappa$. Here, $F_{\nu,0}^{\kappa}$ is the unattenuated emitted specific flux (at the operational reference point $X=0$), $F_{0}^{\kappa}$ is the corresponding band-integrated flux normalization, and $S^{\kappa}(\nu)$ is the normalized emitted SED shape. We write
\begin{equation}
F_{\nu,0}^{\kappa}\;\equiv\;F_{0}^{\kappa}\,S^{\kappa}(\nu),
\qquad
\int d\nu\,S^{\kappa}(\nu)=1,
\label{eq5}
\end{equation}
and define the tracer-level effective coefficient
\begin{equation}
P^{\kappa}\;\equiv\;\frac{\displaystyle \int d\nu\, P(\nu)\,F_{\nu,0}^{\kappa}}{\displaystyle \int d\nu\, F_{\nu,0}^{\kappa}}
\;=\;\int d\nu\,P(\nu)\,S^{\kappa}(\nu),
\label{eq6}
\end{equation}
where $P^{\kappa}$ is the SED-weighted effective coefficient associated with tracer $\kappa$. For compact notation, we associate the same weighting with a tracer-defined \emph{effective wavelength} $\lambda_{\rm eff}^{\kappa}$ and effective frequency $\nu_{\rm eff}^{\kappa}$,
\begin{equation}
\lambda_{\rm eff}^{\kappa}\;\equiv\;\frac{c_{\scriptscriptstyle l}}{\nu_{\rm eff}^{\kappa}},
\qquad
\nu_{\rm eff}^{\kappa}\;\equiv\;\int d\nu\,\nu\,S^{\kappa}(\nu),
\label{eq7}
\end{equation}
so that, when $P(\nu)$ varies slowly across the SED support (or $S^{\kappa}$ is narrow), one may approximate
\begin{equation}
P^{\kappa}\;\simeq\;P\!\bigl(\nu_{\rm eff}^{\kappa}\bigr)\;=\;P\!\bigl(c_{\scriptscriptstyle l}/\lambda_{\rm eff}^{\kappa}\bigr).
\label{eq8}
\end{equation}
In what follows, each tracer is represented by a single parameter $P^{\kappa}$. We consider the set $\kappa\in\{C,S,T,J,M\}$, denoting Cepheids, Type~Ia supernovae, TRGB, JAGB, and the CMB, respectively.  Here $\lambda_{\rm eff}^{\kappa}$ is induced by the emitted spectrum $S^{\kappa}(\nu)$ and is not identified with the nominal central wavelength of any specific instrumental filter.

\subsubsection{Operational Formulation of the WMC}
\label{2.2.4}

With an effective conversion coefficient specified for tracer $\kappa$, and adopting a formalism analogous to the Beer--Lambert law for extinction~\citep{Lambert1760, Beer1852}, we parameterize the incremental flux change over a proper-distance step $dX$ as
\begin{equation}
dF \;\equiv\; -\,F\,P\,dX,
\label{eq9}
\end{equation}
which integrates to
\begin{equation}
F(X)=F(0)\,\exp\!\left(-PX\right),
\label{eq10}
\end{equation}
with $F(0)\equiv F(X=0)$, where $X$ is the proper distance traversed by the light. Hereafter the tracer index $\kappa$ is suppressed in $P$ for notational simplicity.

For a fixed observational framework---specifically, the distance--redshift analysis based on a given tracer---the attenuation factor is mapped to an effective non-metric redshift component $z_q$ via a flux--energy correspondence; under the operational convention of photon-number conservation in the ray bundle,
\begin{equation}
\exp\!\left(PX\right)
\;=\;
\frac{F(0)}{F(X)}
\;=\;
\frac{E(0)}{E(X)}
\;\equiv\;
1+z_{q},
\label{eq11}
\end{equation}
where $E(X)$ denotes the radiative energy per unit observer time per unit collecting area remaining within the tracer-defined effective spectral support after propagation over a proper distance $X$.

The corresponding proper-distance relation is then
\begin{equation}
X_{q}(z_{q})
\;\equiv\;
\frac{1}{P}\,\ln\!\bigl(1+z_{q}\bigr),
\label{eq12}
\end{equation}
and, introducing an effective WMC rate $H_{q}$ defined by $H_{q}\equiv c_{\scriptscriptstyle l}\,P$ with units of ${\rm km\,s^{-1}\,Mpc^{-1}}$, we may rewrite this as
\begin{equation}
X_{q}(z_{q})
\;=\;
\frac{c_{\scriptscriptstyle l}}{H_{q}}\ln\!\bigl(1+z_{q}\bigr).
\label{eq13}
\end{equation}

The resulting logarithmic distance--redshift relation is mathematically isomorphic to the exponential attenuation relations that appear in earlier non-expanding or tired-light models (e.g., \citealt{Browne1962,Gupta2023}). Its interpretation here, however, is different: we introduce $P(\nu)$ as a flux-based, band-limited WMC conversion coefficient that parameterizes a non-metric degree of freedom, and we interpret cosmological observables by having this component coexist with the metric-expansion contribution rather than substituting for it. In other words, $P(\nu)$ is not a replacement term for expansion, but an additional component that enters alongside it.

Operationally, the wavelength dependence can be specified such that the effective conversion is negligible in the CMB microwave band, thereby preserving the observed blackbody spectrum and leaving the thermal history effectively unchanged (Appendix~\ref{AppC}).

\subsection{Flux, Redshift, and Distance Modulus}
\label{2.3}
We consider a hybrid description in which the net flux attenuation inferred for a tracer $\kappa$ arises from two concurrent, multiplicative contributions---metric expansion and a non-metric WMC attenuation---both acting along the same propagation distance.

\subsubsection{Single-path Requirement}
\label{2.3.1}
We adopt the single-path condition of \citetalias{Gupta2023}, namely that the photon follows a unique physical trajectory. Accordingly, the comoving distance associated with the metric-expansion component and the proper distance associated with the WMC component are taken to coincide along that trajectory:
\begin{equation}
X_{\Lambda}(z_{\Lambda}) \;=\; X_{q}(z_{q}) \;\equiv\; X.
\label{eq14}
\end{equation}

\subsubsection{Flux and Luminosity Distance in the Flat $\Lambda$CDM Framework}
\label{2.3.2}
In the metric-expansion component, cosmic expansion introduces two achromatic factors: photon energies are reduced by $(1+z_{\Lambda})^{-1}$ and photon arrival rates are time-dilated by another $(1+z_{\Lambda})^{-1}$. Combining these yields the standard metric relation
\begin{equation}
F_{\Lambda}(X) \;=\; \frac{L^{\kappa}}{4\pi X^{2}(1+z_{\Lambda})^{2} }\;=\; \frac{L^{\kappa}}{4\pi d_{\Lambda}^{\,2}},
\qquad
d_{\Lambda}\;\equiv\;(1+z_{\Lambda})\,X,
\label{eq15}
\end{equation}
where $F_{\Lambda}(X)$ is the metric-expansion (baseline) flux of tracer $\kappa$ received after propagation distance $X$, and $d_{\Lambda}$ is the luminosity distance attributed to metric expansion (i.e., to $z_{\Lambda}$). The quantity $L^{\kappa}$ denotes the intrinsic luminosity of tracer $\kappa$.

\subsubsection{Flux and Luminosity Distance in the Hybrid Framework}
\label{2.3.3}
The WMC contribution is introduced as an additional chromatic attenuation governed by the tracer-dependent effective coefficient $P^{\kappa}$. Along the same distance $X$, this yields a multiplicative suppression of the radiative flux,
\begin{equation}
F_{h}(X) \;=\; F_{\Lambda}(X)\,\exp\!\left(-P^{\kappa}X\right),
\label{eq16}
\end{equation}
where $F_h(X)$ is the hybrid (observed) flux in the tracer-$\kappa$ channel, including both metric expansion and WMC attenuation. Using the operational identification $\exp(P^{\kappa}X)\equiv 1+z_{q}^{\kappa}$, the hybrid flux admits the compact decomposition
\begin{equation}
F_{h}(X) \;\equiv\; \frac{L^{\kappa}}{4\pi d_h^2}
\;=\;\frac{L^{\kappa}}{4\pi d_{\Lambda}^{\,2}\,(1+z_{q}^{\kappa})},
\label{eq17}
\end{equation}
which defines $d_h$ as the effective hybrid luminosity distance inferred from the attenuated flux:
\begin{equation}
d_{h}^{\,2} \;\equiv\; d_{\Lambda}^{\,2}\,(1+z_{q}^{\kappa}),
\qquad
d_{h} \;=\; d_{\Lambda}\,\sqrt{1+z_{q}^{\kappa}}.
\label{eq18}
\end{equation}

\subsubsection{Inverse-Flux Factorization and Redshift Decomposition}
\label{2.3.4}
Combining the flux attenuation relation in equation~\eqref{eq16} with the flux--redshift correspondence defined in equation~\eqref{eq11}, and rewriting the result in terms of inverse flux ratios---the directly observable dimming measure---makes the factorization explicit:
\begin{equation}
\frac{F(0)}{F_{h}(X)} \;=\; \frac{F(0)}{F_{\Lambda}(X)}\,\exp\!\left(P^{\kappa}X\right)
\;=\; \frac{F(0)}{F_{\Lambda}(X)}\,\bigl(1+z_{q}^{\kappa}\bigr).
\label{eq19}
\end{equation}
Consequently, invoking the flux--redshift correspondence, this factorization implies a corresponding decomposition of the effective redshift inferred for tracer $\kappa$:
\begin{equation}
(1+z_{h}) \;=\; (1+z_{\Lambda})(1+z_{q}),
\label{eq20}
\end{equation}
where $z_{h}$ denotes the total effective hybrid redshift, and $z_{q}$ the WMC-induced component (tracer index suppressed for brevity).

By substituting the explicit distance relations for $X_{\Lambda}$ and $X_{q}$ derived from their respective definitions, we obtain the defining equation that implicitly determines the metric redshift $z_{\Lambda}$:
\begin{equation}
\frac{c_{\scriptscriptstyle l}}{H_{\Lambda}} \int_{0}^{z_{\Lambda}} \frac{dz'}{\sqrt{\Omega_m (1+z')^{3} + \Omega_\Lambda}}
\;=\;
\frac{c_{\scriptscriptstyle l}}{H_{q}} \ln\!\left( \frac{1 + z_{h}}{1 + z_{\Lambda}} \right).
\label{eq21}
\end{equation}
This equation establishes the relationship between $z_{\Lambda}$ and the other parameters: the observed hybrid redshift $z_{h}$, the density parameters $\Omega_m$ and $\Omega_\Lambda$, and the associated rate constants $H_{\Lambda}$ and $H_{q}$.

\subsubsection{Luminosity-distance Modulus and Regression Convention}
\label{2.3.5}
Under the assumption that the observed cosmic redshift is entirely induced by cosmic expansion---the flat $\Lambda$CDM interpretation---the distance modulus is obtained from the observed redshift $z_{obs}$,
\begin{equation}
\mu_{\Lambda}(z_{obs}) = 5 \log_{10} \left[ \frac{d_{\Lambda}(z_{obs})}{\rm Mpc} \right] + 25.
\label{eq22}
\end{equation}
In the hybrid framework, the distance modulus is derived from the hybrid luminosity distance $d_{h}$,
\begin{subequations}
\begin{align}
\mu_{h}(z_{h}) &= 5 \log_{10} \left[ \frac{d_{h}(z_{h})}{\rm Mpc} \right] + 25\\[1.5ex]
&= \mu_{\Lambda}(z_{\Lambda}) + \frac{5}{2\ln 10}\,\ln\!\left(1+z_q\right).
\label{eq23}
\end{align}
\end{subequations}
In the redshift duality framework, the observed redshift is equivalent to the hybrid redshift, so $z_{obs}$ and $z_{h}$ are used interchangeably.

Analyzing observed redshifts that contain a WMC component under the assumption that they are entirely due to cosmic expansion can lead to a biased inference of the distance--redshift relation. Specifically, under the operational assumption of DMC, the observational dataset exhibits redshifts that appear elevated relative to the restored distance moduli. Consequently, when a standard metric-only model is regressed against such data, if a nuisance parameter $M$ is introduced to capture modulus residuals, this parameter effectively functions as a \textit{proxy variable}. In this context, $M$ absorbs the apparent modulus discrepancy relative to the observed redshift:
\begin{subequations}
\begin{align}
&\mu^{{\rm reg}}_{\Lambda}(z_{obs}) = \mu_{\Lambda}(z_{obs}) + M, \label{eq24a}\\[2.0ex]
&\mu^{{\rm reg}}_{h}(z_{h}) = \mu_{h}(z_{h}) + M.\label{eq24b}
\end{align}
\end{subequations}

Conversely, the hybrid framework explicitly separates the observed redshift into metric and non-metric components. Under the operational premise that DMC effectively restores the tracer's intrinsic luminosity to its true physical value, the Hybrid model is expected to recover the underlying physical distance--redshift relation. Consequently, we anticipate that the regression will yield a value of $M \approx 0$. Including $M$ in the regression formalism is therefore definitionally essential to verify this prediction and detect any residual offsets.

Considering that the distance ladder technique applied to the Pantheon+SH0ES compilation~\citepalias{Brout2022, Scolnic2022, Riess2022} employs distinct tracers for absolute magnitude calibration and for extending the inference into the Hubble flow, and consistent with the operational definition that the WMC attenuation rate is tracer-specific, a rigorous formulation of the regression modulus would formally include additional WMC-dependent terms beyond the expressions above. Nevertheless, the simplified formulation remains effective for validating the hybrid framework. The mathematical validity of the modulus formulation adopted for this regression analysis is discussed in Appendix~\ref{AppA}.

\subsection{Diagnostic Index}
\label{2.4}
When the WMC contribution is not explicitly separated from the observed redshift, a flat $\Lambda$CDM-only interpretation can systematically overestimate the inferred intrinsic luminosity for a source at fixed observed redshift and flux. We quantify the resulting divergence between the hybrid framework and the metric-only inference by defining a set of diagnostic indices constructed from combinations of luminosity (or related photometric quantities) and redshift.

\subsubsection{Redshift and Stretch}
\label{2.4.1}
To quantify the fractional contribution of the metric expansion versus the WMC attenuation for a specific tracer $\kappa$, we define two diagnostic ratios. The Redshift Ratio quantifies the fraction of the total observed redshift that is assigned to the metric component,
\begin{equation}
\mathrm{RR}(z_{h}) \equiv \frac{z_{\Lambda}}{z_{h}}.
\label{eq25}
\end{equation}
An $\mathrm{RR}\approx 1$ indicates that the observed redshift is largely accounted for by the metric component, while $\mathrm{RR}<1$ indicates a non-metric contribution.

More fundamentally, we define the Metric Stretch Ratio, $\mathrm{SR}$, which compares the metric stretch to the total observed stretch,
\begin{equation}
\mathrm{SR}(z_{h}) \equiv \frac{1+z_{\Lambda}}{1+z_{h}} = \frac{1}{1+z_{q}}.
\label{eq26}
\end{equation}
This ratio measures the fraction of the total wavelength elongation that is attributable to metric expansion; as shown below, it coincides with the time-dilation ratio.

\subsubsection{Luminosity Distance}
\label{2.4.2}
The Luminosity Distance Ratio (LDR) is defined as the ratio between the luminosity distance inferred in the hybrid framework---where the WMC contribution is explicitly separated from the observed redshift---and the distance inferred in the flat $\Lambda$CDM framework, where the entire observed redshift is attributed to cosmic expansion:
\begin{subequations}
\begin{align}
\mathrm{LDR}(z_{h}) \equiv \frac{d_{h}(z_{h})}{d_{\Lambda}(z_{h})}
&= \sqrt{\frac{1+z_{\Lambda}}{1+z_{h}}}\cdot \frac{X_{\Lambda}(z_{\Lambda})}{X_{\Lambda}(z_{h})}\label{eq27a}\\[1.0ex]
&= \sqrt{\mathrm{SR}(z_h)}\cdot \frac{X_{\Lambda}(z_{\Lambda})}{X_{\Lambda}(z_{h})}.\label{eq27b}
\end{align}
\end{subequations}
By construction, $\mathrm{LDR}(z_h)=1$ when $z_{\Lambda}=z_h$. Departures from unity quantify how a non-metric contribution would bias flux-based distance inference under the flat $\Lambda$CDM framework.

Using the inverse-square scaling $F\propto d^{-2}$ for a source of fixed intrinsic luminosity (within the same tracer-defined photometric channel), we define the Diagnostic Flux Ratio (FR) as
\begin{equation}
\mathrm{FR}(z_{h}) \equiv \frac{F_{h}(z_{h})}{F_{\Lambda}(z_{h})}
= \left[ \frac{d_{\Lambda}(z_{h})}{d_{h}(z_{h})} \right]^2
= \left[ \mathrm{LDR}(z_{h}) \right]^{-2}.
\label{eq28}
\end{equation}

\subsubsection{Mass Inference}
\label{2.4.3}
To quantify the impact of the hybrid redshift decomposition on flux-based mass inferences for high-redshift sources, we formulate the rescaling relation for stellar mass within the channel-specific WMC framework. Under a fixed mass-to-light ratio (and otherwise identical population-synthesis assumptions), the inferred intrinsic stellar mass scales approximately as $M_* \propto d^2$, motivating the Mass Ratio (MR):
\begin{equation}
\mathrm{MR}(z_{h}) \equiv \frac{M_{h}(z_{h})}{M_{\Lambda}(z_{h})}
= \left[ \frac{d_{h}(z_{h})}{d_{\Lambda}(z_{h})} \right]^2
= \left[ \mathrm{LDR}(z_{h}) \right]^2.
\label{eq29}
\end{equation}

If $\mathrm{MR}<1$, a metric-only interpretation at fixed observed $z_h$ implies an overestimation of stellar masses relative to the hybrid framework. Furthermore, this mass rescaling can propagate into number-density inferences, because sample completeness is typically mass-dependent and the conversion from counts to densities depends on the differential comoving volume element; in the hybrid framework, the latter is evaluated at the remapped metric coordinate $z_{\Lambda}(z_h)$ rather than at the total observed redshift $z_h$.

\subsubsection{Angular Diameter Distance}
\label{2.4.4}
For a source with the total redshift $z_h$, we define the angular-diameter distance under the standard metric-only inference and under the hybrid framework as
\begin{equation}
D_{\Lambda}(z_{h}) = \frac{X_{\Lambda}(z_{h})}{1 + z_{h}}, \qquad D_{h}(z_{h}) = \frac{X_{\Lambda}(z_{\Lambda})}{1 + z_{\Lambda}} \label{eq30}
\end{equation}
where $D_{\Lambda}$ evaluates the geometric distance at the total observed redshift $z_h$, whereas $D_h$ evaluates the same geometric quantity at the remapped metric component $z_{\Lambda}(z_h)$ implied by the hybrid decomposition. 

The Angular Diameter Distance Ratio (ADDR) is then
\begin{equation}
\mathrm{ADDR}(z_{h}) \equiv \frac{D_{h}(z_{h})}{D_{\Lambda}(z_{h})}
= \frac{1 + z_{h}}{1 + z_{\Lambda}} \cdot \frac{X_{\Lambda}(z_{\Lambda})}{X_{\Lambda}(z_{h})}
= \frac{1}{\mathrm{SR}(z_h)}\cdot \frac{X_{\Lambda}(z_{\Lambda})}{X_{\Lambda}(z_{h})}.
\label{eq31}
\end{equation}

\subsubsection{Local Age and Time Dilation}
\label{2.4.5}

For a source observed at total redshift $z_h$, the cosmic age may be inferred either by treating $z_h$ as purely metric, or by using the remapped metric component $z_{\Lambda}(z_h)$ obtained after explicitly separating the WMC contribution. The resulting difference and ratio between these two age inferences can be formalized as the Age Gain and Age Ratio diagnostics. To this end, we first define the standard cosmic time integral $t_{age}(z)$ in a flat $\Lambda$CDM model with expansion rate $H_{\Lambda}$ as
\begin{equation}
t_{age}(z) = \frac{1}{H_{\Lambda}} \int_{z}^{\infty} \frac{dz'}{(1+z')\sqrt{\Omega_{\rm m}(1+z')^3+\Omega_{\Lambda}}}.
\label{eq32}
\end{equation}
Interpreting this time integral as the cosmic age at redshift $z$ in the flat $\Lambda$CDM baseline, we define the Age Gain (AG) and Age Ratio (AR) diagnostics for a source observed at total redshift $z_h$ as
\begin{subequations}
\begin{align}
\mathrm{AG}(z_{h}) &\equiv t_{age}\!\left(z_{\Lambda}(z_h)\right) - t_{age}(z_{h}), \label{eq34.a}\\[1.2ex]
\mathrm{AR}(z_{h}) &\equiv \frac{t_{age}\!\left(z_{\Lambda}(z_h)\right)}{t_{age}(z_{h})}. \label{eq34.b}
\end{align}
\end{subequations}

Analogously, we compare how the duration of an arbitrary physical process is mapped under the two interpretations. Let $\mathcal{P}_{\rm real}$ denote the rest-frame period of the phenomenon. If the entire observed redshift $z_{h}$ is interpreted as metric, the predicted duration is
\begin{equation}
\mathcal{P}_{\Lambda}(z_{h}) \equiv (1+z_{h})\,\mathcal{P}_{\rm real}.
\label{eq35}
\end{equation}
In the hybrid mapping, only the metric component generates cosmological time dilation, so
\begin{equation}
\mathcal{P}_{h}(z_{h}) \equiv (1+z_{\Lambda})\,\mathcal{P}_{\rm real} = \frac{1+z_{h}}{1+z_{q}}\,\mathcal{P}_{\rm real}.
\label{eq36}
\end{equation}The Time-dilation Ratio, TR, is then
\begin{equation}
\mathrm{TR}(z_{h}) \equiv \frac{\mathcal{P}_{h}(z_{h})}{\mathcal{P}_{\Lambda}(z_{h})}= \frac{1+z_{\Lambda}}{1+z_{h}}= \frac{1}{1+z_{q}} = \mathrm{SR}(z_h).
\label{eq37} 
\end{equation}

Since $z_q > 0$ implies $\mathrm {TR}(z_h) < 1$, metric-only models systematically overestimate the true kinematic time dilation, which is observationally calibrated against the $(1+z_{\rm obs})$ broadening of SNe Ia light curves \citep{Riess1997, Goldhaber2001}. For SNe~Ia, this biases light-curve widths, potentially inducing a redshift-dependent drift in the stretch parameter $x_1$~\citep{Phillips1993, Guy2007}. However, provided $\mathrm{TR} \approx 1$, the intrinsic rest-frame timescales are accurately preserved, leaving the period--luminosity and stretch--luminosity relations intact (see Section~\ref{3.4}). Consequently, examining $\mathrm{TR}(z_h)$ is essential to verify that the Pantheon+SH0ES~\citepalias{Brout2022, Riess2022} distance moduli remain robust under our operational framework.

\subsubsection{Etherington Distance Duality and Tolman Surface Brightness}
\label{2.4.6}
In the hybrid framework with redshift duality, the Etherington distance duality~\citep{Etherington1933} and the Tolman surface-brightness relation~\citep{Tolman1930} are modified through the altered reciprocity between luminosity distance and angular-diameter distance: the standard relation $d=(1+z)^2D$ is replaced by
\begin{equation}
d_{h}(z_h) = (1+z_{h})^2\,D_{h}(z_h)\,(1+z_{q})^{-3/2}.
\label{eq38}
\end{equation}
We then define the Etherington Distance Ratio (EDR) as
\begin{equation}
\mathrm{EDR}(z_{h}) \equiv \frac{d_{h}(z_h)}{(1+z_{h})^2\,D_{h}(z_h)}
= (1+z_q)^{-3/2}
= \bigl[\mathrm{SR}(z_h)\bigr]^{3/2}.
\label{eq39}
\end{equation}

The Tolman surface brightness test evaluates the received flux per unit solid angle, $I$. While standard metric-only inference attributes the $(1+z)^{-4}$ dimming law to the total observed redshift $z_h$, the hybrid framework assigns geometric dimming solely to the metric component $z_\Lambda$, with an additional flux suppression arising from the WMC mechanism. The standard prediction ($I_{\Lambda}$), the hybrid prediction ($I_{h}$), and the resulting Tolman Surface Brightness Ratio (TSBR) are therefore formulated as:
\begin{subequations}
\begin{align}
&I_{\Lambda}(z_h) \propto (1+z_h)^{-4}, \label{eq40a}\\[1.0ex]
&I_{h}(z_h) \propto (1+z_\Lambda)^{-4}\,(1+z_q)^{-1}, \label{eq40b}\\[1.0ex]
&\mathrm{TSBR}(z_h) \equiv \frac{I_h(z_h)}{I_\Lambda(z_h)}
= (1+z_q)^3
= \bigl[\mathrm{SR}(z_h)\bigr]^{-3}. \label{eq40c}
\end{align}
\end{subequations}

While these formulations seemingly conflict with standard Etherington reciprocity and Tolman dimming, such deviations are not fundamental. Since WMC is operationally defined as a wavelength-dependent mechanism, any apparent violation is channel-specific rather than universal. In this sense, EDR and TSBR serve not to indicate a breakdown of spacetime geometry, but to quantify the non-metric attenuation superposed on the standard geometric inference for each observational tracer.

\subsection{Benchmark Models: Dynamical Dark Energy}
\label{2.5}
For comparison, we adopt the CPL parameterization for dynamical dark energy (DDE) \citep{Chevallier2001, Linder2003}, defined as $w(z) = w_0 + w_a z/(1+z)$. With $w_0$ and $w_a$ characterizing the present-day state and time evolution respectively, the corresponding line-of-sight comoving distance is
\begin{subequations}
\begin{align}
X_{\delta}(z)
&= \frac{c_{\scriptscriptstyle l}}{H_{\delta}}
\int_{0}^{z}
\frac{dz'}{\sqrt{
\Omega_m (1+z')^{3}
+ \Omega_\Lambda\,W(z')
}},
\label{eq42.a}\\[1.8ex]
W(z')
&=(1+z')^{3(1+w_0+w_a)}
\exp\!\left(-\frac{3 w_a z'}{1+z'}\right),
\label{eq42.b}
\end{align}
\end{subequations}
where $H_{\delta}$ is the present-day expansion rate and $\Omega_\Lambda = 1 - \Omega_m$ assumes spatial flatness. The standard $w$CDM model is recovered in the limit $w_a \to 0$. The luminosity distance and regression distance modulus are then operationally defined as
\begin{subequations}
\begin{align}
    &d_{\delta}(z) = (1+z)\,X_{\delta}(z),\label{eq43.a}\\[1.3ex]
    & \mu_{\delta}(z)
    = 5\log_{10}\!\left[\frac{d_{\delta}(z)}{\mathrm{Mpc}}\right] + 25, \label{eq43.b}\\[1.3ex]
    &\mu^{\rm reg}_{\delta}(z) = \mu_{\delta}(z) + M.\label{eq43.c}
\end{align}
\end{subequations}

\begin{table}
\setlength{\tabcolsep}{7pt}
\centering
\caption{Thirty configurations used in the regression analysis. The models are categorized into five families: (i) flat $\Lambda$CDM configurations ($\Lambda1$--$\Lambda6$); (ii) hybrid configurations with a free non-metric rate $H_q$ ($Q1$--$Q6$); (iii) hybrid configurations with a best-fit prior on $H_q$ ($Qa$--$Qf$); (iv) $w$CDM benchmark configurations ($w1$--$w6$); and (v) CPL benchmark configurations ($C1$--$C6$).}
\label{tab1}
{\fontsize{7.0pt}{9pt}\selectfont
\begin{tabular}{c c c c c c c c}
\Xhline{0.9pt}
ID & $M$ & $\Omega_m$ & $H_\Lambda$ or $H_{\delta}$& $H_q$ & $w$ & $w_0$ & $w_a$ \\
\Xhline{0.9pt}
$\Lambda1$ & 0    & 0.315 & 67.4 & --    & --   & --   & -- \\
$\Lambda2$ & 0    & 0.315 & free & --    & --   & --   & -- \\
$\Lambda3$ & 0    & free  & 67.4 & --    & --   & --   & -- \\
$\Lambda4$ & 0    & free  & free & --    & --   & --   & -- \\
$\Lambda5$ & free & 0.315 & 67.4 & --    & --   & --   & -- \\
$\Lambda6$ & free & free  & 67.4 & --    & --   & --   & -- \\
\hline
Q1 & 0    & 0.315 & 67.4 & free  & --   & --   & -- \\
Q2 & 0    & 0.315 & free & free  & --   & --   & -- \\
Q3 & 0    & free  & 67.4 & free  & --   & --   & -- \\
Q4 & 0    & free  & free & free  & --   & --   & -- \\
Q5 & free & 0.315 & 67.4 & free  & --   & --   & -- \\
Q6 & free & free  & 67.4 & free  & --   & --   & -- \\
\hline
Qa & 0    & 0.315 & 67.4 & prior & --   & --   & -- \\
Qb & 0    & 0.315 & free & prior & --   & --   & -- \\
Qc & 0    & free  & 67.4 & prior & --   & --   & -- \\
Qd & 0    & free  & free & prior & --   & --   & -- \\
Qe & free & 0.315 & 67.4 & prior & --   & --   & -- \\
Qf & free & free  & 67.4 & prior & --   & --   & -- \\
\hline
w1 & 0    & 0.315 & 67.4 & --  & free   & --   & -- \\
w2 & 0    & 0.315 & free & --  & free   & --   & -- \\
w3 & 0    & free  & 67.4 & --  & free   & --   & -- \\
w4 & 0    & free  & free & --  & free   & --   & -- \\
w5 & free & 0.315 & 67.4 & --  & free   & --   & -- \\
w6 & free & free  & 67.4 & --  & free   & --   & -- \\
\hline
C1 & 0    & 0.315 & 67.4 & --  & --   & free   & free \\
C2 & 0    & 0.315 & free & --  & --   & free   & free \\
C3 & 0    & free  & 67.4 & --  & --   & free   & free \\
C4 & 0    & free  & free & --  & --   & free   & free \\
C5 & free & 0.315 & 67.4 & --  & --   & free   & free \\
C6 & free & free  & 67.4 & --  & --   & free   & free \\
\Xhline{0.9pt}
\end{tabular}
}
\end{table}

\subsection{Validation of the Redshift Duality}
\label{2.6}

\subsubsection{Regression Configuration}
\label{2.6.1}
We adopt the working premise that the expansion of our Universe is well described by a spatially flat $\Lambda$CDM framework, and that the Planck-inferred parameters $(H_0,\Omega_m)$ represent the physically realized metric-expansion sector. We further assume that the Pantheon+SH0ES compilation, under the operational DMC prescription, encodes tracer intrinsic luminosities in a manner that is close to their physical values. In addition, we expect that the hybrid framework can separate, within the observed redshift, the contribution induced by WMC from the contribution attributable to metric expansion. If so, then when the hybrid model is regressed against the Pantheon+SH0ES dataset, it should either (i) remain competitive while coexisting with the Planck-anchored baseline model, or (ii) recover the Planck-based values through the regression itself. In either case, the magnitude offset $M$ should regress toward $M\simeq 0$, or fixing $M=0$ a priori should yield a preferred (or at least not strongly disfavored) fit under the adopted model-selection criterion.

To test these expectations, we consider the regression hybrid distance modulus, $\mu^{\mathrm{reg}}_{h}$, defined above and treat the expansion rate parameter $H_\Lambda$ (or $H_\delta$), $\Omega_m$, and $M$ either as fixed priors or as free regression parameters. Since each parameter admits two choices (fixed versus free), the full design space comprises $2^3=8$ configurations. However, because the expansion rate ($H_\Lambda$ or $H_\delta$) and $M$ enter the Hubble-diagram intercept in a strongly degenerate manner, the two configurations in which both $M$ and $H$ are simultaneously left free are excluded as ill-conditioned. The remaining six configurations define the regression setups summarized in Table~\ref{tab1}.

\subsubsection{Treatment of the Pantheon+SH0ES Dataset}
\label{2.6.2}
We adopt the Pantheon+SH0ES dataset \citepalias{Brout2022, Scolnic2022} in its original form, using the released distance moduli and associated uncertainty products without modification. The released \texttt{Pantheon+SH0ES.dat} file contains 1701 entries spanning $0.00122 \le z_{\mathrm{CMB}} \le 2.26130$. We evaluate the Hubble-diagram fits under three covariance configurations (diagonal-only, statistical covariance, and statistical+systematic covariance) to test robustness to correlated uncertainties and to assess whether any distance-correlated covariance could partially absorb a distance-proportional WMC signature.

Standard analyses typically truncate the local regime (e.g., excluding $z < 0.0233$ \citepalias{Riess2022} or $z < 0.01$ \citepalias{Freedman2025}) to mitigate peculiar-velocity systematics. We depart from this practice and perform a global regression over the entire Pantheon+SH0ES redshift range. Empirically, this choice aligns with the finding by \citet{Scolnic2025} that the Hubble-tension signal is already ``baked-in'' at local cluster scales ($z\approx 0.023$). Theoretically, because our hybrid framework parameterizes the WMC-induced redshift as a path-accumulated effect, it predicts a positive Hubble residual at low redshifts (Appendix~\ref{AppB}; equations~\eqref{eqAppB3}--\eqref{eqAppB5}). We therefore explicitly retain these low-redshift data to securely isolate the anchor baseline.

\subsubsection{Numerical Implementation Details of Global Regression}
\label{2.6.3}
All Hubble--diagram fits are performed via Bayesian Markov chain Monte Carlo (MCMC) sampling using the affine-invariant ensemble sampler \texttt{emcee}. We consider three covariance configurations: (i) \texttt{D0}, a diagonal-only mode in which the likelihood reduces to a WLS form; (ii) \texttt{ST}, which uses the released statistical covariance; and (iii) \texttt{SS}, which uses the full covariance including systematics. In the \texttt{ST} and \texttt{SS} modes, the likelihood is evaluated as a GLS by Cholesky whitening of the residuals. The covariance sub-matrix for the selected redshift range is symmetrized and, if needed, regularized with a small adaptive diagonal jitter to ensure positive definiteness. To account for residual intrinsic scatter or unmodelled variance, we optionally include an intrinsic dispersion term ($\sigma_{\text {int}}$): in \texttt{D0} it is added in quadrature to the diagonal uncertainties, while in GLS modes it is added to the diagonal of the covariance matrix prior to inversion. To avoid overfitting, this intrinsic dispersion term is dynamically tuned.

\texttt{scipy.integrate.quad} is employed to evaluate cosmological line-of-sight integrals with tolerances \texttt{epsrel}=$10^{-10}$ and \texttt{epsabs}=$10^{-12}$. For the flat $\Lambda$CDM baseline, comoving distances are precomputed with cubic-spline interpolation to accelerate sampling, whereas extended models (e.g., $w$CDM, CPL) use direct integration. For the hybrid family, the mapping from observed redshift to the metric component is solved numerically using Brent’s method, with memoization to reduce overhead. 

Gaussian external constraints are implemented as multiplicative priors in the posterior. Model comparison metrics (AIC, BIC) are computed based on the maximum log-likelihood of the data term. Posterior summaries report median values and central credible intervals derived from the converged MCMC chains, capturing the full non-Gaussian structure of the parameter space. 

We sample posteriors with an affine-invariant ensemble MCMC (emcee) using 40 walkers and 7000 steps per walker, discarding the first 1500 steps as burn-in and applying no thinning (thin= 1). Convergence is assessed using integrated autocorrelation times and stability of the marginal posteriors across multiple runs initialized with different seeds (see Appendix~\ref{AppE} for details on the deterministic pre-fit and walker initialization procedure); the reported chains use a fixed seed (19770301) for reproducibility.

\subsubsection{Binned Regressions and Linearity Tests of $H_{\Lambda}$}
\label{2.6.4}

We sort the Pantheon+SH0ES sample by $z_{\mathrm{CMB}}$ and construct redshift bins via equal-count partitioning (see Appendix~\ref{AppD}), repeating the analysis for $N_{\rm bin}\in\{2,4,6,8\}$ to ensure robustness. With the full sample size of \textit{N}=1701, this partitioning yields subsamples of nearly identical size: \{851,850\} for $N_{\rm bin}=2$; $\{426,425,425,425\}$ for $N_{\rm bin}=4$; $\{284,284,284,283,283,283\}$ for $N_{\rm bin}=6$; and $\{213,213,213,213,213,212,212,212\}$ for $N_{\rm bin}=8$. Within each bin, we define the representative redshift coordinate, $z_{\rm center}$, as the median of the constituent $z_{\rm CMB}$ values. For each bin we infer the metric Hubble parameter $H_{\Lambda}$ by sampling the one-dimensional MCMC posterior, where WLS or GLS (depending on the covariance mode) specifies the bin-wise $\chi^2$. Because only $H_{\Lambda}$ is free in each bin after fixing the remaining settings, convergence is rapid; we initialize the sampler at the deterministic WLS/GLS minimizer to accelerate mixing and to provide a consistency check. 

By fixing the nuisance parameters $M$ and $\Omega_m$ to their reference values (e.g., $M=0, \Omega_m=0.315$ for the baseline), we isolate the linearity test from parameter degeneracies within small bins. This constraint effectively reduces the per-bin inference to a single-parameter problem where the likelihood is well-approximated by a Gaussian. In this regime, deterministic minimization yields parameter estimates and uncertainties statistically equivalent to those from full posterior sampling, but with significantly reduced computational cost. 

The null hypothesis implies that the metric mapping $\mu_{\Lambda}(z)$ yields a redshift-independent $H_{\Lambda}$ across bins. Denoting by $H_{\Lambda,n}$ the best-fit value inferred in the $n$th redshift bin, we treat any monotonic trend of $H_{\Lambda,n}$ with bin redshift as an operational signature of unmodeled dimming. We quantify this by a weighted linear regression of $H_{\Lambda,n}$ on bin redshift, using a heteroscedasticity-consistent (HC3) covariance estimator to account for differing uncertainties between bins.

To test whether the apparent $H_{\Lambda}(z)$ drift is specific to the metric-only interpretation, we repeat the same per-bin minimization within the hybrid framework. In this step, we impose a Gaussian prior on the non-metric rate $H_q$ anchored by the global Q1 posterior summary and then fit $H_{\Lambda}$ per bin, holding the remaining settings fixed (corresponding to the Qb in Table~\ref{tab1}), to assess whether the inferred expansion rate stabilizes at the Planck-based reference value.

\subsubsection{Model Selection with the BIC}
\label{2.6.5}
To compare the metric-only framework to the hybrid extension, we use the BIC~\citep{Schwarz1978}, defined as $\mathrm{BIC}=k\ln N-2\ln(\hat{\mathcal{L}})$, which balances goodness of fit against the number of parameters estimated from the Pantheon+SH0ES data. Here, $N$ is the number of SNe used in the fit (e.g., 1701 for Pantheon+SH0ES), $k$ counts only parameters that are actually fitted in that configuration, and $\hat{\mathcal{L}}$ denotes the maximum likelihood achieved by that configuration under the adopted covariance mode. 

Our regression setup includes fixed-reference configurations in which selected cosmological parameters are held to external values (e.g., Planck best-fits); such fixed parameters are not counted in $k$. When $k=0$, the run should be read as a test of a fully specified fixed hypothesis: Pantheon+SH0ES is used only to evaluate the residuals under the adopted covariance model, so the BIC reflects how well that fixed hypothesis reproduces the SN~Ia Hubble diagram. Comparing this $k=0$ benchmark to other configurations that add one or more fitted parameters provides a simple complexity--fit trade-off within the same covariance specification: it asks whether freeing the non-metric term improves the fit enough to justify introducing the additional parameter(s).

\begin{table*}
\setlength{\tabcolsep}{6.75pt}
\centering
\caption{
Global regression results for unbinned fits.
All modes are sampled with MCMC; the data treatments differ by the assumed covariance structure:
D0 uses diagonal-only uncertainties (equivalent to WLS),
whereas ST/SS use GLS with the full statistical covariance (ST) and the full statistical+systematic covariance (SS).
We track the numerical stability of the $z_\Lambda$ inversion via $N_{\rm inv}$, the total number of root-solve calls accumulated over the runs (chains in each mode), summed over the configuration set in each mode (18 configurations per mode; 54 runs total).
Solver outcomes are tallied in the order (brentq\_ok, secant\_ok, bracket\_fail, nan\_eval, fallback\_mid).
All inversions succeeded with \texttt{brentq} (no boundary-contact in any run).
By mode: \textbf{D0} $N_{\rm inv}=4{,}826{,}376{,}653$, outcome $(4{,}826{,}376{,}653,0,0,0,0)$;
\textbf{ST} $N_{\rm inv}=4{,}849{,}373{,}543$, outcome $(4{,}849{,}373{,}543,0,0,0,0)$, mean acceptance $0.692$;
\textbf{SS} $N_{\rm inv}=4{,}839{,}422{,}994$, outcome $(4{,}839{,}422{,}994,0,0,0,0)$, mean acceptance $0.693$.
RMSE (Root Mean Square Error) and MAE (Mean Absolute Error) are quoted in magnitudes $[\rm mag]$.
}

\label{tab3}
{\fontsize{7pt}{10.7}\selectfont
\begin{tabular}{c c c c c c c c c c c}
\Xhline{0.9pt}

ID & $M\,[\mathrm{mag}]$ & $\Omega_m$ & $H_\Lambda~[\mathrm{km\,s^{-1}\,Mpc^{-1}}]$ & $H_q~[\mathrm{km\,s^{-1}\,Mpc^{-1}}]$ & $R^2$ & $\chi^2/{\rm DOF}$ & RMSE & MAE & AIC & BIC \\

\Xhline{0.9pt}

D0-$\Lambda1$ & 0.000 (fixed) & 0.315 (fixed) & 67.400 (fixed) & -- & 0.997 & 1.001 & 0.252 & 0.204 & 1703.000 & \textbf{1708.439} \\
D0-$\Lambda2$ & 0.000 (fixed) & 0.315 (fixed) & $\mathbf{73.090 \pm 0.171}$ & -- & 0.997 & 0.485 & 0.190 & 0.132 & 826.556 & 831.995 \\
D0-$\Lambda3$ & 0.000 (fixed) & $\mathbf{0.732 \pm 0.017}$ & 67.400 (fixed) & -- & 0.996 & 0.690 & 0.213 & 0.156 & 1174.811 & 1180.250 \\
D0-$\Lambda4$ & 0.000 (fixed) & $0.381 \pm 0.020$ & $\mathbf{72.392 \pm 0.264}$ & -- & 0.997 & 0.478 & 0.189 & 0.131 & 816.613 & 827.491 \\
D0-$\Lambda5$ & $\mathbf{-0.176 \pm 0.005}$ & 0.315 (fixed) & 67.400 (fixed) & -- & 0.997 & 0.485 & 0.190 & 0.132 & 826.556 & 831.995 \\
D0-$\Lambda6$ & $\mathbf{-0.155 \pm 0.008}$ & $0.380 \pm 0.020$ & 67.400 (fixed) & -- & 0.997 & 0.478 & 0.189 & 0.131 & 816.613 & 827.491 \\

\Xhline{0.6pt}

D0-Q1 & 0.000 (fixed) & 0.315 (fixed) & 67.400 (fixed) & $\mathbf{4.881 \pm 0.147}$ & 0.997 & 0.477 & 0.189 & 0.131 & 813.311 & \textbf{818.750} \\
D0-Q2 & 0.000 (fixed) & 0.315 (fixed) & $\mathbf{67.438 \pm 1.547}$ & $\mathbf{4.853 \pm 1.321}$ & 0.997 & 0.478 & 0.189 & 0.131 & 815.312 & 826.190 \\
D0-Q3 & 0.000 (fixed) & $\mathbf{0.310 \pm 0.024}$ & 67.400 (fixed) & $4.925 \pm 0.257$ & 0.997 & 0.478 & 0.189 & 0.131 & 815.270 & 826.148 \\
D0-Q4 & 0.000 (fixed) & $\mathit{0.166 \pm 0.100}$ & $\mathit{60.230 \pm 4.576}$ & $\mathit{11.941 \pm 4.469}$ & 0.997 & 0.477 & 0.189 & 0.131 & 816.476 & 832.793 \\
D0-Q5 & $\mathbf{0.002 \pm 0.050}$ & 0.315 (fixed) & 67.400 (fixed) & $4.964 \pm 1.435$ & 0.997 & 0.478 & 0.189 & 0.131 & 815.326 & 826.204 \\
D0-Q6 & $\mathit{0.315 \pm 0.312}$ & $\mathit{0.156 \pm 0.099}$ & 67.400 (fixed) & $\mathit{16.792 \pm 14.938}$ & 0.997 & 0.498 & 0.192 & 0.135 & 851.909 & 868.226 \\

\Xhline{0.6pt}

D0-Qa & 0.000 (fixed) & 0.315 (fixed) & 67.400 (fixed) & 4.881 (Prior) & 0.997 & 0.477 & 0.189 & 0.131 & 813.311 & \textbf{818.750} \\
D0-Qb & 0.000 (fixed) & 0.315 (fixed) & $\mathbf{67.403 \pm 0.242}$ & 4.881 (Prior) & 0.997 & 0.478 & 0.189 & 0.131 & 815.311 & 826.189 \\
D0-Qc & 0.000 (fixed) & $\mathbf{0.313 \pm 0.017}$ & 67.400 (fixed) & 4.881 (Prior) & 0.997 & 0.478 & 0.189 & 0.131 & 815.290 & 826.168 \\
D0-Qd & 0.000 (fixed) & $\mathbf{0.312 \pm 0.021}$ & $\mathbf{67.434 \pm 0.302}$ & 4.881 (Prior) & 0.997 & 0.478 & 0.189 & 0.131 & 817.279 & 833.596 \\
D0-Qe & $\mathbf{-0.000 \pm 0.007}$ & 0.315 (fixed) & 67.400 (fixed) & 4.881 (Prior) & 0.997 & 0.478 & 0.189 & 0.131 & 815.311 & 826.189 \\
D0-Qf & $\mathbf{-0.001 \pm 0.009}$ & $\mathbf{0.312 \pm 0.021}$ & 67.400 (fixed) & 4.881 (Prior) & 0.997 & 0.478 & 0.189 & 0.131 & 817.279 & 833.596 \\

\hline\hline

ST-$\Lambda1$ & 0.000 (fixed) & 0.315 (fixed) & 67.400 (fixed) & -- & 0.997 & 0.955 & 0.252 & 0.204 & 1624.957 & \textbf{1630.396} \\
ST-$\Lambda2$ & 0.000 (fixed) & 0.315 (fixed) & $\mathbf{72.998 \pm 0.144}$ & -- & 0.997 & 0.889 & 0.190 & 0.132 & 1513.060 & 1518.499 \\
ST-$\Lambda3$ & 0.000 (fixed) & $\mathbf{0.706 \pm 0.015}$ & 67.400 (fixed) & -- & 0.996 & 0.911 & 0.212 & 0.156 & 1550.688 & 1556.127 \\
ST-$\Lambda4$ & 0.000 (fixed) & $0.384 \pm 0.016$ & $\mathbf{72.257 \pm 0.218}$ & -- & 0.997 & 0.888 & 0.189 & 0.131 & 1513.221 & 1524.099 \\
ST-$\Lambda5$ & $\mathbf{-0.173 \pm 0.004}$ & 0.315 (fixed) & 67.400 (fixed) & -- & 0.997 & 0.889 & 0.190 & 0.132 & 1513.061 & 1518.500 \\
ST-$\Lambda6$ & $\mathbf{-0.151 \pm 0.007}$ & $0.384 \pm 0.016$ & 67.400 (fixed) & -- & 0.997 & 0.888 & 0.189 & 0.131 & 1513.221 & 1524.099 \\

\Xhline{0.6pt}

ST-Q1 & 0.000 (fixed) & 0.315 (fixed) & 67.400 (fixed) & $\mathbf{4.797 \pm 0.121}$ & 0.997 & 0.888 & 0.189 & 0.131 & 1511.797 & \textbf{1517.236} \\
ST-Q2 & 0.000 (fixed) & 0.315 (fixed) & $\mathbf{66.995 \pm 1.267}$ & $\mathbf{5.141 \pm 1.080}$ & 0.997 & 0.888 & 0.189 & 0.131 & 1513.053 & 1523.931 \\
ST-Q3 & 0.000 (fixed) & $\mathbf{0.317 \pm 0.020}$ & 67.400 (fixed) & $4.784 \pm 0.213$ & 0.997 & 0.888 & 0.189 & 0.131 & 1512.965 & 1523.843 \\
ST-Q4 & 0.000 (fixed) & $\mathit{0.150 \pm 0.095}$ & $\mathit{59.169 \pm 4.264}$ & $\mathit{12.844 \pm 4.161}$ & 0.997 & 0.889 & 0.189 & 0.131 & 1515.098 & 1531.415 \\
ST-Q5 & $\mathbf{0.014 \pm 0.041}$ & 0.315 (fixed) & 67.400 (fixed) & $5.199 \pm 1.184$ & 0.997 & 0.888 & 0.189 & 0.131 & 1513.062 & 1523.940 \\
ST-Q6 & $\mathit{0.354 \pm 0.300}$ & $\mathit{0.136 \pm 0.090}$ & 67.400 (fixed) & $\mathit{18.110 \pm 14.751}$ & 0.997 & 0.916 & 0.191 & 0.134 & 1560.743 & 1577.060 \\

\Xhline{0.6pt}

ST-Qa & 0.000 (fixed) & 0.315 (fixed) & 67.400 (fixed) & 4.797 (Prior) & 0.997 & 0.888 & 0.189 & 0.131 & 1511.797 & \textbf{1517.236} \\
ST-Qb & 0.000 (fixed) & 0.315 (fixed) & $\mathbf{67.391 \pm 0.200}$ & 4.797 (Prior) & 0.997 & 0.888 & 0.189 & 0.131 & 1513.144 & 1524.022 \\
ST-Qc & 0.000 (fixed) & $\mathbf{0.316 \pm 0.014}$ & 67.400 (fixed) & 4.797 (Prior) & 0.997 & 0.888 & 0.189 & 0.131 & 1512.965 & 1523.843 \\
ST-Qd & 0.000 (fixed) & $\mathbf{0.317 \pm 0.017}$ & $\mathbf{67.377 \pm 0.247}$ & 4.797 (Prior) & 0.997 & 0.890 & 0.189 & 0.131 & 1516.994 & 1533.310 \\
ST-Qe & $\mathbf{0.000 \pm 0.006}$ & 0.315 (fixed) & 67.400 (fixed) & 4.797 (Prior) & 0.997 & 0.888 & 0.189 & 0.131 & 1513.144 & 1524.022 \\
ST-Qf & $\mathbf{0.001 \pm 0.007}$ & $\mathbf{0.317 \pm 0.017}$ & 67.400 (fixed) & 4.797 (Prior) & 0.997 & 0.890 & 0.189 & 0.131 & 1516.997 & 1533.314 \\

\hline\hline

SS-$\Lambda1$ & 0.000 (fixed) & 0.315 (fixed) & 67.400 (fixed) & -- & 0.997 & 0.968 & 0.252 & 0.204 & 1646.847 & \textbf{1652.286} \\
SS-$\Lambda2$ & 0.000 (fixed) & 0.315 (fixed) & $\mathbf{72.974 \pm 0.140}$ & -- & 0.997 & 0.932 & 0.190 & 0.132 & 1587.075 & 1592.514 \\
SS-$\Lambda3$ & 0.000 (fixed) & $\mathbf{0.790 \pm 0.018}$ & 67.400 (fixed) & -- & 0.996 & 0.861 & 0.215 & 0.158 & 1465.478 & 1470.917 \\
SS-$\Lambda4$ & 0.000 (fixed) & $0.382 \pm 0.021$ & $\mathbf{72.271 \pm 0.255}$ & -- & 0.997 & 0.939 & 0.189 & 0.131 & 1599.929 & 1610.807 \\
SS-$\Lambda5$ & $\mathbf{-0.173 \pm 0.004}$ & 0.315 (fixed) & 67.400 (fixed) & -- & 0.997 & 0.932 & 0.190 & 0.132 & 1587.075 & 1592.514 \\
SS-$\Lambda6$ & $\mathbf{-0.152 \pm 0.008}$ & $0.382 \pm 0.021$ & 67.400 (fixed) & -- & 0.997 & 0.939 & 0.189 & 0.131 & 1599.929 & 1610.807 \\

\Xhline{0.6pt}

SS-Q1 & 0.000 (fixed) & 0.315 (fixed) & 67.400 (fixed) & $\mathbf{4.802 \pm 0.119}$ & 0.997 & 0.939 & 0.189 & 0.131 & 1598.757 & \textbf{1604.196} \\
SS-Q2 & 0.000 (fixed) & 0.315 (fixed) & $\mathbf{66.850 \pm 1.647}$ & $\mathbf{5.275 \pm 1.416}$ & 0.997 & 0.939 & 0.189 & 0.131 & 1600.086 & 1610.964 \\
SS-Q3 & 0.000 (fixed) & $\mathbf{0.314 \pm 0.026}$ & 67.400 (fixed) & $4.811 \pm 0.246$ & 0.997 & 0.939 & 0.189 & 0.131 & 1599.721 & 1610.598 \\
SS-Q4 & 0.000 (fixed) & $\mathit{0.131 \pm 0.091}$ & $\mathit{58.201 \pm 4.096}$ & $\mathit{13.794 \pm 3.976}$ & 0.997 & 0.940 & 0.189 & 0.131 & 1601.276 & 1617.593 \\
SS-Q5 & $\mathbf{0.022 \pm 0.055}$ & 0.315 (fixed) & 67.400 (fixed) & $5.458 \pm 1.593$ & 0.997 & 0.939 & 0.189 & 0.131 & 1600.131 & 1611.009 \\
SS-Q6 & $\mathit{0.356 \pm 0.192}$ & $\mathit{0.120 \pm 0.088}$ & 67.400 (fixed) & $\mathit{17.662 \pm 7.869}$ & 0.997 & 0.942 & 0.189 & 0.132 & 1605.494 & 1621.811 \\

\Xhline{0.6pt}

SS-Qa & 0.000 (fixed) & 0.315 (fixed) & 67.400 (fixed) & 4.802 (Prior) & 0.997 & 0.939 & 0.189 & 0.131 & 1598.757 & \textbf{1604.196} \\
SS-Qb & 0.000 (fixed) & 0.315 (fixed) & $\mathbf{67.392 \pm 0.196}$ & 4.802 (Prior) & 0.997 & 0.939 & 0.189 & 0.131 & 1600.190 & 1611.068 \\
SS-Qc & 0.000 (fixed) & $\mathbf{0.315 \pm 0.016}$ & 67.400 (fixed) & 4.802 (Prior) & 0.997 & 0.939 & 0.189 & 0.131 & 1599.722 & 1610.600 \\
SS-Qd & 0.000 (fixed) & $\mathbf{0.316 \pm 0.023}$ & $\mathbf{67.389 \pm 0.275}$ & 4.802 (Prior) & 0.997 & 0.941 & 0.189 & 0.131 & 1604.335 & 1620.652 \\
SS-Qe & $\mathbf{0.000 \pm 0.006}$ & 0.315 (fixed) & 67.400 (fixed) & 4.802 (Prior) & 0.997 & 0.939 & 0.189 & 0.131 & 1600.192 & 1611.070 \\
SS-Qf & $\mathbf{0.000 \pm 0.008}$ & $\mathbf{0.316 \pm 0.023}$ & 67.400 (fixed) & 4.802 (Prior) & 0.997 & 0.941 & 0.189 & 0.131 & 1604.334 & 1620.651 \\

\Xhline{0.9pt}
\end{tabular}
} 
\end{table*}

\begin{table*}
\centering
\caption{
Organized redshift-binned estimates of $H_\Lambda(z)$ from the equal-count partitioning analyses with $N_{\mathrm bin}\in\{2,4,6,8\}$ (20 bins total). Rows are grouped by the binning configuration. Columns report the covariance modes (D0, ST, SS) under the flat $\Lambda$CDM framework (i.e., the $\Lambda2$ configuration) and under the hybrid framework (i.e., the Qb configuration, with $H_q$ fixed to the global best-fit value for each mode shown in the column headers). In the tomographic setup, only $H_\Lambda$ is estimated per bin while the nuisance parameter $M$ is held fixed as configured. Entries are $H_\Lambda \pm 1\sigma$, where $1\sigma$ denotes the standard error from the WLS/GLS fit; all Hubble parameters reported in this table, including the fixed $H_q$ values in the column headers, are in km\,s$^{-1}$\,Mpc$^{-1}$.
}
\label{tab5}

\setlength{\tabcolsep}{8.0pt}
\renewcommand{\arraystretch}{1.5}
{\fontsize{7.2}{7.85}\selectfont
\begin{tabular}{c c c c c c c c}
\toprule
\multicolumn{1}{c}{} &
\multicolumn{3}{c}{Flat $\Lambda$CDM framework ($\Omega_m=0.315$, $H_q=0$)} &
\multicolumn{3}{c}{Hybrid framework ($H_q$ fixed to the best-fit value)} &
\multicolumn{1}{c}{} \\
\cmidrule(lr){2-4}\cmidrule(lr){5-7}
$z_{\rm center}$ &
D0 & ST & SS &
D0 ($H_q=\mathbf{4.881}$) & ST ($H_q=\mathbf{4.797}$) & SS ($H_q=\mathbf{4.802}$) &
Bin ID \\
\midrule

% --- 2 Bins ---
0.027210 & $72.537 \pm 0.183$ & $72.335 \pm 0.228$ & $72.415 \pm 0.254$ & $67.404 \pm 0.205$ & $67.307 \pm 0.255$ & $67.343 \pm 0.281$ & 2Bin-1 \\
0.328595 & $73.573 \pm 0.151$ & $73.575 \pm 0.151$ & $73.522 \pm 0.233$ & $67.400 \pm 0.180$ & $67.512 \pm 0.207$ & $67.505 \pm 0.271$ & 2Bin-2 \\
\midrule
% --- 4 Bins ---
0.016190 & $72.355 \pm 0.327$ & $72.119 \pm 0.429$ & $72.000 \pm 0.512$ & $\mathbf{67.382 \pm 0.340}$ & $\mathbf{67.238 \pm 0.444}$ & $\mathbf{67.110 \pm 0.525}$ & 4Bin-1 \\
0.048390 & $72.631 \pm 0.216$ & $72.648 \pm 0.222$ & $72.921 \pm 0.244$ & $\mathbf{67.418 \pm 0.233}$ & $\mathbf{67.524 \pm 0.254}$ & $\mathbf{67.769 \pm 0.274}$ & 4Bin-2 \\
0.243160 & $73.295 \pm 0.215$ & $73.316 \pm 0.215$ & $73.341 \pm 0.300$ & $\mathbf{67.423 \pm 0.237}$ & $\mathbf{67.545 \pm 0.259}$ & $\mathbf{67.558 \pm 0.333}$ & 4Bin-3 \\
0.495600 & $73.921 \pm 0.211$ & $73.897 \pm 0.211$ & $73.725 \pm 0.310$ & $\mathbf{67.365 \pm 0.231}$ & $\mathbf{67.453 \pm 0.253}$ & $\mathbf{67.327 \pm 0.338}$ & 4Bin-4 \\
\midrule
% --- 6 Bins ---
0.012815 & $71.081 \pm 0.464$ & $70.949 \pm 0.608$ & $70.716 \pm 0.705$ & $66.132 \pm 0.474$ & $66.090 \pm 0.619$ & $65.854 \pm 0.713$ & 6Bin-1 \\
0.027225 & $72.622 \pm 0.286$ & $72.752 \pm 0.308$ & $72.779 \pm 0.360$ & $67.606 \pm 0.299$ & $67.823 \pm 0.331$ & $67.844 \pm 0.379$ & 6Bin-2 \\
0.072485 & $73.044 \pm 0.257$ & $73.004 \pm 0.260$ & $73.157 \pm 0.286$ & $67.758 \pm 0.271$ & $67.809 \pm 0.289$ & $67.952 \pm 0.312$ & 6Bin-3 \\
0.217810 & $73.398 \pm 0.267$ & $73.402 \pm 0.267$ & $73.487 \pm 0.366$ & $67.613 \pm 0.285$ & $67.713 \pm 0.302$ & $67.798 \pm 0.393$ & 6Bin-4 \\
0.328600 & $73.253 \pm 0.246$ & $73.251 \pm 0.247$ & $73.327 \pm 0.310$ & $67.084 \pm 0.264$ & $67.188 \pm 0.284$ & $67.262 \pm 0.339$ & 6Bin-5 \\
0.579590 & $74.245 \pm 0.272$ & $74.222 \pm 0.272$ & $74.120 \pm 0.386$ & $67.486 \pm 0.290$ & $67.581 \pm 0.309$ & $67.491 \pm 0.410$ & 6Bin-6 \\
\midrule
% --- 8 Bins ---
0.009800 & $71.323 \pm 0.612$ & $71.007 \pm 0.799$ & $70.959 \pm 0.882$ & $66.387 \pm 0.621$ & $66.159 \pm 0.807$ & $66.108 \pm 0.889$ & 8Bin-1 \\
0.022450 & $72.857 \pm 0.360$ & $72.921 \pm 0.391$ & $72.887 \pm 0.456$ & $67.867 \pm 0.369$ & $68.018 \pm 0.409$ & $67.976 \pm 0.471$ & 8Bin-2 \\
0.033370 & $71.781 \pm 0.317$ & $71.800 \pm 0.332$ & $72.082 \pm 0.384$ & $66.730 \pm 0.329$ & $66.834 \pm 0.353$ & $67.111 \pm 0.404$ & 8Bin-3 \\
0.103380 & $73.407 \pm 0.285$ & $73.393 \pm 0.287$ & $73.532 \pm 0.321$ & $68.057 \pm 0.297$ & $68.134 \pm 0.314$ & $68.274 \pm 0.344$ & 8Bin-4 \\
0.203230 & $73.408 \pm 0.315$ & $73.375 \pm 0.317$ & $73.405 \pm 0.411$ & $67.664 \pm 0.332$ & $67.727 \pm 0.347$ & $67.753 \pm 0.436$ & 8Bin-5 \\
0.286890 & $73.169 \pm 0.292$ & $73.200 \pm 0.293$ & $73.222 \pm 0.359$ & $67.146 \pm 0.308$ & $67.279 \pm 0.325$ & $67.292 \pm 0.385$ & 8Bin-6 \\
0.384940 & $73.553 \pm 0.283$ & $73.527 \pm 0.283$ & $73.608 \pm 0.369$ & $67.214 \pm 0.299$ & $67.297 \pm 0.317$ & $67.382 \pm 0.395$ & 8Bin-7 \\
0.635815 & $74.465 \pm 0.318$ & $74.435 \pm 0.318$ & $74.564 \pm 0.444$ & $67.572 \pm 0.335$ & $67.663 \pm 0.350$ & $67.783 \pm 0.466$ & 8Bin-8 \\
\hline
\end{tabular}
}
\end{table*}

\begin{table*}
\centering
\caption{
Per-bin $H_\Lambda(z)$ fit summaries from the 2-, 4-, 6-, and 8-bin analyses as shown in Table~\ref{tab5}, reordered into a single list by increasing $z_{\rm center}$. Here $z_{\rm center}$ is the representative redshift assigned to each bin (taken as the bin median of $z_{\rm CMB}$ under the equal-count partition), where $z_{\rm CMB}$ denotes the redshift corrected to the CMB rest frame. The Bin ID encodes the originating binning configuration. Reported values are $H_\Lambda \pm 1\sigma$, where $1\sigma$ denotes the WLS/GLS standard error; all expansion rates ($H_\Lambda$ in entries and fixed $H_q$ in headers) are in units of km\,s$^{-1}$\,Mpc$^{-1}$.
}
\label{tab6}

\setlength{\tabcolsep}{8.0pt}
\renewcommand{\arraystretch}{1.5}
{\fontsize{7.2}{7.6}\selectfont
\begin{tabular}{c c c c c c c c}
\toprule
\multicolumn{1}{c}{} &
\multicolumn{3}{c}{Flat $\Lambda$CDM framework ($\Omega_m=0.315$, $H_q=0$)} &
\multicolumn{3}{c}{Hybrid framework ($H_q$ fixed to the best-fit value)} &
\multicolumn{1}{c}{} \\
\cmidrule(lr){2-4}\cmidrule(lr){5-7}
$z_{\rm center}$ &
D0 & ST & SS &
D0 ($H_q=\mathbf{4.881}$) & ST ($H_q=\mathbf{4.797}$) & SS ($H_q=\mathbf{4.802}$) &
Bin ID \\
\midrule

0.009800 & $71.323 \pm 0.612$ & $71.007 \pm 0.799$ & $70.959 \pm 0.882$ & $66.387 \pm 0.621$ & $66.159 \pm 0.807$ & $66.108 \pm 0.889$ & 8Bin-1 \\
0.012815 & $71.081 \pm 0.464$ & $70.949 \pm 0.608$ & $70.716 \pm 0.705$ & $66.132 \pm 0.474$ & $66.090 \pm 0.619$ & $65.854 \pm 0.713$ & 6Bin-1 \\
0.016190 & $\mathbf{72.355 \pm 0.327}$ & $\mathbf{72.119 \pm 0.429}$ & $\mathbf{72.000 \pm 0.512}$ & $\mathbf{67.382 \pm 0.340}$ & $\mathbf{67.238 \pm 0.444}$ & $\mathbf{67.110 \pm 0.525}$ & 4Bin-1 \\
0.022450 & $72.857 \pm 0.360$ & $72.921 \pm 0.391$ & $72.887 \pm 0.456$ & $67.867 \pm 0.369$ & $68.018 \pm 0.409$ & $67.976 \pm 0.471$ & 8Bin-2 \\
0.027210 & $72.537 \pm 0.183$ & $72.335 \pm 0.228$ & $72.415 \pm 0.254$ & $67.404 \pm 0.205$ & $67.307 \pm 0.255$ & $67.343 \pm 0.281$ & 2Bin-1 \\
0.027225 & $72.622 \pm 0.286$ & $72.752 \pm 0.308$ & $72.779 \pm 0.360$ & $67.606 \pm 0.299$ & $67.823 \pm 0.331$ & $67.844 \pm 0.379$ & 6Bin-2 \\
0.033370 & $71.781 \pm 0.317$ & $71.800 \pm 0.332$ & $72.082 \pm 0.384$ & $66.730 \pm 0.329$ & $66.834 \pm 0.353$ & $67.111 \pm 0.404$ & 8Bin-3 \\
0.048390 & $72.631 \pm 0.216$ & $72.648 \pm 0.222$ & $72.921 \pm 0.244$ & $67.418 \pm 0.233$ & $67.524 \pm 0.254$ & $67.769 \pm 0.274$ & 4Bin-2 \\
0.072485 & $73.044 \pm 0.257$ & $73.004 \pm 0.260$ & $73.157 \pm 0.286$ & $67.758 \pm 0.271$ & $67.809 \pm 0.289$ & $67.952 \pm 0.312$ & 6Bin-3 \\
0.103380 & $73.407 \pm 0.285$ & $73.393 \pm 0.287$ & $73.532 \pm 0.321$ & $68.057 \pm 0.297$ & $68.134 \pm 0.314$ & $68.274 \pm 0.344$ & 8Bin-4 \\
0.203230 & $73.408 \pm 0.315$ & $73.375 \pm 0.317$ & $73.405 \pm 0.411$ & $67.664 \pm 0.332$ & $67.727 \pm 0.347$ & $67.753 \pm 0.436$ & 8Bin-5 \\
0.217810 & $73.398 \pm 0.267$ & $73.402 \pm 0.267$ & $73.487 \pm 0.366$ & $67.613 \pm 0.285$ & $67.713 \pm 0.302$ & $67.798 \pm 0.393$ & 6Bin-4 \\
0.243160 & $73.295 \pm 0.215$ & $73.316 \pm 0.215$ & $73.341 \pm 0.300$ & $67.423 \pm 0.237$ & $67.545 \pm 0.259$ & $67.558 \pm 0.333$ & 4Bin-3 \\
0.286890 & $73.169 \pm 0.292$ & $73.200 \pm 0.293$ & $73.222 \pm 0.359$ & $67.146 \pm 0.308$ & $67.279 \pm 0.325$ & $67.292 \pm 0.385$ & 8Bin-6 \\
0.328595 & $73.573 \pm 0.151$ & $73.575 \pm 0.151$ & $73.522 \pm 0.233$ & $67.400 \pm 0.180$ & $67.512 \pm 0.207$ & $67.505 \pm 0.271$ & 2Bin-2 \\
0.328600 & $73.253 \pm 0.246$ & $73.251 \pm 0.247$ & $73.327 \pm 0.310$ & $67.084 \pm 0.264$ & $67.188 \pm 0.284$ & $67.262 \pm 0.339$ & 6Bin-5 \\
0.384940 & $73.553 \pm 0.283$ & $73.527 \pm 0.283$ & $73.608 \pm 0.369$ & $67.214 \pm 0.299$ & $67.297 \pm 0.317$ & $67.382 \pm 0.395$ & 8Bin-7 \\
0.495600 & $73.921 \pm 0.211$ & $73.897 \pm 0.211$ & $73.725 \pm 0.310$ & $67.365 \pm 0.231$ & $67.453 \pm 0.253$ & $67.327 \pm 0.338$ & 4Bin-4 \\
0.579590 & $74.245 \pm 0.272$ & $74.222 \pm 0.272$ & $74.120 \pm 0.386$ & $67.486 \pm 0.290$ & $67.581 \pm 0.309$ & $67.491 \pm 0.410$ & 6Bin-6 \\
0.635815 & $\mathbf{74.465 \pm 0.318}$ & $\mathbf{74.435 \pm 0.318}$ & $\mathbf{74.564 \pm 0.444}$ & $\mathbf{67.572 \pm 0.335}$ & $\mathbf{67.663 \pm 0.350}$ & $\mathbf{67.783 \pm 0.466}$ & 8Bin-8 \\
\hline
\end{tabular}
}
\end{table*}

\begin{figure*}
    \centering
    \includegraphics[width=\linewidth]{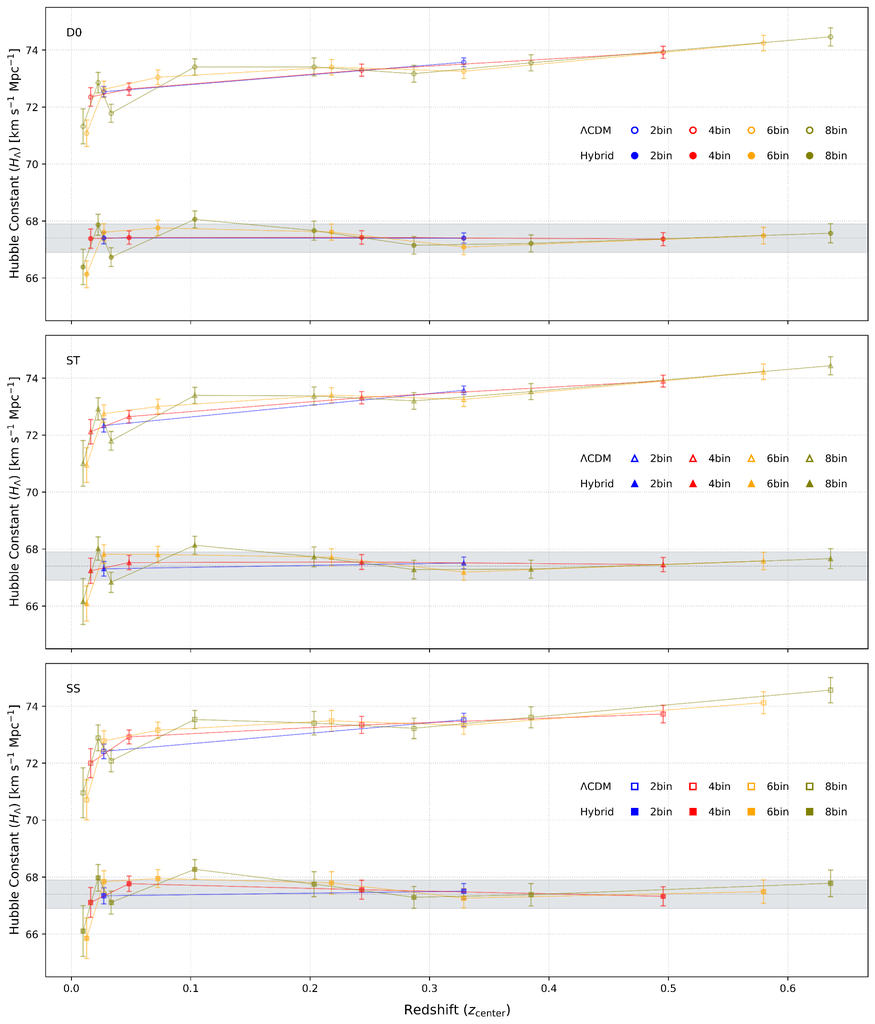}
 \caption{
Tomographic estimates of $H_\Lambda$ in redshift bins ($N_{\rm bin}\in\{2,4,6,8\}$). The binned results in Table~\ref{tab5} are shown for three covariance treatments: D0 (top), ST (middle), and SS (bottom). In all tomographic fits, $\Omega_m$ is fixed to 0.315 so that only the expansion rate $H_\Lambda$ is varied in each bin. Open markers denote the flat $\Lambda$CDM tomographic fits corresponding to the $\Lambda$2 configuration ($H_q=0$), while filled markers denote the hybrid fits corresponding to the Qb configuration, with $H_q$ fixed to the global best-fit value derived from Q1. Each binning scheme ($N_{\rm bin}$) is represented by a distinct set of markers and lines as specified in the legend. The shaded region marks the Planck reference range~\citepalias{Planck2020}, and the horizontal dotted line indicates the central Planck Hubble constant of $67.4\,\rm km\,s^{-1}\,Mpc^{-1}$. Across covariance treatments and binning schemes, the metric-only estimates remain systematically above the Planck range and show a mild positive drift with redshift, whereas the hybrid estimates remain close to the Planck range with substantially weaker drift.
}
\label{fig1}
\end{figure*}

\begin{table*}
\centering
\caption{
Weighted linear fits $y(z)=a+bz$ to the per-bin estimates within each binning configuration (2, 4, 6, and 8 bins) and the pooled ``all-bins'' sample.
For each mode (D0, ST, SS), we report results for (i) flat $\Lambda$CDM (the $\Lambda2$ configuration), (ii) hybrid (the Qb configuration), and (iii) $\Delta H \equiv H_\Lambda^{\rm \Lambda CDM}-H_\Lambda^{\rm hybrid}$.
We report the intercept $a$ and slope $b$ with ${\rm SE}$ (standard errors from the weighted fit), and the two-sided $p$-value for the slope $b$.
All reported intercepts $a$ and slopes $b$ are in units of km\,s$^{-1}$\,Mpc$^{-1}$; the slope $b$ is expressed per unit redshift.
}
\label{tab7}

\setlength{\tabcolsep}{5.9pt}
\renewcommand{\arraystretch}{1.18}
{\fontsize{7.1}{7.5}\selectfont
\begin{tabular}{c c c c c c c c c c c}
\toprule
\multirow{2}{*}{Bin scheme} & \multirow{2}{*}{Mode}
& \multicolumn{3}{c}{Flat $\Lambda$CDM ($H_q=0$)}
& \multicolumn{3}{c}{Hybrid ($H_q$ fixed)}
& \multicolumn{3}{c}{Difference $\Delta H$} \\
\cmidrule(lr){3-5}\cmidrule(lr){6-8}\cmidrule(lr){9-11}
& & Intercept $a$ & Slope $b$ & $p$-value
  & Intercept $a$ & Slope $b$ & $p$-value
  & Intercept $a$ & Slope $b$ & $p$-value \\
\midrule

\multirow{3}{*}{2 Bins}
& D0 & $72.443 \pm 0.200$ & $3.437 \pm 0.787$ & -- & $67.404 \pm 0.224$ & $-0.013 \pm 0.905$ & -- & $5.039 \pm 0.300$ & $3.451 \pm 1.200$ & -- \\
& ST & $72.223 \pm 0.249$ & $4.114 \pm 0.907$ & -- & $67.288 \pm 0.279$ & $\mathord{+}0.680 \pm 1.090$ & -- & $4.935 \pm 0.374$ & $3.434 \pm 1.418$ & -- \\
& SS & $72.315 \pm 0.278$ & $3.673 \pm 1.144$ & -- & $67.328 \pm 0.307$ & $\mathord{+}0.538 \pm 1.295$ & -- & $4.987 \pm 0.414$ & $3.136 \pm 1.728$ & -- \\
\midrule

\multirow{3}{*}{4 Bins}
& D0 & $72.464 \pm 0.183$ & $3.029 \pm 0.606$ & $0.038$ & $67.418 \pm 0.196$ & $-0.084 \pm 0.655$ & $0.910$ & $5.046 \pm 0.268$ & $3.120 \pm 0.893$ & $0.073$ \\
& ST & $72.463 \pm 0.199$ & $3.011 \pm 0.635$ & $0.042$ & $67.475 \pm 0.225$ & $\mathord{+}0.021 \pm 0.736$ & $0.980$ & $4.988 \pm 0.300$ & $3.035 \pm 0.973$ & $0.089$ \\
& SS & $72.680 \pm 0.231$ & $2.253 \pm 0.833$ & $0.114$ & $67.658 \pm 0.254$ & $-0.594 \pm 0.910$ & $0.581$ & $5.022 \pm 0.368$ & $2.903 \pm 1.234$ & $0.143$ \\
\midrule

\multirow{3}{*}{6 Bins}
& D0 & $72.462 \pm 0.179$ & $3.071 \pm 0.586$ & $0.006$ & $67.421 \pm 0.188$ & $-0.050 \pm 0.620$ & $0.939$ & $5.040 \pm 0.258$ & $3.155 \pm 0.853$ & $0.021$ \\
& ST & $72.572 \pm 0.190$ & $2.774 \pm 0.606$ & $0.010$ & $67.595 \pm 0.208$ & $-0.234 \pm 0.675$ & $0.747$ & $4.976 \pm 0.283$ & $3.071 \pm 0.907$ & $0.028$ \\
& SS & $72.674 \pm 0.224$ & $2.518 \pm 0.783$ & $0.032$ & $67.695 \pm 0.239$ & $-0.492 \pm 0.834$ & $0.587$ & $4.980 \pm 0.342$ & $3.083 \pm 1.144$ & $0.054$ \\
\midrule

\multirow{3}{*}{8 Bins}
& D0 & $72.417 \pm 0.179$ & $3.252 \pm 0.576$ & $0.001$ & $67.399 \pm 0.185$ & $\mathord{+}0.079 \pm 0.602$ & $0.900$ & $5.018 \pm 0.258$ & $3.184 \pm 0.834$ & $0.009$ \\
& ST & $72.465 \pm 0.188$ & $3.097 \pm 0.592$ & $0.002$ & $67.511 \pm 0.201$ & $\mathord{+}0.037 \pm 0.643$ & $0.956$ & $4.954 \pm 0.275$ & $3.104 \pm 0.874$ & $0.012$ \\
& SS & $72.569 \pm 0.222$ & $3.070 \pm 0.761$ & $0.007$ & $67.626 \pm 0.233$ & $-0.049 \pm 0.800$ & $0.953$ & $4.944 \pm 0.332$ & $3.159 \pm 1.104$ & $0.029$ \\
\midrule

\multirow{3}{*}{All bins}
& D0 & $72.453 \pm 0.092$ & $\mathbf{3.161 \pm 0.311}$ & $\mathbf{<0.001}$ & $67.410 \pm 0.098$ & $\mathbf{-0.013 \pm 0.334}$ & $\mathbf{0.971}$ & $\mathbf{5.043 \pm 0.131}$ & $3.186 \pm 0.457$ & $<0.001$ \\
& ST & $72.462 \pm 0.100$ & $\mathbf{3.111 \pm 0.328}$ & $\mathbf{<0.001}$ & $67.490 \pm 0.111$ & $\mathbf{\mathord{+}0.037 \pm 0.369}$ & $\mathbf{0.921}$ & $\mathbf{4.956 \pm 0.150}$ & $3.110 \pm 0.495$ & $<0.001$ \\
& SS & $72.581 \pm 0.117$ & $\mathbf{2.798 \pm 0.423}$ & $\mathbf{<0.001}$ & $67.600 \pm 0.127$ & $\mathbf{-0.218 \pm 0.455}$ & $\mathbf{0.637}$ & $\mathbf{4.966 \pm 0.173}$ & $3.061 \pm 0.622$ & $<0.001$ \\
\bottomrule
\end{tabular}
}
\end{table*}

\begin{table*}
\setlength{\tabcolsep}{4.58pt}
\centering

\caption{
Global regression results for $w$CDM (w1--w6) and CPL (C1--C6).
Entries show best-fit values $\pm1\sigma$ when free and ``(fixed)'' when held fixed.
The column ``$w$ (or $w_0$)'' contains $w$ for $w$CDM rows and $w_0$ for CPL rows; $w_a$ is omitted (--) for $w$CDM.
All fits are performed via MCMC posterior sampling under a Gaussian likelihood; the data treatments differ only by the assumed covariance:
D0: diagonal-only uncertainties (equivalent to WLS);
ST: full statistical covariance (GLS);
SS: full statistical+systematic covariance (GLS).
We report the Root Mean Square Error (RMSE) and the Mean Absolute Error (MAE); both metrics are in units of magnitudes [mag].
}

\label{tab8}
{\fontsize{7.0pt}{9.8}\selectfont
\begin{tabular}{c c c c c c c c c c c c}
\Xhline{0.9pt}
ID & $M\,[\rm mag]$ & $\Omega_m$ & $H_{\delta}\,[\mathrm{km\,s^{-1}\,Mpc^{-1}}]$ & $w$ or $w_0$ & $w_a$ & $R^2$ & $\chi^2/{\rm DOF}$ & RMSE & MAE & AIC & BIC \\

\Xhline{0.9pt}

D0-w1 & 0.000 (fixed) & 0.315 (fixed) & 67.400 (fixed) & $\mathbf{-0.325 \pm 0.021}$ & -- & 0.996 & 0.669 & 0.210 & 0.153 & 1138.861 & 1144.300 \\
D0-w2 & 0.000 (fixed) & 0.315 (fixed) & $\mathbf{72.223 \pm 0.285}$ & $-0.844 \pm 0.041$ & -- & 0.997 & 0.477 & 0.189 & 0.131 & 814.607 & 825.484 \\
D0-w3 & 0.000 (fixed) & $\mathbf{0.124 \pm 0.164}$ & 67.400 (fixed) & $\mathbf{-0.203 \pm 0.283}$ & -- & 0.996 & 0.672 & 0.211 & 0.155 & 1146.269 & 1157.147 \\
D0-w4 & 0.000 (fixed) & $\mathbf{0.221 \pm 0.104}$ & $\mathbf{72.107 \pm 0.301}$ & $-0.724 \pm 0.154$ & -- & 0.997 & 0.478 & 0.189 & 0.131 & 817.526 & 833.843 \\
D0-w5 & $\mathbf{-0.150 \pm 0.009}$ & 0.315 (fixed) & 67.400 (fixed) & $-0.844 \pm 0.041$ & -- & 0.997 & 0.477 & 0.189 & 0.131 & 814.606 & 825.484 \\
D0-w6 & $\mathbf{-0.147 \pm 0.009}$ & $\mathbf{0.223 \pm 0.104}$ & 67.400 (fixed) & $-0.727 \pm 0.153$ & -- & 0.997 & 0.478 & 0.189 & 0.131 & 817.552 & 833.868 \\
\hline
D0-C1 & 0.000 (fixed) & 0.315 (fixed) & 67.400 (fixed) & $\mathbf{0.471 \pm 0.078}$ & $\mathbf{-7.558 \pm 0.767}$ & 0.996 & 0.587 & 0.203 & 0.145 & 1000.470 & 1011.348 \\
D0-C2 & 0.000 (fixed) & 0.315 (fixed) & $\mathbf{72.016 \pm 0.347}$ & $-0.735 \pm 0.114$ & $-0.834 \pm 0.808$ & 0.997 & 0.477 & 0.189 & 0.131 & 815.659 & 831.976 \\
D0-C3 & 0.000 (fixed) & $\mathbf{0.659 \pm 0.014}$ & 67.400 (fixed) & $\mathbf{3.466 \pm 0.151}$ & $\mathbf{-57.459 \pm 2.378}$ & 0.997 & 0.529 & 0.196 & 0.137 & 903.701 & 920.018 \\
D0-C4 & 0.000 (fixed) & $0.390 \pm 0.132$ & $\mathbf{71.886 \pm 0.394}$ & $-0.694 \pm 0.204$ & $\mathbf{-3.721 \pm 3.941}$ & 0.997 & 0.485 & 0.189 & 0.132 & 831.819 & 853.575 \\
D0-C5 & $\mathbf{-0.144 \pm 0.011}$ & 0.315 (fixed) & 67.400 (fixed) & $-0.736 \pm 0.112$ & $-0.817 \pm 0.790$ & 0.997 & 0.477 & 0.189 & 0.131 & 815.651 & 831.968 \\
D0-C6 & $\mathbf{-0.140 \pm 0.012}$ & $0.397 \pm 0.124$ & 67.400 (fixed) & $-0.691 \pm 0.203$ & $\mathbf{-3.836 \pm 3.961}$ & 0.997 & 0.484 & 0.189 & 0.132 & 829.948 & 851.704 \\

\hline\hline

ST-w1 & 0.000 (fixed) & 0.315 (fixed) & 67.400 (fixed) & $\mathbf{-0.352 \pm 0.019}$ & -- & 0.996 & 0.908 & 0.210 & 0.153 & 1545.214 & 1550.653 \\
ST-w2 & 0.000 (fixed) & 0.315 (fixed) & $\mathbf{72.067 \pm 0.238}$ & $-0.834 \pm 0.033$ & -- & 0.997 & 0.888 & 0.189 & 0.131 & 1512.961 & 1523.839 \\
ST-w3 & 0.000 (fixed) & $\mathbf{0.103 \pm 0.160}$ & 67.400 (fixed) & $\mathbf{-0.216 \pm 0.267}$ & -- & 0.996 & 0.915 & 0.211 & 0.154 & 1559.349 & 1570.227 \\
ST-w4 & 0.000 (fixed) & $\mathbf{0.208 \pm 0.090}$ & $\mathbf{71.939 \pm 0.247}$ & $-0.690 \pm 0.119$ & -- & 0.997 & 0.889 & 0.189 & 0.131 & 1515.292 & 1531.609 \\
ST-w5 & $\mathbf{-0.145 \pm 0.007}$ & 0.315 (fixed) & 67.400 (fixed) & $-0.834 \pm 0.033$ & -- & 0.997 & 0.888 & 0.189 & 0.131 & 1512.961 & 1523.839 \\
ST-w6 & $\mathbf{-0.141 \pm 0.008}$ & $\mathbf{0.208 \pm 0.090}$ & 67.400 (fixed) & $-0.690 \pm 0.121$ & -- & 0.997 & 0.889 & 0.189 & 0.131 & 1515.405 & 1531.722 \\
\hline
ST-C1 & 0.000 (fixed) & 0.315 (fixed) & 67.400 (fixed) & $\mathbf{0.396 \pm 0.069}$ & $\mathbf{-6.835 \pm 0.662}$ & 0.997 & 0.897 & 0.203 & 0.145 & 1528.336 & 1539.214 \\
ST-C2 & 0.000 (fixed) & 0.315 (fixed) & $\mathbf{71.839 \pm 0.287}$ & $-0.716 \pm 0.092$ & $-0.871 \pm 0.637$ & 0.997 & 0.888 & 0.189 & 0.131 & 1514.149 & 1530.466 \\
ST-C3 & 0.000 (fixed) & $\mathbf{0.658 \pm 0.011}$ & 67.400 (fixed) & $\mathbf{3.464 \pm 0.130}$ & $\mathbf{-57.917 \pm 1.993}$ & 0.997 & 0.891 & 0.196 & 0.137 & 1518.385 & 1534.702 \\
ST-C4 & 0.000 (fixed) & $0.363 \pm 0.121$ & $\mathbf{71.751 \pm 0.325}$ & $-0.686 \pm 0.153$ & $\mathbf{-2.720 \pm 3.084}$ & 0.997 & 0.896 & 0.189 & 0.132 & 1528.955 & 1550.711 \\
ST-C5 & $\mathbf{-0.138 \pm 0.009}$ & 0.315 (fixed) & 67.400 (fixed) & $-0.715 \pm 0.090$ & $-0.872 \pm 0.629$ & 0.997 & 0.888 & 0.189 & 0.131 & 1514.148 & 1530.465 \\
ST-C6 & $\mathbf{-0.136 \pm 0.010}$ & $0.361 \pm 0.125$ & 67.400 (fixed) & $-0.684 \pm 0.156$ & $\mathbf{-2.733 \pm 3.101}$ & 0.997 & 0.897 & 0.189 & 0.132 & 1530.130 & 1551.886 \\

\hline\hline

SS-w1 & 0.000 (fixed) & 0.315 (fixed) & 67.400 (fixed) & $\mathbf{-0.253 \pm 0.021}$ & -- & 0.996 & 0.867 & 0.212 & 0.155 & 1475.626 & 1481.065 \\
SS-w2 & 0.000 (fixed) & 0.315 (fixed) & $\mathbf{72.030 \pm 0.279}$ & $-0.825 \pm 0.044$ & -- & 0.997 & 0.939 & 0.189 & 0.131 & 1599.895 & 1610.772 \\
SS-w3 & 0.000 (fixed) & $\mathbf{0.155 \pm 0.122}$ & 67.400 (fixed) & $\mathbf{-0.209 \pm 0.042}$ & -- & 0.996 & 0.869 & 0.212 & 0.155 & 1480.908 & 1491.786 \\
SS-w4 & 0.000 (fixed) & $\mathbf{0.169 \pm 0.090}$ & $\mathbf{71.874 \pm 0.280}$ & $-0.636 \pm 0.111$ & -- & 0.997 & 0.939 & 0.189 & 0.131 & 1600.312 & 1616.629 \\
SS-w5 & $\mathbf{-0.144 \pm 0.008}$ & 0.315 (fixed) & 67.400 (fixed) & $-0.824 \pm 0.044$ & -- & 0.997 & 0.939 & 0.189 & 0.131 & 1599.894 & 1610.772 \\
SS-w6 & $\mathbf{-0.140 \pm 0.008}$ & $\mathbf{0.167 \pm 0.090}$ & 67.400 (fixed) & $-0.633 \pm 0.110$ & -- & 0.997 & 0.939 & 0.189 & 0.131 & 1600.257 & 1616.574 \\
\hline
SS-C1 & 0.000 (fixed) & 0.315 (fixed) & 67.400 (fixed) & $\mathbf{0.488 \pm 0.075}$ & $\mathbf{-6.797 \pm 0.719}$ & 0.996 & 0.878 & 0.204 & 0.147 & 1495.080 & 1505.958 \\
SS-C2 & 0.000 (fixed) & 0.315 (fixed) & $\mathbf{71.609 \pm 0.339}$ & $-0.610 \pm 0.110$ & $-1.526 \pm 0.737$ & 0.997 & 0.937 & 0.189 & 0.131 & 1597.680 & 1613.997 \\
SS-C3 & 0.000 (fixed) & $\mathbf{0.675 \pm 0.015}$ & 67.400 (fixed) & $\mathbf{3.573 \pm 0.198}$ & $\mathbf{-56.173 \pm 3.376}$ & 0.997 & 0.894 & 0.198 & 0.139 & 1524.447 & 1540.764 \\
SS-C4 & 0.000 (fixed) & $\mathbf{0.450 \pm 0.112}$ & $\mathbf{71.233 \pm 0.481}$ & $-0.348 \pm 0.354$ & $\mathbf{-7.895 \pm 6.155}$ & 0.997 & 0.943 & 0.189 & 0.132 & 1607.533 & 1629.289 \\
SS-C5 & $\mathbf{-0.132 \pm 0.010}$ & 0.315 (fixed) & 67.400 (fixed) & $-0.612 \pm 0.109$ & $-1.511 \pm 0.734$ & 0.997 & 0.937 & 0.189 & 0.131 & 1597.678 & 1613.995 \\
SS-C6 & $\mathbf{-0.119 \pm 0.015}$ & $\mathbf{0.462 \pm 0.095}$ & 67.400 (fixed) & $-0.337 \pm 0.351$ & $\mathbf{-8.187 \pm 6.007}$ & 0.997 & 0.941 & 0.189 & 0.132 & 1604.351 & 1626.107 \\

\Xhline{0.9pt}
\end{tabular}
}
\end{table*}

\begin{figure*}
    \centering
    \includegraphics[width=\linewidth]{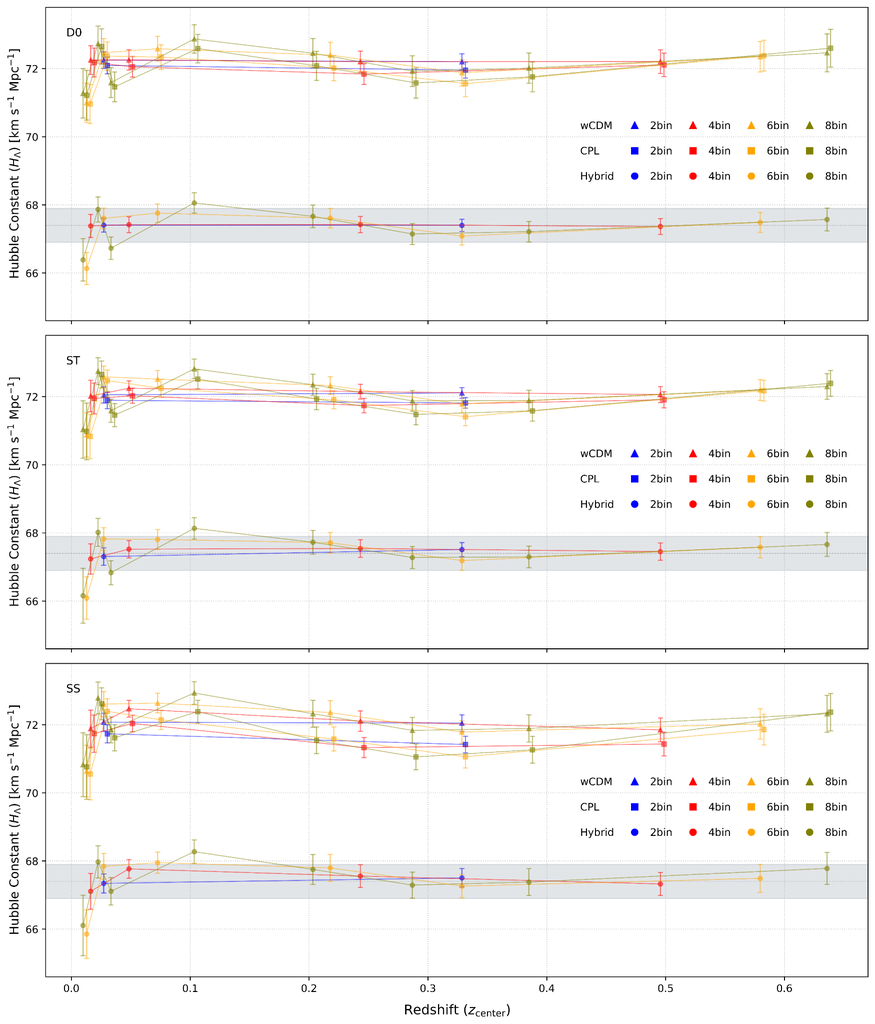}
 \caption{
Tomographic estimates of $H_\delta$ (or $H_\Lambda$) in redshift bins ($N_{\mathrm{bin}} \in \{2, 4, 6, 8\}$) for three model families: Hybrid, $w$CDM, and CPL. For the $w$CDM and CPL tomographic fits (namely, the wb and Cb configurations), the equation-of-state parameters are fixed to their global best-fit values from the corresponding configurations (i.e., w2 and C2)---which phenomenologically align with recent observational reports \citep{Adame2025, Rubin2025}---with $\Omega_m = 0.315$, so that only the expansion rate $H_\delta$ is varied in each bin. This tests whether the phenomenological shape flexibility of the DDE models can absorb the redshift drift and simultaneously recover the Planck-anchored level. Filled circles denote the Hybrid fits (representing the Qb), filled triangles denote the $w$CDM fits (representing the wb), and filled squares denote the CPL fits (representing the Cb). The shaded region marks the Planck reference range~\citepalias{Planck2020}. Across covariance treatments and binning schemes, the Hybrid estimates remain close to the Planck range, whereas the $w$CDM and CPL estimates are systematically higher.
}
\label{fig wCDM CPL Bin}
\end{figure*}

\begin{table}

\centering
\setlength{\tabcolsep}{5.75pt}
\renewcommand{\arraystretch}{1.1}

\caption{
Model selection with the BIC across three covariance modes (D0, ST, SS). 
We compare metric-only baselines against their hybrid counterparts in four groups: 
$\Lambda1$ vs. $\rm Q1$, $\rm Qa$, w1, C1; 
$\Lambda2$ vs. $\rm Q2$, $\rm Qb$, w2, C2; 
$\Lambda3$ vs. $\rm Q3$, $\rm Qc$, w3, C3; 
$\Lambda4$ vs. $\rm Q4$, $\rm Qd$, w4, C4. 
$\Delta{\rm BIC}$ is measured relative to the $\Lambda$ baseline in each group 
(negative values favor the alternative model).
}

\label{tab4}

{\fontsize{7.1}{10.5}\selectfont
\begin{tabular}{c r r r r r r}
\toprule
& \multicolumn{3}{c}{BIC} & \multicolumn{3}{c}{$\Delta{\rm BIC}$} \\
\cmidrule(lr){2-4}\cmidrule(lr){5-7}
ID &
D0 & ST & SS &
D0 & ST & SS \\
\midrule

$\Lambda1$ & 1708.439 & 1630.396 & 1652.286 & 0.000 & 0.000 & 0.000 \\
Q1         &  818.750 & 1517.236 & 1604.196 & \textbf{-889.689} & \textbf{-113.160} & \textbf{-48.090} \\
Qa         &  818.750 & 1517.236 & 1604.196 & \textbf{-889.689} & \textbf{-113.160} & \textbf{-48.090} \\
w1         & 1144.300 & 1550.653 & 1481.065 & -564.139 & -79.743 & \textit{-171.221} \\
C1         & 1011.348 & 1539.214 & 1505.958 & -697.091 & -91.182 & \textit{-146.328} \\
\hline

$\Lambda2$ &  831.995 & 1518.499 & 1592.514 & 0.000 & 0.000 & 0.000 \\
Q2         &  826.190 & 1523.931 & 1610.964 & -5.805 & 5.432 & 18.450 \\
Qb         &  826.189 & 1524.022 & 1611.068 & -5.806 & 5.523 & 18.554 \\
w2         &  825.484 & 1523.839 & 1610.772 & \textit{-6.511}& \textit{5.340} & \textit{18.258} \\
C2         &  831.976 & 1530.466 & 1613.997 & -0.019 & 11.967 & 21.483 \\
\hline

$\Lambda3$ & 1180.250 & 1556.127 & 1470.917 & 0.000 & 0.000 & 0.000 \\
Q3         &  826.148 & 1523.843 & 1610.598 & -354.102 & -32.284 & 139.681 \\
Qc         &  826.168 & 1523.843 & 1610.600 & -354.082 & -32.284 & 139.683 \\
w3         & 1157.147 & 1570.227 & 1491.786 & -23.103 & 14.100 & \textit{20.869} \\
C3         &  920.018 & 1534.702 & 1540.764 & -260.232 & -21.425 & \textit{69.847} \\
\hline

$\Lambda4$ &  827.491 & 1524.099 & 1610.807 & 0.000 & 0.000 & 0.000 \\
Q4         &  832.793 & 1531.415 & 1617.593 & 5.302 & 7.316 & 6.786 \\
Qd         &  833.596 & 1533.310 & 1620.652 & 6.105 & 9.211 & 9.845 \\
w4         &  833.843 & 1531.609 & 1616.629 & 6.352 & 7.510& \textit{5.822} \\
C4         &  853.575 & 1550.711 & 1629.289 & 26.084 & 26.612 & 18.482 \\
\bottomrule
\end{tabular}
}

\end{table}

\section{Results \& Analysis}
\label{3}

\subsection{Global Regression Results from the $\Lambda$CDM Framework}
\label{3.1}

\noindent\textbf{Results.} As summarized in Table~\ref{tab3}, the fully fixed Planck-SH0ES baseline configuration $\Lambda1$ (fixed $\Omega_m=0.315$, $H_\Lambda=67.4\,\rm km\,s^{-1}\,Mpc^{-1}$, $M=0\,\mathrm{mag}$) records the highest BIC in every covariance treatment, ranging from $1630.396$ (ST) to $1708.439$ (D0). Within the $\Lambda$-family, the statistically preferred model varies by covariance mode: D0 selects $\Lambda4$ (joint intercept--shape adjustment), ST selects $\Lambda2$ (intercept-only), and SS selects $\Lambda3$ (shape-only).

Regarding the Hubble parameter, the intercept-driven configuration $\Lambda2$ yields $H_\Lambda = 73.090\pm0.171$ (D0), $72.998\pm0.144$ (ST), and $72.974\pm0.140~\mathrm{km\,s^{-1}\,Mpc^{-1}}$ (SS). Relative to the Pantheon+SH0ES baseline, $H_0=73.04\pm1.04$~\citepalias{Riess2022}, these values show agreement at the sub-$\sigma$ level, $+0.05\sigma$ in D0, $-0.04\sigma$ in ST, and $-0.06\sigma$ in SS.

For the curvature (shape-driven) configuration $\Lambda3$, the inferred matter density exceeds $\Omega_m\approx0.7$ in all covariance modes and therefore departs from the Planck anchor $\Omega_m=0.315\pm0.007$~\citepalias{Planck2020} at the $\sim 23$--$25\sigma$ level: $\Omega_m=0.732\pm0.017$ (D0; $22.68\sigma$), $0.706\pm0.015$ (ST; $23.62\sigma$), and $0.790\pm0.018$ (SS; $24.59\sigma$). The corresponding BIC values are $1180.250$ (D0), $1556.127$ (ST), and $1470.917$ (SS), implying that $\Lambda3$ is the second-most disfavored $\Lambda$ configuration in D0 and ST (second-highest BIC within the $\Lambda$ family), while it is the most favored $\Lambda$ configuration in SS (lowest BIC within the $\Lambda$ family).

For the joint intercept--shape configuration $\Lambda4$ (free $\Omega_m$ and $H_\Lambda$ with $M=0$ fixed), the fitted $\Omega_m$ departs from the Planck anchor $\Omega_m=0.315\pm0.007$~\citepalias{Planck2020} at the $\sim 3$--$4\sigma$ level: $\Omega_m=0.381\pm0.020$ (D0; $3.11\sigma$), $0.384\pm0.016$ (ST; $3.95\sigma$), and $0.382\pm0.021$ (SS; $3.03\sigma$). The fitted Hubble parameter departs from the Planck anchor $H_{0}^{\rm \scriptscriptstyle CMB}=67.4\pm0.5~{\rm km\,s^{-1}\,Mpc^{-1}}$~\citepalias{Planck2020} at the $\sim 9\sigma$ level: $H_\Lambda=72.392\pm0.264$ (D0; $8.83\sigma$), $72.257\pm0.218$ (ST; $8.91\sigma$), and $72.271\pm0.255~{\rm km\,s^{-1}\,Mpc^{-1}}$ (SS; $8.68\sigma$). The corresponding BIC values are $827.491$ (D0), $1524.099$ (ST), and $1610.807$ (SS), so that $\Lambda4$ is the most favored $\Lambda$ configuration in D0 (lowest BIC within the $\Lambda$ family), whereas it is among the least favored $\Lambda$ configurations in SS (second-highest BIC within the $\Lambda$ family, tied with $\Lambda6$).

Due to the strong $M$--$H_\Lambda$ degeneracy, $\Lambda5$ is numerically equivalent to $\Lambda2$ and $\Lambda6$ is numerically equivalent to $\Lambda4$ in terms of information criteria: $\Lambda5$ matches $\Lambda2$ in BIC in each mode (D0: $831.995$; ST: $1518.500$; SS: $1592.514$) and yields $M=-0.176\pm0.005$ (D0), $-0.173\pm0.004$ (ST), and $-0.173\pm0.004\,\mathrm{mag}$ (SS), while $\Lambda6$ matches $\Lambda4$ in BIC in each mode (D0: $827.491$; ST: $1524.099$; SS: $1610.807$) and yields $M=-0.155\pm0.008$ (D0), $-0.151\pm0.007$ (ST), and $-0.152\pm0.008\,\mathrm{mag}$ (SS); in these configurations, fixing $H_\Lambda$ to the Planck value leaves the remaining $H$-residual to be absorbed by $M$.

\vspace{2.0mm}

\noindent\textbf{Analysis.} The standard $\Lambda$CDM framework alone is fundamentally incapable of providing a consistent joint description of the Planck cosmological priors \citepalias{Planck2020} and the Pantheon+SH0ES sample \citepalias{Brout2022, Scolnic2022, Riess2022}. This is evidenced by the consistently high BIC values recorded for the fully fixed baseline configuration ($\Lambda1$).

The Hubble tension is reproduced within the $\Lambda$CDM framework. The sub-$\sigma$ agreement between our derived $H_\Lambda$ values and the SH0ES reported tension value ($73.04 \pm 1.04~\mathrm{km\,s^{-1}\,Mpc^{-1}}$) \citepalias{Riess2022} validates the numerical robustness and methodological integrity of our regression pipeline.

Within the $\Lambda$CDM architecture, the statistical preference for adjusting the intercept ($H_\Lambda$ or $M$) versus the curve shape ($\Omega_m$) proves covariance-dependent. The favored configuration shifts distinctively across modes: D0 favors a joint adjustment ($\Lambda4$), ST an intercept-only shift ($\Lambda2$), and SS a shape-only modification ($\Lambda3$). 

The Planck priors \citepalias{Planck2020} are irreconcilable with the Pantheon+SH0ES sample \citepalias{Brout2022, Scolnic2022, Riess2022} under $\Lambda$CDM without invoking a significant calibration offset. Specifically, anchoring the expansion rate to the Planck value forces the residual $H$-tension to be effectively absorbed as a proxy  by the magnitude, resulting in a substantial and non-physical shift ($M < -0.15~\mathrm{mag}$).

\subsection{Global Regression Results from the Hybrid Framework}
\label{3.2}

\noindent\textbf{Results.} Excluding Q4 and Q6, the hybrid configurations recover $(\Omega_m, H_\Lambda)$ values consistent with the Planck anchors $\Omega_m=0.315\pm0.007$ and $H_{0}^{\rm \scriptscriptstyle CMB}=67.4\pm0.5~\mathrm{km\,s^{-1}\,Mpc^{-1}}$ \citepalias{Planck2020}, where the quoted $\sigma$-level agreements use the quadrature sum of the Planck and fitted $1\sigma$ uncertainties. For $\Omega_m$, Q3 yields $\Omega_m=0.310\pm0.024$ (D0; $0.20\sigma$), $0.317\pm0.020$ (ST; $0.09\sigma$), and $0.314\pm0.026$ (SS; $0.04\sigma$). For $H_\Lambda$, Q2 yields $H_\Lambda=67.438\pm1.547$ (D0; $0.02\sigma$), $66.995\pm1.267$ (ST; $0.30\sigma$), and $66.850\pm1.647~\mathrm{km\,s^{-1}\,Mpc^{-1}}$ (SS; $0.32\sigma$).

Relative to $\Lambda1$, Q1 achieves lower BIC in every covariance mode: ${\rm BIC}=818.750$ in D0 ($\Delta{\rm BIC}=-889.689$), $1517.236$ in ST ($\Delta{\rm BIC}=-113.160$), and $1604.196$ in SS ($\Delta{\rm BIC}=-48.090$), indicating a uniform improvement in statistical preference across all treatments.

In the $M$-free configuration Q5, the regression returns an absolute-magnitude offset consistent with zero: $M=0.002\pm0.050$ (D0), $0.014\pm0.041$ (ST), and $0.022\pm0.055~\mathrm{mag}$ (SS), i.e., a nearly vanishing calibration shift is reported when the hybrid channel is allowed.

For the inferred hybrid-channel scale, $H_q$ is recovered as a consistent value of approximately $5~\mathrm{km\,s^{-1}\,Mpc^{-1}}$ once Q4 and Q6 are excluded: across Q1--Q3 and Q5, the best-fit central values span $H_q\simeq4.784$--$5.458~\mathrm{km\,s^{-1}\,Mpc^{-1}}$ (mode-dependent), with Q1 specifically yielding $H_q=4.881\pm0.147$ (D0), $4.797\pm0.121$ (ST), and $4.802\pm0.119~\mathrm{km\,s^{-1}\,Mpc^{-1}}$ (SS).

For the prior-informed series (using the Q1-derived $H_q$ as a fixed prior in each mode), the recovered parameters remain concentrated near the Planck anchors: $\Omega_m$ lies in the range $0.312\lesssim\Omega_m\lesssim0.317$ (Qc, Qd, Qf) (e.g., D0: $0.312$--$0.313$; ST: $0.316$--$0.317$; SS: $0.315$--$0.316$), corresponding to agreement at $\lesssim0.15\sigma$ relative to $\Omega_m=0.315\pm0.007$~\citepalias{Planck2020}, and $H_\Lambda$ lies in the range $67.377\lesssim H_\Lambda\lesssim67.434~\mathrm{km\,s^{-1}\,Mpc^{-1}}$ (Qb, Qd), corresponding to agreement at $\lesssim0.06\sigma$ relative to $H_{0}^{\rm \scriptscriptstyle CMB}=67.4\pm0.5~\mathrm{km\,s^{-1}\,Mpc^{-1}}$~\citepalias{Planck2020}. For the Qe--Qf fits, $M$ remains near zero, with representative constraints such as $M=-0.000\pm0.007$ (D0-Qe), $0.000\pm0.006$ (ST-Qe), and $0.000\pm0.006~\mathrm{mag}$ (SS-Qe), as well as $M=-0.001\pm0.009$ (D0-Qf), $0.001\pm0.007$ (ST-Qf), and $0.000\pm0.008~\mathrm{mag}$ (SS-Qf), i.e., $|M|\lesssim10^{-3}$ at the best-fit level with sub-$10^{-2}$ uncertainties.

\vspace{2.0mm}

\noindent\textbf{Analysis.} Q1 attains the minimum BIC within the Q-family, indicating statistical preference for the hybrid description under the Planck-anchored $M=0$ configurations. The stability of $H_q \approx 5~\mathrm{km\,s^{-1}\,Mpc^{-1}}$ across D0, ST, and SS shows that this additional channel is consistently selected by the data. The Q2 recovery of a Planck-adjacent $H_\Lambda$ may likewise be interpreted as a non-trivial consequence of the hybrid framework (Appendix~\ref{AppG}).

Across all modes, the $\rm Q4$ configuration in Table~\ref{tab3} closely reproduces the $\Lambda$CDM+TL results in Table 1 of \citetalias{Gupta2023} ($\Omega_{m,0}=0.1351\pm0.0109$, $H_{\rm x}=60.48\pm1.06~\mathrm{km\,s^{-1}\,Mpc^{-1}}$), with differences $\le 0.31\sigma$ in $\Omega_m$ and $\le 0.54\sigma$ in $H_\Lambda$, thereby cross-validating the operational consistency of our regression pipeline. However, because our scope is to test whether the hybrid framework can recover the Planck-anchored baseline, the interpretive use of this agreement differs from \citetalias{Gupta2023}. Specifically, when the metric curvature freedom ($\Omega_m$) is activated alongside the WMC channel, the shape adjustment in $\Omega_m$ becomes partially degenerate with the WMC-induced curvature, producing a coupled $H_\Lambda$--$H_q$--$\Omega_m$ degeneracy that substantially reduces identifiability (see Appendix~\ref{AppF}); accordingly, $\rm Q4$ yields markedly weaker constraints than $\rm Q1$--$\rm Q3$ and $\rm Q5$.

Moreover, although $H_\Lambda$ is anchored in the $\rm Q6$ configuration, the parameter estimation is destabilized by a compound degeneracy between the curve-shaping effects of $\Omega_m$ and the WMC distance--redshift relation, together with the strong $M$--$H_\Lambda$ and $H_\Lambda$--$H_q$ degeneracies (see Appendix~\ref{AppF}). Consequently, the remaining intercept budget is effectively transferred to the absolute magnitude via the $M$--$H_q$ coupling, and $\rm Q6$ reports an excessively large offset (e.g., $M=0.315\pm0.312$ in D0, $0.354\pm0.300$ in ST, and $0.356\pm0.192~\mathrm{mag}$ in SS). Therefore, these regression artifacts cannot be definitively attributed to the intrinsic properties of the WMC channel itself.

\subsection{Tomographic Drift and Apparent Phantom Crossing}
\label{3.3}

\noindent\textbf{Results.} The tomographic (redshift-binned) regressions in Tables~\ref{tab5}--\ref{tab6} and Fig.~\ref{fig1} show that, under metric-only flat $\Lambda$CDM with $(\Omega_m=0.315,\;H_q=0)$, the per-bin estimates $H_\Lambda(z)$ are systematically elevated relative to the Planck reference band and exhibit an apparent positive redshift dependence across binning schemes and covariance treatments. This redshift drift is quantified by the weighted linear fits in Table~\ref{tab7}: for the pooled ``all-bins'' sample, the slopes are $b=3.161\pm0.311$ (D0), $3.111\pm0.328$ (ST), and $2.798\pm0.423~\mathrm{km\,s^{-1}\,Mpc^{-1}}$ per unit redshift (SS), with $p<0.001$ in all three modes. 

In contrast, under the hybrid framework with $H_q$ fixed to the global best-fit value in each mode, the inferred $H_\Lambda(z)$ remains largely consistent with the Planck range with substantially reduced drift (Fig.~\ref{fig1}); correspondingly, the ``all-bins'' slopes are consistent with zero, $b=-0.013\pm0.334$ (D0; $p=0.971$), $+0.037\pm0.369$ (ST; $p=0.921$), and $-0.218\pm0.455~\mathrm{km\,s^{-1}\,Mpc^{-1}}$ (SS; $p=0.637$).

\noindent\textbf{Analysis.} Interpreted within a metric-only paradigm, the statistically significant drift in $H_\Lambda(z)$ under $\Lambda$CDM is naturally read as an apparent redshift dependence in the late-time expansion history, i.e., a signature commonly associated with evolving dark energy. Consistently, the DDE benchmark fits in Table~\ref{tab8} favor departures from $(w_0,w_a)=(-1,0)$ toward the $w_0>-1$ and $w_a<0$ sector across all covariance modes: the CPL (C2) configuration yields $w_0=-0.735\pm0.114$, $w_a=-0.834\pm0.808$ (D0); $w_0=-0.716\pm0.092$, $w_a=-0.871\pm0.637$ (ST); and $w_0=-0.610\pm0.110$, $w_a=-1.526\pm0.737$ (SS). Similarly, the $w$CDM (w2) benchmarks consistently yield $w>-1$ ($w=-0.844\pm0.041$ in D0; $-0.834\pm0.033$ in ST; $-0.825\pm0.044$ in SS).

The qualitative alignment with recent large-scale analyses \citep{Adame2025,Rubin2025} suggests that the tomographic $H_\Lambda(z)$ drift is operationally degenerate with apparent DDE evolution. Empirically, whereas metric-only models redistribute this discrepancy into phantom-crossing-like parameter shifts, the hybrid framework suppresses the drift and restores the Planck-anchored constancy of $H_\Lambda$. This entails a theoretical trade-off: in the hybrid model, avoiding phenomenological complexities in the dark energy sector is premised upon a fundamental modification of the standard understanding of cosmological redshift and light propagation.

\subsection{Regression Results from the DDE Parameterization}
\label{3.5}

\noindent\textbf{Results.} As shown in Table~\ref{tab8}, for the $w$CDM configurations, the fully anchored case (w1) yields $w = -0.325 \pm 0.021$ in D0, $-0.352 \pm 0.019$ in ST, and $-0.253 \pm 0.021$ in SS, corresponding to absolute offsets of $0.675$, $0.648$, and $0.747$, respectively, from the cosmological-constant reference $w=-1$. In w2, where $\Omega_m$ is fixed to the Planck prior and $H_\delta$ is left free, the inferred expansion rate differs from the Planck reference value $67.4 \pm0.5~\mathrm{km\,s^{-1}\,Mpc^{-1}}$~\citepalias{Planck2020} by $8.1\sigma$ to $8.4\sigma$ across all modes, where the significance is evaluated from the quadrature sum of the quoted uncertainties. In w3, the fitted $\Omega_m$ differs from the Planck baseline $0.315 \pm0.007$~\citepalias{Planck2020} by $1.16\sigma$ to $1.32\sigma$ across the three modes, owing to the relatively broad uncertainties; however, the preferred central values remain consistently low, $\Omega_m=0.103$ to $0.155$, well below the Planck reference. At the same time, the corresponding fitted $w$ values cluster around $-0.2$ rather than $-1$. In w4, the inferred $H_\delta$ differs from the Planck reference value $67.4 \pm0.5~\mathrm{km\,s^{-1}\,Mpc^{-1}}$ by approximately $7.8\sigma$ to $8.1\sigma$ across the three modes. For w5 and w6, the fitted magnitude offset differs from the Pantheon-calibrated baseline $M=0$ by $16.3\sigma$ to $20.7\sigma$ across the three modes.

For the CPL models, the C1 configuration yields \textit{positive} $w_0$ values of $(w_0, w_a) = (0.471 \pm 0.078, -7.558 \pm 0.767)$ in D0, $(0.396 \pm 0.069, -6.835 \pm 0.662)$ in ST, and $(0.488 \pm 0.075, -6.797 \pm 0.719)$ in SS, implying absolute parameter offsets of more than 1.1 in $w_0$ and 6.0 in $w_a$ relative to the representative joint constraints of $(w_0, w_a) \approx (-0.83, -0.75)$ \citep{Adame2025} and $(-0.74, -0.79)$ \citep{Rubin2025}. In C2, the inferred $H_\delta$ differs from the Planck baseline $67.4 \pm0.5~\mathrm{km\,s^{-1}\,Mpc^{-1}}$ by $7.0\sigma$ to $7.7\sigma$ across all modes. In C3, the fitted $\Omega_m$ differs from the Planck baseline $0.315 \pm 0.007$ by $21.6\sigma$--$26.1\sigma$ across all modes, based on quadrature-combined uncertainties. In C4, the inferred $H_\delta$ differs from the Planck baseline by $7.1\sigma$ to $7.3\sigma$ in the D0 and ST modes, and by $5.6\sigma$ in the SS mode. For C5 and C6, the fitted magnitude offset differs from the Pantheon-calibrated baseline $M=0$ by $7.9\sigma$ to $15.3\sigma$ across all modes.

As shown in Figure~\ref{fig wCDM CPL Bin}, the binned $H_\delta$ estimates span $71.00$--$72.87$, $70.88$--$72.82$, and $70.64$--$72.93~\mathrm{km\,s^{-1}\,Mpc^{-1}}$ for the wb ($w$CDM) in the D0, ST, and SS modes, respectively, and $70.96$--$72.65$, $70.83$--$72.65$, and $70.55$--$72.62~\mathrm{km\,s^{-1}\,Mpc^{-1}}$ for the Cb (CPL).

\noindent\textbf{Analysis.} Metric-only DDE models redistribute the Planck-anchored residual into shifts in $\Omega_m$, $M$, or $H_\delta$, depending on the fixed quantities. Notably, the C2 configuration yields $(w_0, w_a) = (-0.735 \pm 0.114, -0.834 \pm 0.808)$ in D0, $(-0.716 \pm 0.092, -0.871 \pm 0.637)$ in ST, and $(-0.610 \pm 0.110, -1.526 \pm 0.737)$ in SS. Evaluated separately, these values remain within $0.06\sigma$--$1.71\sigma$ of the published one-parameter constraints on $w_0$ and $w_a$, namely $(-0.827 \pm 0.063, -0.75^{+0.29}_{-0.25})$ for DESI+CMB+Pantheon+, $(-0.744^{+0.097}_{-0.100}, -0.79^{+0.38}_{-0.35})$ for Union3+CMB+BAO, and $(-0.64 \pm 0.11, -1.27^{+0.40}_{-0.34})$ for the combined DESI+CMB+Union3 analysis, as reported by~\citet{Adame2025} and~\citet{Rubin2025}.

This statistical proximity motivates a phenomenological interpretation: the $(w_0, w_a)$ fits from the C2 configuration effectively serve as a metric-based parameterization to account for a non-metric WMC signature. This reasoning is grounded in the tomographic behavior (Table~\ref{tab7}) and by Figure~\ref{fig wCDM CPL Bin}. (i) Under the flat $\Lambda$CDM baseline (i.e., the $\Lambda2$ configuration), a residual structure grows proportionally with line-of-sight distance, manifesting as a positive redshift drift. However, this distance-proportional drift is absorbed differently depending on the parameterized framework. (ii) When the hybrid framework is applied (Qb), the WMC channel absorbs this residual, reducing the drift to near zero and stabilizing $H_\Lambda$ to be consistent with the Planck baseline. (iii) Conversely, when the DDE framework is applied (C2), the residual is absorbed by $w_0$ and $w_a$, manifesting as phantom crossing. Therefore, the phantom-crossing signature in DDE may act as an operational proxy for the WMC effect.

Since the Qb configuration absorbs the residual structure to recover a bin-by-bin $H_\Lambda$ consistent with the Planck baseline, and yields a lower BIC than the C2 configuration across all modes (Table~\ref{tab4}), the observational reports of evolving dark energy \citep{Adame2025, Rubin2025} may be reinterpreted phenomenologically as reflecting an observational degeneracy with the WMC framework.

\subsection{BIC Comparison and Parameter Stability}
\label{3.6}

\noindent\textbf{Results.} Table~\ref{tab4} shows that the hybrid model is not uniformly preferred over the DDE alternatives across all covariance treatments. Within the strongly anchored group ($\Lambda1$), Q1 has the lowest BIC in D0 and ST, whereas in SS the ordering reverses and w1 yields the lowest BIC, followed by C1 and then Q1. For the $H_\delta$-free group ($\Lambda2$), w2 gives the lowest BIC in D0, but in ST and SS the metric-only baseline $\Lambda2$ is preferred, with w2, Q2, and Qb remaining nearly degenerate. In the $\Omega_m$-free, $H_\delta$-fixed group ($\Lambda3$), Q3 is preferred in D0 and ST, whereas in SS both w3 and C3 outperform Q3, although $\Lambda3$ remains the lowest-BIC model in that mode. Among the jointly free group ($\Lambda4$), no alternative is preferred over $\Lambda4$ in any covariance mode, and Q4 and w4 remain close while C4 is less favored.

\vspace{2.0 mm}

\noindent\textbf{Analysis.} When a DDE model yields a lower BIC than the corresponding Q model, this is associated with a substantial shift in the fitted parameters. As summarized in Table~\ref{tab10}, SS-w1 and SS-C1 move to $w=-0.253\pm0.021$ and $(w_0,w_a)=(0.488\pm0.075,\,-6.797\pm0.719)$, the w2 cases remain at $H_\delta=72.03$--$72.22~\mathrm{km\,s^{-1}\,Mpc^{-1}}$, and SS-w3 and SS-C3 shift to $\Omega_m=0.155\pm0.122$ and $0.675\pm0.015$, respectively. In the corresponding SS comparisons, Q1 and Q3 remain comparatively stable, with $H_q=4.802\pm0.119$ and $(\Omega_m,H_q)=(0.314\pm0.026,\;4.811\pm0.246)$. The Q4 case, by contrast, deviates from the Planck baseline, but this configuration exhibits parameter degeneracy (Appendix~\ref{AppF}). Therefore, the statistical capacity of DDE parameterizations to absorb the Planck--Pantheon+SH0ES residual is distinct from their physical consistency with the Planck-anchored flat $\Lambda$CDM background.

\begin{table}

\centering
\setlength{\tabcolsep}{5.3pt}
\renewcommand{\arraystretch}{1.1}

\caption{
DDE cases with lower BIC than the corresponding hybrid Q model. Here $\Delta{\rm BIC}_{\rm DDE-Q}\equiv {\rm BIC}_{\rm DDE}-{\rm BIC}_{\rm Q}$ within the same index set and covariance mode, so negative values favor the DDE model. Only the fitted parameters most relevant to the associated displacement are listed.
}

\label{tab10}

{\fontsize{7.0}{9.4}\selectfont

\newcommand{\ccell}[2]{\parbox[c][#1][c]{\linewidth}{\centering #2}}

\begin{tabular}{
>{\centering\arraybackslash}m{0.10\columnwidth}
>{\centering\arraybackslash}m{0.10\columnwidth}
>{\centering\arraybackslash}m{0.17\columnwidth}
>{\centering\arraybackslash}m{0.45\columnwidth}
}
\toprule
Mode & ID & $\Delta{\rm BIC}_{\rm DDE-Q}$ & \shortstack[c]{Relevant fitted values} \\
\midrule
\ccell{1.35\baselineskip}{SS} &
\ccell{1.35\baselineskip}{w1} &
\ccell{1.35\baselineskip}{$-123.131$} &
\ccell{1.35\baselineskip}{$w=-0.253\pm0.021$} \\
\hline
\ccell{2.70\baselineskip}{SS} &
\ccell{2.70\baselineskip}{C1} &
\ccell{2.70\baselineskip}{$-98.238$} &
\ccell{2.70\baselineskip}{$w_0=0.488\pm0.075$\\$w_a=-6.797\pm0.719$} \\
\hline
\ccell{1.35\baselineskip}{D0} &
\ccell{1.35\baselineskip}{w2} &
\ccell{1.35\baselineskip}{$-0.706$} &
\ccell{1.35\baselineskip}{$H_\delta=72.223\pm0.285~\mathrm{km\,s^{-1}\,Mpc^{-1}}$} \\
\hline
\ccell{1.35\baselineskip}{ST} &
\ccell{1.35\baselineskip}{w2} &
\ccell{1.35\baselineskip}{$-0.092$} &
\ccell{1.35\baselineskip}{$H_\delta=72.067\pm0.238~\mathrm{km\,s^{-1}\,Mpc^{-1}}$} \\
\hline
\ccell{1.35\baselineskip}{SS} &
\ccell{1.35\baselineskip}{w2} &
\ccell{1.35\baselineskip}{$-0.192$} &
\ccell{1.35\baselineskip}{$H_\delta=72.030\pm0.279~\mathrm{km\,s^{-1}\,Mpc^{-1}}$} \\
\hline
\ccell{2.70\baselineskip}{SS} &
\ccell{2.70\baselineskip}{w3} &
\ccell{2.70\baselineskip}{$-118.812$} &
\ccell{2.70\baselineskip}{$\Omega_m=0.155\pm0.122$\\$w=-0.209\pm0.042$} \\
\hline
\ccell{4.05\baselineskip}{SS} &
\ccell{4.05\baselineskip}{C3} &
\ccell{4.05\baselineskip}{$-69.834$} &
\ccell{4.05\baselineskip}{$\Omega_m=0.675\pm0.015$\\$w_0=3.573\pm0.198$\\$w_a=-56.173\pm3.376$} \\
\hline
\ccell{2.70\baselineskip}{SS} &
\ccell{2.70\baselineskip}{w4} &
\ccell{2.70\baselineskip}{$-0.964$} &
\ccell{2.70\baselineskip}{$\Omega_m=0.169\pm0.090$\\$H_\delta=71.874\pm0.280~\mathrm{km\,s^{-1}\,Mpc^{-1}}$} \\
\bottomrule
\end{tabular}
}

\end{table}

\begin{figure*}
    \centering
    \includegraphics[width=\linewidth]{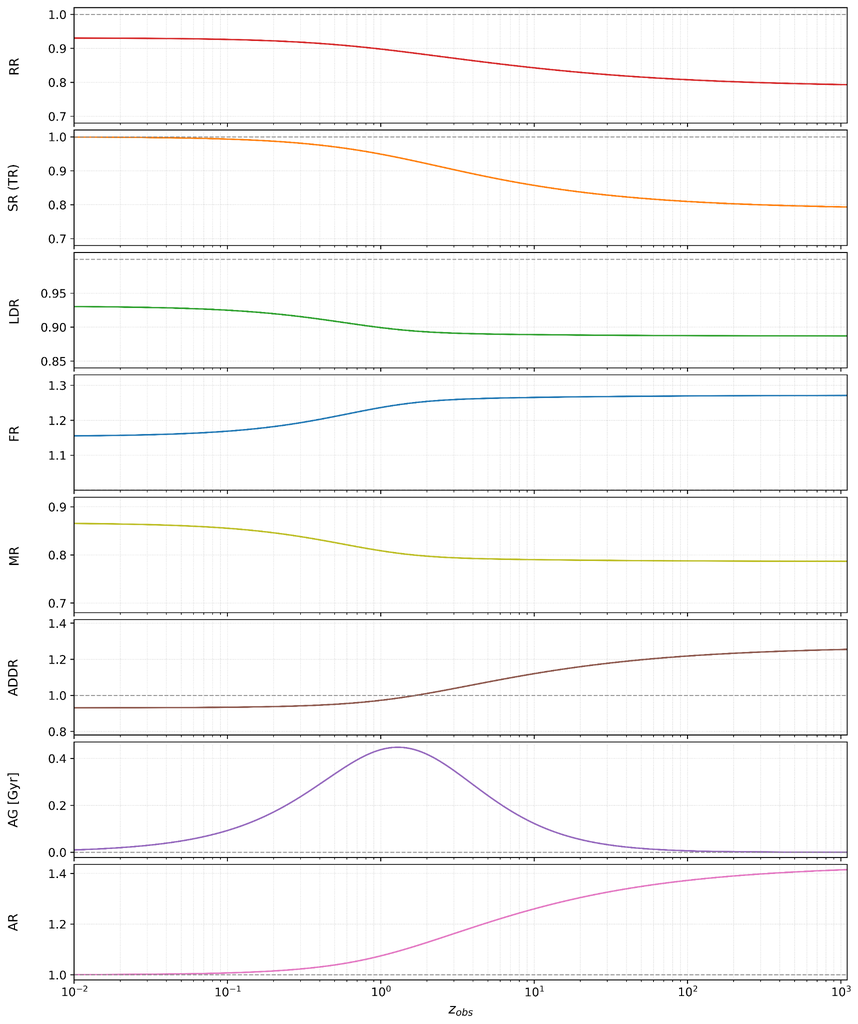}
\caption{
The trends of diagnostic indices are displayed, applying the global regression result of $H_q=5.0~{\rm km\,s^{-1}\,Mpc^{-1}}$ derived within the hybrid framework.
The horizontal physical quantity $z_{\rm obs}$ represents the observed redshift, defined as $z_{\rm obs}=(1+z_\Lambda)(1+z_q)-1$.
The local age gain ($\mathrm{AG}$) reaches a maximum of $\approx 0.448~\mathrm{Gyr}$ at $z_{\rm obs} \approx 1.293$, and the angular diameter distance ratio ($\mathrm{ADDR}$) crosses unity at $z_{\rm obs} \approx 1.688$.
Characteristic inflection points for each diagnostic metric are identified at $z_{\rm obs} \approx 0.556$ ($\mathrm{MR}$), $0.565$ ($\mathrm{LDR}$), $0.593$ ($\mathrm{FR}$), $2.443$ ($\mathrm{SR}$), $2.496$ ($\mathrm{RR}$), $3.019$ ($\mathrm{AR}$), $3.908$ ($\mathrm{ADDR}$), and $4.009$ ($\mathrm{AG}$).
}
\label{fig2}

\end{figure*}

\begin{table*}
\setlength{\tabcolsep}{5.5pt}
\centering
\caption{
Selected values of the hybrid diagnostics at representative observed redshifts $z_{\rm obs}$, evaluated for a flat $\Lambda$CDM background with $H_\Lambda=67.4~{\rm km\,s^{-1}\,Mpc^{-1}}$, $\Omega_m=0.315$, $\Omega_\Lambda=0.685$, and a WMC rate $H_q=5.0~{\rm km\,s^{-1}\,Mpc^{-1}}$. Over the sampled range, $\mathrm{ADDR}$ crosses unity at $z_{\rm obs}\simeq 1.688$, and $\mathrm{AG}$ reaches a maximum of $\simeq 0.448$~Gyr at $z_{\rm obs}\simeq 1.293$.
}
\label{tab9}
{\fontsize{7.2pt}{12.0pt}\selectfont
\begin{tabular}{c c c c c c c c c c c c c}
\Xhline{0.9pt}
$z_{\rm obs}$ & $z_\Lambda$ & $z_q$ & RR & TR(SR) & LDR & FR & MR & ADDR & AG [Gyr] & AR & EDR & TSBR \\
\Xhline{0.9pt}
0.010 & 0.009305 & 0.000688 & 0.930462 & 0.999312 & 0.930295 & 1.155469 & 0.865449 & 0.931257 & 0.009946 & 1.000729 & 0.998967 & 1.002068 \\
0.050 & 0.046430 & 0.003412 & 0.928591 & 0.996600 & 0.927816 & 1.161654 & 0.860842 & 0.932568 & 0.048275 & 1.003686 & 0.994904 & 1.010271 \\
0.100 & 0.092634 & 0.006744 & 0.926337 & 0.993303 & 0.924926 & 1.168924 & 0.855488 & 0.934295 & 0.092939 & 1.007467 & 0.989972 & 1.020362 \\
0.500 & 0.455636 & 0.030481 & 0.911271 & 0.970424 & 0.908733 & 1.210953 & 0.825796 & 0.950592 & 0.333920 & 1.038890 & 0.955965 & 1.094248 \\
1.000 & 0.897960 & 0.053751 & 0.897959 & 0.948980 & 0.899301 & 1.236488 & 0.808742 & 0.972792 & 0.437590 & 1.074857 & 0.924454 & 1.170117 \\
1.293 & 1.153412 & 0.064822 & 0.892043 & 0.939124 & 0.896473 & 1.244303 & 0.803663 & 0.985037 & 0.447819 & 1.092818 & 0.910090 & 1.207344 \\
1.500 & 1.399187 & 0.071730 & 0.888458 & 0.933075 & 0.895104 & 1.248109 & 0.801212 & 0.993114 & 0.444175 & 1.104164 & 0.901311 & 1.230980 \\
1.688 & 1.494797 & 0.077442 & 0.885543 & 0.928124 & 0.894161 & 1.250745 & 0.799523 & 1.000015 & 0.436297 & 1.113599 & 0.894147 & 1.250784 \\
2.000 & 1.762579 & 0.085948 & 0.881290 & 0.920860 & 0.893020 & 1.253944 & 0.797484 & 1.010580 & 0.417864 & 1.127655 & 0.883670 & 1.280618 \\
3.000 & 2.613041 & 0.107135 & 0.871014 & 0.903260 & 0.891149 & 1.259214 & 0.794146 & 1.038080 & 0.348278 & 1.162593 & 0.858459 & 1.356942 \\
5.000 & 4.291942 & 0.133785 & 0.858388 & 0.881990 & 0.889886 & 1.262792 & 0.791896 & 1.074332 & 0.241545 & 1.206352 & 0.828315 & 1.457500 \\
8.000 & 6.780317 & 0.156725 & 0.847540 & 0.864480 & 0.889211 & 1.264709 & 0.790696 & 1.106299 & 0.155523 & 1.243798 & 0.803771 & 1.547874 \\
10.000 & 8.427278 & 0.166822 & 0.842728 & 0.857025 & 0.888965 & 1.265409 & 0.790259 & 1.120454 & 0.122870 & 1.260202 & 0.793397 & 1.588617 \\
15.000 & 12.519206 & 0.183502 & 0.834614 & 0.844950 & 0.888588 & 1.266483 & 0.789588 & 1.144072 & 0.077389 & 1.287443 & 0.776688 & 1.657702 \\
50.000 & 40.777215 & 0.220629 & 0.815548 & 0.819164 & 0.887798 & 1.268737 & 0.788185 & 1.197451 & 0.016502 & 1.348784 & 0.741407 & 1.819227 \\
\Xhline{0.9pt}
\end{tabular}
} 
\end{table*}

\subsection{Diagnostic Parameters in the Hybrid Model}
\label{3.4}

\noindent\textbf{Results.} Table~\ref{tab9} and Figure~\ref{fig2} summarize the quantitative behavior of the diagnostic parameters over the evaluated redshift range (Table~\ref{tab9}: $z_{\rm obs}=0.01$--$50$; Figure~\ref{fig2}: extended to higher $z_{\rm obs}$). The $\mathrm{AG}$ exhibits a non-monotonic evolution with a distinct peak at $z_{\rm obs} \approx 1.29$, where the absolute age difference reaches a maximum of approximately $0.448~\mathrm{Gyr}$. The $\mathrm{AR}$ increases monotonically with redshift, rising from near unity at low redshift to approximately $1.26$ at $z_{\rm obs}=10$.

The $\mathrm{ADDR}$ shows a crossing behavior relative to unity. It remains below unity at low redshifts and crosses $1.0$ at $z_{\rm obs} \approx 1.69$. For redshifts $z_{\rm obs} \gtrsim 1.69$, $\mathrm{ADDR}$ exceeds unity, indicating that the hybrid angular diameter distance is larger than the standard metric prediction in this regime. 

The $\mathrm{LDR}$ stays below unity ($\mathrm{LDR} < 1$) across the entire sampled redshift range and exhibits a decreasing trend as redshift increases. Conversely, the Flux Ratio ($\mathrm{FR}$) consistently exceeds unity, ranging from $\sim 1.15$ at $z_{\rm obs}=0.01$ to a saturation level near $\sim 1.27$ at high redshift. The $\mathrm{MR}$, defined as $\left[\mathrm{LDR}\right]^2$, consistently falls below unity, decreasing from $\sim 0.86$ at low redshift to $\sim 0.79$ at high redshift.

In the low-redshift limit, the $\mathrm{RR}$ converges to a constant value of approximately $0.93$. As redshift increases, $\mathrm{RR}$ decreases monotonically from this initial value. The $\mathrm{TR}$ exhibits a value near unity in the low-redshift ($z_{\rm obs}=0$) approximation and shows a monotonically decreasing trend.

The $\mathrm{EDR}$ exhibits a value near unity in the low-redshift ($z_{\rm obs}=0$) approximation, remains below unity ($\mathrm{EDR} < 1$) throughout the redshift range, and shows a monotonically decreasing trend. In contrast, the $\mathrm{TSBR}$ consistently exceeds unity ($\mathrm{TSBR} > 1$) and increases monotonically with redshift.

\vspace{2.0mm}

\noindent\textbf{Analysis.} The condition $\mathrm{TR} < 1$ (equivalent to $\mathrm{SR} < 1$) reveals that the standard model applies an excessive time-dilation correction by attributing the full redshift to expansion, as cautioned in Section~\ref{2.4.5}. In principle, this over-correction can propagate into luminosity standardization relations that depend on rest-frame timescales, including the Phillips relation for SNe~Ia and the Leavitt law for Cepheids \citep[e.g.,][]{Phillips1993,Leavitt1912}, thereby biasing the inferred intrinsic luminosities. However, the absolute magnitude calibration in the Pantheon+SH0ES analysis is strictly anchored by Cepheid-hosting galaxies located at $z < 0.01$ \citepalias{Riess2022,Brout2022}. In this ultra-low redshift regime, the deviation from unity is negligible ($1-\mathrm{TR} < 0.01\%$, Table~\ref{tab9}), ensuring that the baseline calibration of $M$ remains immune to the proposed time-dilation modification. Consequently, our working assumption that the Pantheon+SH0ES dataset provides a valid baseline for regression is methodologically sound. Nevertheless, as the redshift increases, the cumulative effect of the $\mathrm{TR}$ divergence becomes significant, suggesting that the intrinsic luminosities inferred by the standard pipeline at high redshifts may require progressively larger corrections.

Because the luminosity-distance ratio remains below unity ($\mathrm{LDR}<1$) and approaches a quasi-saturated regime beyond the first transition in the diagnostics, a metric-only analysis systematically overestimates the luminosity distance and must compensate by inflating intrinsic source properties. This is reflected in $\mathrm{MR}\equiv \mathrm{LDR}^2<1$: under the standard mapping, the same observed flux is matched by assigning a larger distance and hence a larger stellar mass, so that otherwise normal high-redshift galaxies are biased toward being inferred as over-massive. In observational inferences that leverage the empirical link between galaxy stellar mass and evolutionary stage (age), such mass overestimation can in turn bias the interpretation toward a more mature (older) galaxy population than is warranted. In parallel, the age diagnostics show that metric-only inference compresses the cosmic timeline: the non-monotonic $\mathrm{AG}$ (peaking near $z_{\rm obs}\simeq 1.29$) and the monotonic rise of $\mathrm{AR}$ together imply that the standard pipeline tends to underestimate ages increasingly toward higher redshift, thereby enhancing the apparent prevalence of overly mature stellar populations. Finally, the angular-diameter distance ratio exhibits a sign reversal across its unity crossing (near $z_{\rm obs}\simeq 1.69$): below this threshold, the standard mapping overestimates physical scales, whereas above it the mapping underestimates angular-diameter distances and infers radii that are artificially small, making high-redshift systems appear anomalously compact. Taken together, if the WMC channel is unmodeled, high-redshift sources are pushed toward a coherent inference bias in which galaxies are inferred to be younger, smaller, and more massive, more prematurely evolved, and denser than in the hybrid mapping, implying systematically overestimated mean densities at high redshift.

The apparently large deviation $\mathrm{LDR}(z_{\rm obs}\!\to\!0)\simeq0.93$ cannot serve as evidence that the Pantheon+SH0ES dataset inherently misrepresents the luminosity (or the distance--luminosity relation). In the low-redshift limit where $d_\Lambda(z)\propto z$, the definition yields
\begin{subequations}
\begin{align}
\mathrm{LDR} \equiv \frac{d_h(z_{\rm obs})}{d_\Lambda(z_{\rm obs})}
&= \frac{d_\Lambda(z_\Lambda)\sqrt{1+z_q}}{d_\Lambda(z_{\rm obs})}
\label{eq:42a}\\[1.3ex]
&\simeq \frac{z_\Lambda}{z_{\rm obs}}\sqrt{1+z_q}\,\simeq \left(1+\frac{z_q}{z_\Lambda}\right)^{-1}\sqrt{1+z_q}
\label{eq:42b}\\[1.5ex]
&\simeq\; \left[1+\frac{H_q}{H_\Lambda}\right]^{-1}\simeq\left[1+\frac{5.0}{67.4}\right]^{-1} 
\label{eq:42c}\\[2.1ex]
&\approx 0.93.
\label{eq:42d}
\end{align}
\end{subequations}
where $\sqrt{1+z_q}$ is almost equal to 1 (differing by less than 0.1\% at $z_{\rm obs}\sim 0.01$). Consequently, the $\sim7\%$ deviation in luminosity distance is strictly a mathematical consequence of the additional redshift component postulated by the WMC framework (essentially, a parameter derived relative to the total redshift). Meanwhile, we adopt an operational definition of the dust-mimicking magnitude compensation (DMC); this operational premise can explain why no anomalies have been reported in luminosity observations to date, despite the theoretical presence of an offset at the level of $\mathrm{LDR}\simeq0.93$. In summary, due to the DMC, the luminosity in the Pantheon+ dataset is effectively restored to its intrinsic value, and the offset of $\mathrm{LDR} \simeq 0.93$ is analyzed as a direct consequence of the WMC-derived redshift component present in the observed data. Applying this same analytical framework, the magnitude offsets ($\lesssim -0.15~\mathrm{mag}$) observed in regression configuration IDs $\Lambda$5 and $\Lambda$6 discussed in Section~\ref{3.1} should be interpreted not as an intrinsic luminosity tension, but as a signature of the additional redshift induced by WMC (or equivalently, as a proxy for $H_q$).

Notably, analyzed effects are wavelength-dependent by operational definition. This implies that the violation of the Etherington reciprocity relation \citep{Etherington1933} and the non-compliance with the Tolman surface brightness law \citep{Tolman1930}, as manifested in the $\mathrm{EDR}$ and $\mathrm{TSBR}$ diagnostics, are strictly determined by the wavelength of the observed electromagnetic radiation.

\section{DISCUSSION}
\label{4}

\subsection{Implications for Hubble Tension}
\label{4.1}

As analyzed and presented in the Results and Analysis sections~\ref{3.1}--\ref{3.3}, it has been confirmed that the hybrid framework allows for the compatibility of the Planck baseline~\citepalias{Planck2020} and the Pantheon+SH0ES compilation~\citepalias{Brout2022, Scolnic2022, Riess2022}. In the redshift-binned regression ($z_{\rm center} > 0.016$), the hybrid model recovered a metric Hubble constant, $H_\Lambda$, that is nearly constant across bins and consistent with the Planck value ($67.4 \pm 0.5~\mathrm{km\,s^{-1}\,Mpc^{-1}}$) within $\lesssim 0.33\sigma$. In contrast, under the flat $\Lambda$CDM framework, models fixed to the Planck or Pantheon+SH0ES baselines were statistically disfavored based on the Bayesian Information Criterion (BIC), or failed to yield consistent values through regression. Furthermore, the metric-only binned analysis exhibited a ``drift'' phenomenon where higher $H_\Lambda$ values are inferred at higher redshift bins. This drift and the regression results presented in Table~\ref{tab8} were analyzed as being phenomenologically equivalent to the ``phantom crossing'' signature ($w_0 > -1$, $w_a < 0$) observed in CPL parameterizations in recent large-scale studies~\citep{Adame2025, Rubin2025}.

Regarding the diagnostic indices discussed in Section~\ref{3.4}, under the working hypothesis of the hybrid framework where the intrinsic magnitude (luminosity) is effectively restored by the Dust-mimicking Magnitude Compensation (DMC) mechanism while the WMC-induced excess redshift remains, an offset of $M \approx 0.15~\mathrm{mag}$ is observed in the low-redshift approximation. However, this is analyzed not as an intrinsic calibration error in the Pantheon+SH0ES dataset, but as a relative magnitude difference arising from the excess redshift component ($z_q$) added by the WMC effect.

Additionally, Appendix~\ref{AppC} demonstrates that when WMC manifests in a wavelength-dependent manner, the DMC mechanism can naturally operate through the color-correction terms in the Tripp Estimator~\citep{Tripp1998, Guy2007, Betoule2014}. Notably, as discussed in Appendix~\ref{AppB}, the wavelength dependence of WMC provides a physical alternative that allows for the coexistence of the diverse Hubble constant values reported based on the extension of the SNe~Ia Hubble flow, such as the JAGB and TRGB based values ($H_0 \approx 67.8$ and $68.8~\mathrm{km\,s^{-1}\,Mpc^{-1}}$) reported by \citetalias{Freedman2025} and the Cepheid-based value ($H_0 \approx 73.0~\mathrm{km\,s^{-1}\,Mpc^{-1}}$) by \citetalias{Riess2022}.

Synthesizing these results, our hybrid framework can be considered a viable candidate capable of alleviating the Hubble tension. A definitive judgment on its universality requires extending the analysis to determine whether other tension-level Hubble measurements can be similarly alleviated by accounting for the WMC effect. As introduced in Section~\ref{1}, the high Hubble values inferred from various probes can be discussed as potential manifestations of the WMC effect. Because the WMC-induced non-metric redshift systematically alters the inferred distance--redshift mapping across all scales, we suggest that a re-analysis within the WMC framework is warranted for independent local measurements, such as the Tully-Fisher relation reported using spiral galaxies~\citep{Schombert2020, Kourkchi2020}, the time-delay cosmography results from H0LiCOW~\citep{Wong2020, Shajib2020}, the IR-SBF measurements calibrated by Cepheids \citep{Blakeslee2021} and TRGB \citep{Jensen2025}, and the independent distance measurements using SNe~II~\citep{deJaeger2020, Vogl2025}.

Beyond alleviating the Hubble tension, the WMC framework may also offer a common theoretical basis for interpreting other high-redshift anomalies. As detailed in Section~\ref{3.4}, metric-only inferences that omit WMC can induce structural biases through distorted distance--redshift mappings. The broader implications of correcting these distortions are examined in Section~\ref{4.2}.

\subsection{Implications for Observational Inferences}
\label{4.2}

As analyzed in Section~\ref{3.4}, if observational inferences are conducted within the flat $\Lambda$CDM framework without accounting for WMC, specific biases arise in the high-redshift regime (equivalent to the early Universe): objects are inferred to be brighter (lower magnitude relative to redshift), more massive, smaller, denser, younger, and more prematurely evolved than they are in the hybrid framework. Conversely, in the low-redshift regime ($z \lesssim 1.7$), physical dimensions and volumes can be systematically overestimated. Furthermore, as proposed in Appendix~\ref{AppC}, the wavelength dependence of WMC systematically causes the SEDs of targets observed in the high-redshift regime to appear redder. Consequently, this can induce an aggravated misinterpretation of the evolutionary state of these targets based on color--evolution relationships \citep[e.g.,][]{Bell2001, Taylor2011}. Notably, at the bright or massive end of luminosity and stellar mass functions—regions characterized by a steep decline in distribution—even modest shifts in inferred mass or luminosity estimates can result in exponential changes in the inferred number density (abundance) \citep{Schechter1976}. In other words, a slight overestimation of mass or luminosity shifts targets into the exponential tail of the distribution, thereby leading to the misclassification of intrinsically common galaxies as statistically improbable anomalies.

Applying this perspective to the anomalous interpretations reported in high-redshift observations---those that conflict with or challenge the standard framework---provides a critical cross-validation test. Examining whether the degree of tension in these anomalies can be alleviated under the hybrid framework serves as a major criterion for assessing the plausibility of the WMC hypothesis beyond the alleviation of the Hubble tension.

\subsubsection{JWST Anomalies}
\label{4.2.1}

JWST observations have reported high-redshift galaxy candidates that appear to be in tension with the standard $\Lambda$CDM framework. These anomalies can be broadly categorized into three phenomenological classes: extreme compactness, excessive stellar mass, and premature evolution. First, regarding \textit{size and compactness}, high-redshift galaxies exhibit a systematic trend of decreasing size. \citet{Ormerod2024} found that galaxy sizes scale as $(1+z)^{-0.71}$ out to $z \sim 8$. Furthermore, \citet{Baggen2023} showed that the massive candidates identified by \citet{Labbe2023} are extremely compact, with mean effective radii $\langle r_e \rangle \approx 150$~pc, making their overall sizes significantly smaller than their putative local descendants despite exhibiting comparable central stellar densities. Second, regarding \textit{mass and baryon efficiency}, the discovery of massive candidates at $z \sim 7.5$--$9.1$ challenges standard formation efficiency limits. \citet{Labbe2023} reported stellar mass densities significantly exceeding extrapolations from UV-selected samples. \citet{BoylanKolchin2023} demonstrated that reconciling these inferred masses with the standard halo abundance requires a baryon-to-star conversion efficiency of $\epsilon \approx 1$, a value in strong tension with the canonical limit ($\epsilon \le 0.2$). Third, regarding \textit{age and formation epochs}, spectroscopic confirmation of massive quiescent galaxies at $z \sim 3.2$, such as ZF-UDS-7329, implies formation epochs as early as $z \sim 11$ \citep{Glazebrook2024, Nanayakkara2024}. This rapid quenching challenges the timescale available for halo assembly and stellar mass buildup within the standard concordance cosmology.

In summary, these anomalies qualitatively align with the systematic biases expected when the WMC effect is neglected in the standard $\Lambda$CDM framework.

\subsubsection{$S_8$ Tension}
\label{4.3}

The $S_8$ tension reflects a discrepancy where low-redshift observations, such as weak gravitational lensing, infer a lower ``clustering amplitude'' than predicted by the \textit{Planck} baseline~\citep{Hildebrandt2017,Dalal2023,GarciaGarcia2024}. Specifically, while the \textit{Planck} primary CMB analysis yields $S_8 \approx 0.83$~\citepalias{Planck2020}, local weak-lensing surveys report systematically lower values, typically in the range of $S_8 \approx 0.77\text{--}0.78$~\citep{Abbott2022, Heymans2021}. This parameter, defined as $S_8 \equiv \sigma_8 (\Omega_m / 0.3)^{1/2}$, quantifies matter fluctuations on physical scales of $\sim 8\,h^{-1}\,\text{Mpc}$. Since weak-lensing analyses transform observed angular shear signals into physical density fluctuations, they are fundamentally dependent on the distance-redshift mapping. Consequently, even a minor bias in the standard distance scale can shift the interpretation of lensing signals toward an apparent deficit in clustering.

In our hybrid framework, as analyzed in Section~\ref{3.4}, the $\mathrm{ADDR}$ systematically drops below unity in the low-redshift regime ($z \lesssim 1.7$). This implies that calculating distances using the standard metric while neglecting the WMC effect leads to an overestimation of the angular diameter distance and, consequently, of the physical volumes inferred from observations. By interpreting a fixed lensing signal within an artificially inflated volume, the standard framework naturally yields a reduced clustering amplitude. Thus, the $S_8$ tension should not be viewed solely as a physical lack of matter clustering; rather, it may reflect a systematic error in the scale transformation that biases the results in that direction.

To a first-order approximation, the ratio between these disparate values ($0.77 / 0.83 \approx 0.93$) is consistent with the $\mathrm{ADDR}$ values obtained in our analysis for the low-redshift regime ($z \lesssim 1.7$). Given that $\sigma_8$ is fundamentally defined over a specific linear spatial scale ($8\,h^{-1}\,\text{Mpc}$), this numerical correspondence suggests that the observed tension may be a direct consequence of a metric-induced scale shift. Determining the precise impact of this geometric effect requires rigorous cross-validation by re-evaluating the entire weak-lensing pipeline under the hybrid distance mapping.

\subsection{Future Tests of the WMC Signatures}
\label{4.3}

Deep Space Optical Communications (DSOC) offers a viable platform for experimental verification of WMC signatures. While the inferred WMC rate, $H_q \sim 5\,\mathrm{km\,s^{-1}\,Mpc^{-1}}$, is minute, the extreme precision of modern metrology relaxes the need for cosmological path lengths. For a spacecraft at a distance comparable to Pluto's orbit ($d \sim 40\,\mathrm{AU}$), the predicted cumulative fractional frequency shift is $\Delta \nu / \nu \approx 3 \times 10^{-15}$. Since state-of-the-art optical lattice clocks have demonstrated fractional instabilities reaching the $10^{-18}$ level \citep{McGrew2018, Bothwell2022}, this signal provides a measurement margin of nearly three orders of magnitude, while coherent multi-band heterodyne (beat-note) comparisons may render even shorter baselines experimentally relevant. By establishing phase-coherent laser links across multiple wavelength bands, the resulting differential beat-note offsets can isolate the wavelength-dependent WMC signature from achromatic relativistic Doppler shifts from spacecraft motion and gravitational redshifts.

Notably, validating the WMC mechanism would enable direct cosmic distance measurement from the observed redshift alone:
\begin{equation}
X = \frac{c_{l}}{H_\Lambda} \int_{0}^{\left[(1+z_{\rm obs})\exp\left(-\frac{H_q}{c_{\scriptscriptstyle l}}X\right)-1\right]} \frac{dz'}{\sqrt{\Omega_m(1+z')^3 + (1-\Omega_m)}}.
\label{eq43}
\end{equation}
The distance $X$ is determinable provided that the WMC rate $H_q$, $H_\Lambda, \Omega_m$, and the observed redshift $z_{\rm obs}$ are specified.

\subsection{Implications for the Foundations of the Vacuum}
\label{4.4}

The WMC framework operationally defines the redshift mechanism as a partial transition of photon energy from a radiative to a non-radiative state during propagation, manifesting as an effective energy attenuation of the propagating radiation. From the perspective of classical optics, such attenuation is formally represented by the imaginary component of the complex refractive index, which encodes energy dissipation in a medium~\citep{Born1999}. Furthermore, in the electrodynamics of continuous media, optical response functions are coupled to the medium's mechanical response, including elasticity- and viscosity-related behavior~\citep{Landau1984}. Motivated by this analogy, we introduce effective vacuum parameters---an effective elastic modulus ($K_{\rm eff}$) and an effective viscosity-like coefficient ($V_{\rm eff}$)---as an operational parametrization of the WMC mechanism.

By analogy with General Relativity and classical mechanics, these effective quantities can be estimated as order-of-magnitude scales rather than literal material constants. The vacuum's resistance to geometric deformation (curvature) is governed by the Einstein field equations, suggesting an large effective stiffness scale of order
\begin{equation}
    K_{\rm eff} \sim \frac{c^4}{8\pi G} \approx 4.8 \times 10^{42} \, \mathrm{N},
\end{equation}
where $G$ denotes the gravitational constant~\citep{Sakharov1968}. Conversely, the inferred WMC attenuation rate implies a finite, but extremely small, effective viscosity (or damping rate):
\begin{equation}
    V_{\rm eff} \sim H_q \approx 1.6 \times 10^{-19} \, \mathrm{s}^{-1},
\end{equation}
based on the representative WMC rate $H_q \approx 5 \, \mathrm{km \, s^{-1} \, Mpc^{-1}}$. Consistent with the operational definition of WMC, this effective damping scale is wavelength-dependent, analogous to the chromatic dispersion and absorption behavior of classical dispersive media.

Recalling Einstein's 1920 assertion that ``space without ether is unthinkable'' in the sense that the metric field itself possesses physical qualities~\citep{Einstein1920}, and noting later effective-medium and emergent-vacuum approaches~\citep{Sakharov1968, Pendry1997, Volovik2003}, we treat the vacuum here as an operational physical entity rather than a purely geometric abstraction.

\subsection{Implications for Cosmology}
\label{4.5}

\subsubsection{Thermal Fate of Our Universe}
\label{4.5.1}

If propagating light continuously transitions into non-radiative degrees of freedom through interaction with the vacuum, and if these are interpreted as matter possessing equivalent mass, this process inevitably results in a cumulative increase in matter density. Concurrently, as the Universe expands and the physical distance between baryons grows, the volume of spacetime (the vacuum) increases commensurately. If the generation of WMC non-radiative degrees of freedom occurs at a constant ratio relative to this increasing volume---specifically, in a scenario where the total WMC generation rate balances the volumetric expansion rate---this implies a state of dynamic equilibrium. This mechanism is analogous to a vessel being replenished with water at the exact rate its capacity expands, thereby maintaining a constant level.

Such a scenario predicts the emergence of a thermal equilibrium within the Universe, under the premise that the intrinsic energy of the vacuum~\citep{Weinberg1989} is incorporated into the effective mass of the non-radiative degrees of freedom, offering a potential explanatory pathway for the observation that the cosmic microwave background exhibits a nearly perfect blackbody radiation spectrum~\citep{Fixsen1996}.

\subsubsection{The Origin of Dark Matter}
\label{4.5.2}

We explore whether the non-radiative degrees of freedom---operationally, the effective mass generated by WMC---can be associated with Dark Matter. Because WMC is cumulative along the propagation path and wavelength-dependent, one might argue that it merely mimics uniform absorption by an intervening medium; however, any such absorber must either re-emit (at least partially) the absorbed energy or absorb it without re-emission.

The first hypothesis---absorption followed by re-emission---is effectively excluded by observations: re-emission implies scattering and would blur astronomical images, yet high-redshift point sources remain sharp even at cosmological distances~\citep[e.g.,][]{Windhorst2023}. The second case---absorption without re-emission---is phenomenologically indistinguishable from WMC; the distinction is only whether the agent is a particulate medium or the vacuum.

If light transfers part of its energy to degrees of freedom operationally defined as non-radiative via interaction with the vacuum, it leaves no direct electromagnetic signature while in that state, aside from the redshift and color change in the residual radiative component. Hence the non-radiative degrees of freedom implied by WMC---initially introduced to alleviate the Hubble tension---naturally emerge as a Dark Matter candidate, motivating a reinterpretation of its origin, epoch-dependent accumulation, and gravitational distribution; consequently, the mechanisms governing large-scale structure formation and galaxy evolution may require fundamental re-examination. If the accumulation of dark matter is a dynamically evolving process, this temporal reinterpretation provides an alternative perspective on the severe timescale conflicts observed in early massive galaxies. As highlighted by \citet{Glazebrook2024}, systems such as ZF-UDS-7329 imply formation epochs ($z \approx 11$) when dark matter halos of sufficient hosting mass had not yet assembled under the standard scenario.

\section{Conclusion}

We report that the Pantheon+SH0ES data and the Planck-based $\Lambda$CDM baseline are compatible within our hybrid framework, demonstrating that the Hubble tension is alleviated by accounting for a cumulative, wavelength-dependent non-metric attenuation.

Navigating the intricacies of modern cosmology requires a definitive anchor. As articulated by Harry Collins \citep{Collins1985}, resolving the ``experimenter's regress'' demands a stabilizing criterion to break the circularity of validating a measurement against an unknown true value. In this study, we adopted the Planck-based $\Lambda$CDM model as our reference point. It is crucial to emphasize, however, that validating the physical coherence of the Planck baseline must not be misconstrued as implying that the Pantheon compilation is flawed or incorrect. 

Confronted with contrasting observational inferences, we are liberated from the necessity of a dichotomous choice within our hybrid framework. Elevating any single pipeline as the exclusive bearer of truth risks the \emph{epistemic fallacy} warned of by \citet{Bhaskar1975}---conflating our fragmented empirical access with reality itself. Taking to heart the lesson versified by \citet{Saxe1872}, we can instead appreciate that each observational channel---whether the Planck baseline~\citepalias{Planck2020,Tristram2024}, the Pantheon compilation~\citepalias{Brout2022,Scolnic2022}, the ``baked-in'' imprint~\citepalias{Scolnic2025}, independent distance ladders~\citepalias{Freedman2025,Riess2022}, or recent cosmic surveys~\citep{Adame2025,Rubin2025,Abbott2024}---likely captures a valid but partial projection of a greater whole. They serve not as mutually exclusive competitors, but as deeply valid reference points.

From the outset, distinguishing kinematic recession from other propagation-induced energy-loss mechanisms has been recognized as an open empirical question \citep{HubbleTolman1935}. Rather than elevating the expansion-centered interpretation to an immutable axiom, placing this foundational ambiguity back on the bench with modern precision data---whether it ultimately strengthens the prevailing framework or demands its modification---constitutes a scientific mandate.

Paradoxically, entering the high-redshift era, our results suggest that precision cosmology can function as a sensitive probe of microscopic quantum effects whose signatures emerge once integrated over cosmological path lengths.

\section*{Acknowledgements}

%The acknowledgements are presented in approximate chronological order, reflecting the development of the ideas and analyses underlying this work.

\vspace{0.35mm}

The author expresses deep appreciation to Professor Jim Peebles for his insights into the cosmological challenges that can arise when considering frameworks beyond flat $\Lambda$CDM. He emphasized that, while independent approaches are valuable, departures from the prevailing framework should also address how the established observational successes of the standard model are to be retained or replaced. This perspective motivated the view that some apparent discrepancies may be most productively approached by introducing additional physical ingredients that can coexist with the standard framework.

\vspace{0.35mm}

Heartfelt thanks are extended to Professor Changbom Park, whose response to the early conceptual idea of this work---concerning the possibility that photons might, albeit with low probability, convert into matter during propagation---emphasized the necessity of observational verification and thereby motivated the subsequent use of Pantheon+SH0ES data as regression baselines.  

\vspace{0.35mm}

Sincere gratitude is extended to Professor Rajendra Gupta for kindly sharing prior research on redshift decomposition and hybrid formulations. The dual-redshift model, conceived independently by the author from an information-theoretic perspective, was refined into a more rigorous physical formulation by building upon the pioneering hybrid framework developed by Professor Gupta. Furthermore, his work provided a valuable foundation for developing the diagnostic indices essential to the validation of the Hybrid framework.

\vspace{0.35mm}

The author is indebted to Professor Hyunsun Park. Her lectures on research methodology---specifically the conceptual metaphor of the \textit{window and the flower}---illuminated the epistemic importance of the chosen observational frame. This insight motivated the reinterpretation of discordant Hubble measurements not merely as conflicting systematic errors, but as physical outcomes derived from \textit{distinct observational windows}, each possessing its own intrinsic validity.

\vspace{0.35mm}

Special thanks are due to the reviewer, Patrick Armstrong, for suggesting the inclusion of full systematic covariance and extended statistical benchmarks, which motivated additional analyses across multiple covariance modes and allowed us to verify that the main results remain consistent under different covariance treatments. The author also thanks Patrick Armstrong for his meticulous efforts in refining the presentation of the manuscript, thereby helping its central message be conveyed with greater clarity and objectivity.

\vspace{0.35mm}

Particular appreciation is extended to Chansol Noh for providing computational resources that accelerated the regression calculations and for serving as a thoughtful sounding board through attentive listening and constructive discussions, which helped the author clarify the conceptual framing of this work\footnote{The acknowledgements have been updated in this author version.}.

%%%%%%%%%%%%%%%%%%%%%%%%%%%%%%%%%%%%%%%%%%%

\appendix

\section{Distance Modulus for Regression}
\label{AppA}

As introduced in Sections~\ref{1.4} and~\ref{1.5}, our study operationally assumes that the Pantheon+SH0ES dataset encodes both cosmic expansion and the WMC effect. Because this dataset provides the empirical baseline for fitting the Hubble diagram, the regression distance modulus defined in our hybrid framework (equation~\eqref{eq24b}) requires mathematical scrutiny before the regression results can be interpreted robustly. This is particularly important because WMC is operationally defined as wavelength-dependent, so the additional WMC-induced redshift is expected to vary with the effective emission wavelengths of the distance-ladder tracers, notably Cepheids (absolute-magnitude calibration) and SNe~Ia (Hubble-flow extension). We therefore examine how these observational heterogeneities propagate through the defined regression modulus in order to validate the overall regression results.

Let $L_{0}$ represent the true, intrinsic luminosity of a Cepheid. In practice, however, calibrating this absolute luminosity from observations inevitably incorporates some degree of error. We account for this by introducing a multiplicative factor, $\epsilon_{\rm cep}$, which represents the overall calibration error. A perfectly accurate calibration corresponds to $\epsilon_{\rm cep}=1$. The inferred Cepheid luminosity, $L_{\rm cep}$, is thus simply written as
\begin{equation}
    L_{\rm cep} \equiv \epsilon_{\rm cep}\,L_{0}.
    \label{eqAppA1}
\end{equation}

Once $L_{\rm cep}$ is calibrated using a geometric distance anchor, we can infer the distance to a host galaxy by observing the Cepheid flux, $F_{\rm cep}$:
\begin{equation}
    d_{\rm host} = \sqrt{\frac{L_{\rm cep}}{4 \pi F_{\rm cep}}}
    = \sqrt{\frac{\epsilon_{\rm cep}\,L_{0}}{4 \pi F_{\rm cep}}}.
        \label{eqAppA2}
\end{equation}

When a Type~Ia supernova occurs in one of these host galaxies, we measure its flux at Earth, $F_{\rm SN,host}$. Using the host distance derived from the Cepheids, we can infer the supernova's luminosity:
\begin{equation}
    L_{\rm SN,host}
    = F_{\rm SN,host}\,4\pi d_{\rm host}^2
    = F_{\rm SN,host}\,4\pi \frac{\epsilon_{\rm cep}\,L_{0}}{4\pi F_{\rm cep}}
    = L_{\rm cep}\,\frac{F_{\rm SN,host}}{F_{\rm cep}}.
        \label{eqAppA3}
\end{equation}
This step links (or anchors) the SN~Ia luminosity scale directly to the Cepheid scale.

If all supernovae had the exact same peak luminosity, measuring distances would be trivial. In reality, SNe~Ia vary in brightness and require standardization (e.g., via light-curve shape). Just as with Cepheids, this standardization process introduces error. We represent this overall standardization error with a factor $\epsilon_{\rm SN}$ (where $\epsilon_{\rm SN}=1$ means perfectly accurate standardization). The inferred, standardized SN~Ia luminosity becomes:
\begin{equation}
    L_{\rm adj}
    \equiv \epsilon_{\rm SN}\,L_{\rm SN,host}
    = \epsilon_{\rm SN}\,L_{\rm cep}\,\frac{F_{\rm SN,host}}{F_{\rm cep}}
    = \epsilon_{\rm SN}\,\epsilon_{\rm cep}\,L_{0}\,\frac{F_{\rm SN,host}}{F_{\rm cep}}.
        \label{eqAppA4}
\end{equation}

We must also account for dust extinction and color correction. Similarly, we introduce a factor $\epsilon_{\rm cal}$ to represent any error remaining after these corrections ($\epsilon_{\rm cal}=1$ for a flawless correction). The final inferred luminosity is:
\begin{equation}
    L_{\rm calibrated}
    \equiv \epsilon_{\rm cal}\,L_{\rm adj}
    = L_{\rm cal,0}\,\frac{F_{\rm SN,host}}{F_{\rm cep}},
        \label{eqAppA5}
\end{equation}
where we have bundled all intrinsic luminosity and calibration errors into a single zero-point term:
\begin{equation}
    L_{\rm cal,0} \equiv \epsilon_{\rm cal}\,\epsilon_{\rm SN}\,\epsilon_{\rm cep}\,L_{0}.
        \label{eqAppA6}
\end{equation}
Using this fully calibrated luminosity and the observed flux $F_{\rm obs}$ of a supernova in the Hubble flow, the distance is established as:
\begin{equation}
    d_{\rm SN,obs}
    = \sqrt{\frac{L_{\rm calibrated}}{4\pi F_{\rm obs}}}
    = \sqrt{\frac{L_{\rm cal,0}}{4\pi F_{\rm obs}}\cdot \frac{F_{\rm SN,host}}{F_{\rm cep}} }.
        \label{eqAppA7}
\end{equation}

The standard distance ladder assumes light only loses intensity due to geometric expansion (encoded in $d_\Lambda$). Under this assumption, inferred luminosities match the true physical luminosities. To introduce our WMC attenuation effect, we keep the standard operational definitions above, but now use a superscript $\star$ to denote the true, attenuation-free intrinsic luminosities.

With WMC, propagating light suffers an additional exponential attenuation. For an observation channel $\kappa$, we express this attenuation as a redshift-like factor $(1+z_q^\kappa)^{-1}$. The observed flux is therefore:
\begin{equation}
    F(X) = \frac{L^\star}{4\pi [d_\Lambda(X)]^2\,(1+z_q^\kappa(X))}.
        \label{eqAppA8}
\end{equation}

Applying this to a host galaxy at distance $X_0$ (which contains both Cepheids and a supernova), the observed fluxes at Earth are:
\begin{subequations}
\begin{align}
    F_{\rm SN,host}(X_0)
    &= \frac{L_{\rm SN,host}^\star}{4\pi [d_\Lambda(X_0)]^2\,(1+z_q^{\rm S}(X_0))},\label{eqAppA9a} \\[1.3ex]
    F_{\rm cep}(X_0)
    &= \frac{L_{\rm cep}^\star}{4\pi [d_\Lambda(X_0)]^2\,(1+z_q^{\rm C}(X_0))}.\label{eqAppA9b}
\end{align}
\end{subequations}
For a Hubble-flow supernova at distance $X$, we have:
\begin{equation}
    F_{\rm obs}(X)
    = \frac{L_{\rm calibrated}^\star}{4\pi [d_\Lambda(X)]^2\,(1+z_q^{\rm S}(X))}.
    \label{eqAppA10}
\end{equation}
Here, the superscripts ${\rm S}$ and ${\rm C}$ denote the specific attenuation for the supernova and Cepheid bands, respectively. Taking the ratio of the two host fluxes immediately gives:
\begin{equation}
    \frac{F_{\rm SN,host}}{F_{\rm cep}}
    = \frac{L_{\rm SN,host}^\star}{L_{\rm cep}^\star}\cdot \frac{1+z_q^{\rm C}(X_0)}{1+z_q^{\rm S}(X_0)}.
    \label{eqAppA11}
\end{equation}

During the anchoring step, the true luminosity ratio $L_{\rm SN,host}^\star/L_{\rm cep}^\star$ is naturally absorbed into the calibration zero-point ($L_{\rm cal,0}$). Therefore, operationally, the measured flux ratio simply isolates the difference in WMC attenuation between the Cepheid and supernova bands at distance $X_0$:
\begin{equation}
    \frac{F_{\rm SN,host}}{F_{\rm cep}}
    \equiv \frac{1+z_q^{\rm C}(X_0)}{1+z_q^{\rm S}(X_0)}.
        \label{eqAppA12}
\end{equation}

Next, we substitute the attenuated flux $F_{\rm obs}(X)$ back into the distance estimator. This introduces a normalization ratio $L_{\rm cal,0}/L_{\rm calibrated}^{\star}$. Crucially, this ratio does not depend on the Hubble-flow distance $X$. It only adds a constant offset to the distance modulus, which we can safely absorb into the regression intercept $M$. The distance estimator then becomes:
\begin{subequations}
\begin{align}
d_{\rm SN,obs}
&=\sqrt{\frac{L_{\rm cal,0}}{4\pi F_{\rm obs}}\cdot \frac{F_{\rm SN,host}}{F_{\rm cep}}}
    \label{eqAppA13a}\\[1.0ex]
&=\sqrt{\frac{L_{\rm cal,0}}{L_{\rm calibrated}^{\star}}\cdot [d_\Lambda(X)]^2\,(1+z_q^{\rm S}(X))\cdot \frac{F_{\rm SN,host}}{F_{\rm cep}}}
\label{eqAppA13b}\\[1.0ex]
&=
d_\Lambda(X)\,\sqrt{1+z_q^{\rm S}(X)}\cdot \sqrt{\frac{L_{\rm cal,0}}{L_{\rm calibrated}^{\star}}\cdot \frac{1+z_q^{\rm C}(X_0)}{1+z_q^{\rm S}(X_0)}}.\label{eqAppA13c}
\end{align}
\end{subequations}
For compactness, we define this constant offset as:
\begin{equation}
M_{cal}\equiv \frac{L_{\rm cal,0}}{L_{\rm calibrated}^{\star}} \cdot \frac{1+z_q^{\rm C}(X_0)}{1+z_q^{\rm S}(X_0)}.\label{eqAppA14}
\end{equation}
Following our hybrid modulus definition, the regression distance modulus is:
\begin{subequations}
\begin{align}
\mu_h^{\rm reg}
&\equiv 5\log_{10}\!\left(\frac{d_{\rm SN,obs}}{{\rm Mpc}}\right)+25+M
\label{eqAppA15a}\\[1.5ex]
&= \mu_\Lambda +\frac{5}{2}\log_{10}\!\left[1+z_q^{\rm S}(X)\right]+\frac{5}{2}\log_{10}\!\left[M_{cal}\right]+M
\label{eqAppA15b}\\[1.5ex]
&= \mu_h +\frac{5}{2}\log_{10}\!\left[M_{cal}\right]+M.\label{eqAppA15c}
\end{align}
\end{subequations}
Using our WMC parameters, the anchor-dependent term simplifies to:
\begin{align}
\frac{5}{2}\log_{10}\!\left[\frac{1+z_q^{\rm C}(X_0)}{1+z_q^{\rm S}(X_0)}\right]
&\simeq \frac{5}{2\ln10}\left[(H^{\rm C}-H^{\rm S})\frac{X_0}{c_{\scriptscriptstyle l}}\right],\label{eqAppA16}
\end{align}
where the sign of the shift depends on $(H^{\rm C}-H^{\rm S})$.

As an illustrative example, applying the anchor galaxy $\mathrm{NGC~4258}$ (Messier~106) at $X_0 = 7.58~{\rm Mpc}$ \citep{Reid2019}, and adopting a conservative bound $|H^{\rm C}-H^{\rm S}|\lesssim (H_{\max}-H_{\min})$ from Appendix~\ref{AppB}, we obtain:
\begin{equation}
(H_{\max}-H_{\min})\frac{X_0}{c_{\scriptscriptstyle l}}\simeq \frac{(75.257-67.400)\times 7.58}{299792.458}\approx 2.0\times10^{-4},\label{eqAppA17}
\end{equation}
meaning the corresponding intercept shift is negligibly small:
\begin{equation}
\left|\frac{5}{2}\log_{10}\!\left[\frac{1+z_q^{\rm C}(X_0)}{1+z_q^{\rm S}(X_0)}\right]\right|
\lesssim 2.2\times10^{-4}\ {\rm mag}.\label{eqAppA18}
\end{equation}

Finally, we evaluate $M_{cal}$. If the calibration pipeline successfully recovers the WMC attenuation via standard color corrections (the DMC effect), the ratio $L_{\rm cal,0}/L_{\rm calibrated}^{\star}$ stays close to 1. In this case, $\log_{10}[M_{cal}] \approx 0$, yielding our simple regression form:
\begin{equation}
\mu_h^{\rm reg} \simeq \mu_h + M. \label{eqAppA19}
\end{equation}
Under these conditions, setting a fixed prior like $M=0$ is valid. Even if $M_{cal}$ deviates from 1, it remains completely independent of $X$ for the Hubble-flow sample. It merely acts as a constant offset absorbed by $M$. In either scenario, the hybrid distance modulus (equation~\eqref{eq24b}) we use for our regression is mathematically justified.

\section{Derivation of Channel-Dependent Effective Expansion Rates}
\label{AppB}

In this Appendix, we connect the channel-level WMC parametrization from Appendix~\ref{AppA} to the regression behavior observed in our tomographic fits (Tables~\ref{tab5}--\ref{tab7}). The observable redshift in the Hubble diagram is a combination of cosmic expansion and propagation attenuation. If the data contains cumulative WMC attenuation but the regression model ignores it (by enforcing $H_q=0$), the fit artificially absorbs this missing effect into the expansion rate. This creates an apparent redshift dependence in the binned estimates $H_\Lambda^{\rm(bin)}(z)$. For this reason, we treat the empirical drift slope $b\equiv dH_\Lambda^{\rm(bin)}/dz$ (Table~\ref{tab7}) not as a direct physical parameter, but as a regression diagnostic to estimate the underlying path-rate $H_q^{\rm S}$.

\subsection{Identity and Low-$z$ Linearization}
\label{AppB1}
For a Hubble-flow SN at distance $X$, the effective observed redshift $z_{\rm obs}(X)$ is the product of the metric expansion, the cumulative SN-channel attenuation along the path $X$, and the calibration ratio fixed at the anchor distance $X_0$:
\begin{equation}
(1+z_{\rm obs}) = (1+z_{\Lambda}(X))(1+z_q^{\rm S}(X))\left[\frac{1+z_q^{\rm C}(X_0)}{1+z_q^{\rm S}(X_0)}\right].
\label{eqAppB1}
\end{equation}
Assuming an exponential accumulation $1+z_q^\kappa(X)=\exp\!\left(H_q^\kappa X/c_{\scriptscriptstyle l}\right)$, taking the logarithm yields the exact identity:
\begin{equation}
\ln(1+z_{\rm obs})=\ln\!\bigl(1+z_{\Lambda}(X)\bigr)+\frac{H_q^{\rm S}}{c_{\scriptscriptstyle l}}X+\frac{H_q^{\rm C}-H_q^{\rm S}}{c_{\scriptscriptstyle l}}X_0.
\label{eqAppB2}
\end{equation}
In the low-redshift regime ($z\ll1$), applying the approximations $\ln(1+z)\simeq z$ and $z_\Lambda(X)\simeq (H_\Lambda/c_{\scriptscriptstyle l})X$ simplifies this to:
\begin{equation}
z_{\rm obs}(X)\simeq \frac{H_\Lambda + H_q^{\rm S}}{c_{\scriptscriptstyle l}}X+\frac{H_q^{\rm C} - H_q^{\rm S}}{c_{\scriptscriptstyle l}}X_0.
\label{eqAppB3}
\end{equation}
Equation~\eqref{eqAppB3} clearly shows how the two effects separate at low $z$: the SN attenuation grows proportionally with distance $X$, while the anchor difference acts as a constant, $X$-independent offset. This distance-independent anchor-scale offset is qualitatively consistent with the ``baked-in'' characterization of the low-$z$ Hubble-diagram tension discussed by \citet{Scolnic2025}.

\subsection{A Compact Weighted Form for the Effective Rate}
\label{AppB2}
We can express equation~\eqref{eqAppB3} as an effective expansion rate $H_{\rm eff}(X)\equiv c_{l}\,z_{\rm obs}(X)/X$. Dividing equation~\eqref{eqAppB3} by $X/c_{l}$ gives:
\begin{subequations}
\begin{align}
c_{l}\frac{z_{\rm obs}(X)}{X}
&\simeq (H_\Lambda + H_q^{\rm S}) + (H_q^{\rm C}-H_q^{\rm S})\left(\frac{X_0}{X}\right),
\label{eqAppB4a}\\[1.0ex]
H_{\rm eff}(X)
\;&\equiv  (H_\Lambda + H_q^{\rm C})\,\zeta + (H_\Lambda + H_q^{\rm S})\bigl(1-\zeta\bigr).
\label{eqAppB4b}
\end{align}
\end{subequations}
Here, the weighting factor
\begin{equation}
\zeta \;\equiv\; \frac{X_0}{X},\qquad 0<\zeta<1
\label{eqAppB5}
\end{equation}
represents the relative influence of the anchor offset on a source at distance $X$. In this form, $H_{\rm eff}(X)$ is simply a weighted average of the two channel-dependent rates, $(H_\Lambda+H_q^{\rm C})$ and $(H_\Lambda+H_q^{\rm S})$. Naturally, as the observation distance extends far beyond the anchor ($X\gg X_0$), the anchor's influence $\zeta$ fades away.

\subsection{Connection to Tomographic Regressions and the Drift Slope}
\label{AppB3}
Our tomographic analysis (Tables~\ref{tab5}--\ref{tab7}) does not fit equation~\eqref{eqAppB3} directly. Instead, it fits the distance modulus within distinct redshift bins, holding other parameters fixed. Therefore, the observed linear trend $H_\Lambda^{\rm(bin)}(z)\approx a+bz$ is a mathematical artifact of forcing a standard model ($H_q=0$) onto data that inherently contains WMC attenuation.

This artifact, however, provides a diagnostic tool. When $H_q=0$ is strictly enforced, the regression absorbs the missing cumulative attenuation into an apparent redshift dependence, yielding positive slopes ($b>0$) with high statistical significance. Conversely, when the hybrid WMC term is included, the artificial drift disappears ($b\simeq 0$) and the binned expansion rates collapse stably toward the Planck-anchored value (Tables~\ref{tab5}--\ref{tab7}; Fig.~\ref{fig1}). This contrast confirms that the drift slope is indeed the regression signature of the omitted propagation effect.

\subsection{Estimates of Channel Scales}
\label{AppB4}

Using the weighted decomposition (equation~\eqref{eqAppB4b}), we infer channel-specific rates $H_q^{\rm C}$ and $H_q^{\rm S}$ from the empirical linear fits ($a+bz$) of the ``All bins'' samples, adopting the Planck baseline $H_{0}^{\scriptscriptstyle \rm CMB}=67.4~{\rm km\,s^{-1}\,Mpc^{-1}}$ as the true metric expansion rate: as $z\to 0$, the anchor scale dominates so $H_q^{\rm C}\simeq a-H_{0}^{\scriptscriptstyle \rm CMB}$, while at the high-redshift edge ($z_{\max}\approx 0.64$; Table~\ref{tab6}) the cumulative Hubble flow dominates so $H_{\rm eff}(z_{\max})\simeq a+b\,z_{\max}$ and $H_q^{\rm S}\simeq H_{\rm eff}(z_{\max})-H_{0}^{\scriptscriptstyle \rm CMB}$.

Applying Table~\ref{tab7} across covariance modes yields a consistent parallel structure: Mode D0 ($a=72.45$, $b=3.16$) gives $H_q^{\rm C}\simeq 5.05~{\rm km\,s^{-1}\,Mpc^{-1}}$ and, at $z_{\max}=0.64$, $H_{\rm eff}\simeq 74.47$ so $H_q^{\rm S}\simeq 7.07~{\rm km\,s^{-1}\,Mpc^{-1}}$; Mode ST ($a=72.46$, $b=3.11$) gives $H_q^{\rm C}\simeq 5.06~{\rm km\,s^{-1}\,Mpc^{-1}}$ and $H_{\rm eff}\simeq 74.45$ so $H_q^{\rm S}\simeq 7.05~{\rm km\,s^{-1}\,Mpc^{-1}}$; Mode SS ($a=72.58$, $b=2.80$) gives $H_q^{\rm C}\simeq 5.18~{\rm km\,s^{-1}\,Mpc^{-1}}$ and $H_{\rm eff}\simeq 74.37$ so $H_q^{\rm S}\simeq 6.97~{\rm km\,s^{-1}\,Mpc^{-1}}$.

Operationally, these diagnostics suggest a consistent hierarchy across all modes: the cumulative Hubble-flow excess ($H_q^{\rm S}\approx 7~{\rm km\,s^{-1}\,Mpc^{-1}}$) generally exceeds the anchor-scale offset ($H_q^{\rm C}\approx 5~{\rm km\,s^{-1}\,Mpc^{-1}}$). We interpret these values as statistically weighted averages over the entire sample. Since individual Cepheids and SNe~Ia may possess varying effective emission wavelengths, their specific WMC attenuations are expected to vary accordingly. This variance offers a plausible way to understand why the inferred Hubble constant can exhibit scatter or sample-dependent fluctuations.

\subsection{Bridging the Distance Ladder: Towards a Wavelength-Dependent WMC}
\label{AppB5}

As derived above, because the final inferred expansion rate can be expressed as a weighted synthesis across the distance ladder, the net effective Hubble value is expected to be bounded by the anchor-side and Hubble-flow-side WMC components:
\begin{equation}
H_{0}^{\scriptscriptstyle \mathrm{CMB}} + H_{q}^{\scriptscriptstyle \mathrm{Anchor}} \;<\; H_{\mathrm{eff}} \;<\; H_{0}^{\scriptscriptstyle \mathrm{CMB}} + H_{q}^{\scriptscriptstyle \mathrm{Flow}}.
\label{eqB6}
\end{equation}

This boundary suggests an operational route to method-dependent $H_0$ stratification: lower values may arise when the anchor-side channel dominates (e.g., JAGB or TRGB), whereas higher values may arise when the flow-tracer side (e.g., SNe~Ia) dominates the inferred slope. Drawing from the benchmark measurements discussed in Section~\ref{1.2} and the empirical estimates derived in Appendix~\ref{AppB4}, we adopt representative channel components for JAGB, TRGB, Cepheids, and SNe~Ia as $0.4$, $2.4$\footnote{The adopted $H_q \approx 2.4~{\rm km\,s^{-1}\,Mpc^{-1}}$ maps to $H_0 \approx 69.8~{\rm km\,s^{-1}\,Mpc^{-1}}$ \citep{Freedman2019} rather than the recent JWST-based $68.8~{\rm km\,s^{-1}\,Mpc^{-1}}$ \citepalias{Freedman2025}. We retain the former because its larger calibrator sample and shallower redshift range ($z \le 0.08$ vs. $2.3$) yield lower statistical uncertainty and maximize the local anchor weight $\zeta$ as defined in equation~\eqref{eqAppB4b}.}, $5.0$, and $7.0~\mathrm{km\,s^{-1}\,Mpc^{-1}}$, respectively. 

The variation of $H_q$ across different tracers further suggests that the effective conversion coefficient $P$ may be tracer-dependent. Operationally, each tracer population can be associated with a characteristic effective temperature range, empirically motivated by the cited observational and evolutionary literature: JAGB ($\sim 2000$--$3000$~K) \citep{Marigo2003, Ventura2018, HofnerOlofsson2018,Freedman2020}, TRGB ($\sim 3000$--$4500$~K) \citep{Iben1967,Pols1998, Saltas2022}, Cepheids ($\sim 5000$--$7000$~K) \citep{Pel1978, Cox1980,Anderson2016, EspinozaArancibia2024}, and SNe~Ia photospheric emission ($\sim 10000$--$15000$~K) \citep{Patat1996, Silverman2011}, with the CMB fixed at $2.725$~K \citep{Fixsen2009}.

Taken together, these considerations are consistent with the possibility that a tracer's specific attenuation rate—and its corresponding effective Hubble rate—may depend on its effective temperature (or equivalently, its characteristic emission wavelength). In this sense, wavelength dependence provides a plausible mechanism for understanding why distinct distance-ladder implementations can report systematically different Hubble constants \citepalias{Riess2022, Freedman2025}.

This perspective is also consistent with the operational premise introduced in Section~\ref{1.4}: if the WMC mechanism is wavelength-dependent, it may operationally mimic standard dust extinction. We examine candidate mathematical and physical models that satisfy this dust-mimicking condition in Appendix~\ref{AppC}.

\section{Wavelength Dependence of WMC}
\label{AppC}

This appendix investigates which functional form of the wavelength dependence of WMC operationally assumed in Section~\ref{1.4} is compatible with the hybrid framework, under the simultaneous requirements of tracer-dependent $H_q$ stratification, dust-like color behavior in the optical band, and negligible conversion in the microwave band.

%%%%%%%%%%%%%%%%%%%%%%%%%%
\begin{table*}
\centering
\caption{Constrained solution summary for the exponential-cutoff response family under the Appendix~\ref{AppC} joint conditions (fixed tracer-stratified channel targets, bounded latent tracer temperatures, and the imposed optical color-law analogue $\beta(R_{\rm \scriptscriptstyle WMC})=3.0$). Here, $\lambda_c$ denotes the critical wavelength (transition) scale and $T_c$ denotes the corresponding critical temperature scale, related through the Wien mapping $\lambda=b/T_{\rm eff}$ (equivalently $T_c=b/\lambda_c$), with $b=2.897771955\times10^{-3}~\mathrm{m\,K}$. Because the anchor amplitudes are imposed constraints, the reported goodness-of-fit statistics are diagnostic. The constrained solution yields $(H_{q,\max},\lambda_c,p)=(8.095,\,623.899~\mathrm{nm},\,2.108)$, corresponding to $T_c=4644.61~\mathrm{K}$, with a negligible present-microwave contribution $H_q(2.725~\mathrm{K})\approx 0$ and a diagnostic extrapolation $H_q(3000.225~\mathrm{K})=0.656~\mathrm{km\,s^{-1}\,Mpc^{-1}}$.}
\label{tabA1}
\setlength{\tabcolsep}{8.2pt}
\renewcommand{\arraystretch}{1.1}
{\fontsize{7.0pt}{8.05pt}\selectfont
\begin{tabular}{lcccccccc}
\toprule
Max.&
$p$ &
$\lambda_c~[\mathrm{nm}]$ &
$T_c~[\mathrm{K}]$ &
$R^2$ &
RMSE &
$\chi^2$ &
$H_q(2.725~\mathrm{K})~[\mathrm{km\,s^{-1}\,Mpc^{-1}}]$ &
$H_q(3000.225~\mathrm{K})~[\mathrm{km\,s^{-1}\,Mpc^{-1}}]$ \\
\midrule
$8.095$ &
$2.108$ &
$623.899$ &
$4644.61$ &
$1.000$ &
$0.000$ &
$0.000$ &
$0.000$ &
$0.656$ \\
\bottomrule
\end{tabular}
}
\end{table*}

\begin{figure*}
\centering
\includegraphics[height=0.22\textheight,keepaspectratio]{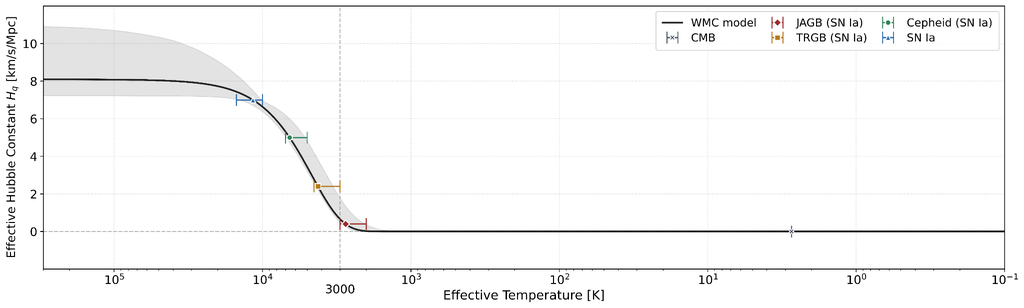}
\caption{Tracer-stratified channel components $H_q^\kappa$ are plotted against effective temperature, with distinct geometric markers and horizontal bars indicating the adopted tracer-temperature intervals; the CMB point is displayed only as a diagnostic reference marker (not as a fitted anchor). The solid curve shows the constrained exponential-cutoff solution enforcing $\beta_{\rm target}=3.0$ while reproducing the adopted anchor amplitudes within the tracer-temperature bounds, and the vertical dashed line marks the diagnostic recombination-scale temperature, $T(z=1100)=3000.225~\mathrm{K}$, at which the constrained curve gives $H_q=0.656~\mathrm{km\,s^{-1}\,Mpc^{-1}}$. The shaded envelope shows the family of valid forward Monte Carlo realizations (6,406 out of $5\times10^{6}$ draws), illustrating a finite admissible solution set and a comparatively broader spread across the transition region and high-temperature plateau.}
\label{figA1}
\end{figure*}

\begin{table*}
\centering
\caption{Constrained solution summary for the Hill/logistic saturation response family under the Appendix~\ref{AppC} joint conditions (fixed tracer-stratified channel targets, bounded latent tracer temperatures, and the imposed optical color-law analogue $\beta(R_{\rm \scriptscriptstyle WMC})=3.0$). Here, $T_c$ denotes the critical temperature (half-response) scale and $\lambda_c$ denotes the corresponding critical wavelength scale, reported via the Wien mapping $\lambda=b/T_{\rm eff}$ (equivalently $\lambda_c=b/T_c$), with $b=2.897771955\times10^{-3}~\mathrm{m\,K}$. Because the anchor amplitudes are imposed constraints, the reported goodness-of-fit statistics are diagnostic. The constrained solution yields $(H_{q,\max},\lambda_c,p)=(7.134,\,598.670~\mathrm{nm},\,4.177)$, corresponding to $T_c=4840.35~\mathrm{K}$, with a negligible present-microwave contribution $H_q(2.725~\mathrm{K})\approx 1.91\times10^{-13}~\mathrm{km\,s^{-1}\,Mpc^{-1}}$ and a diagnostic extrapolation $H_q(3000.225~\mathrm{K})=0.852~\mathrm{km\,s^{-1}\,Mpc^{-1}}$.}
\label{tabA2}
\setlength{\tabcolsep}{8.2pt}
\renewcommand{\arraystretch}{1.1}
{\fontsize{7.0pt}{8.05pt}\selectfont
\begin{tabular}{lcccccccc}
\toprule
Max.&
$p$&
$\lambda_c~[\mathrm{nm}]$&
$T_c~[\mathrm{K}]$&
$R^2$&
RMSE&
$\chi^2$&
$H_q(2.725~\mathrm{K})~[\mathrm{km\,s^{-1}\,Mpc^{-1}}]$&
$H_q(3000.225~\mathrm{K})~[\mathrm{km\,s^{-1}\,Mpc^{-1}}]$ \\
\midrule
$7.134$&
$4.177$&
$598.670$&
$4840.35$&
$1.000$&
$0.000$&
$0.000$&
$0.000$&
$0.852$ \\
\bottomrule
\end{tabular}
}
\end{table*}

\begin{figure*}
\centering
\includegraphics[height=0.22\textheight,keepaspectratio]{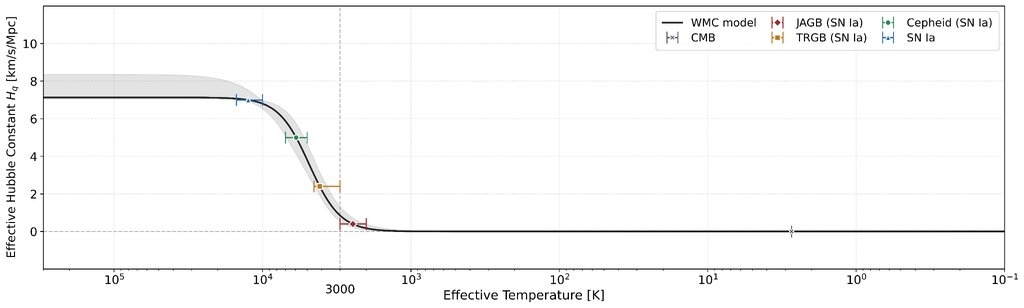}
\caption{Tracer-stratified channel components $H_q^\kappa$ are plotted against effective temperature, with distinct geometric markers and horizontal bars indicating the adopted tracer-temperature intervals; the CMB point is displayed only as a diagnostic reference marker (not as a fitted anchor). The solid curve shows the constrained Hill/logistic saturation solution enforcing $\beta_{\rm target}=3.0$ while reproducing the adopted anchor amplitudes within the tracer-temperature bounds, and the vertical dashed line marks the diagnostic recombination-scale temperature, $T(z=1100)=3000.225~\mathrm{K}$, at which the constrained curve gives $H_q=0.852~\mathrm{km\,s^{-1}\,Mpc^{-1}}$. The shaded envelope shows the family of valid forward Monte Carlo realizations (4,484 out of $5\times10^{6}$ draws), illustrating a finite admissible solution set and a comparatively tighter concentration around the constrained transition profile.}
\label{figA2}
\end{figure*}

\subsection{Operational Model for the WMC}
\label{AppC1}

We model WMC operationally as a radiative-to-non-radiative transition and seek a response function that increases toward higher photon frequency while remaining suppressed in the low-frequency regime.

Drawing an analogy to the Schwinger mechanism in quantum electrodynamics~\citep{Schwinger1951}, where vacuum pair production is exponentially suppressed below a characteristic field scale:
\begin{equation}
P_{\rm \scriptscriptstyle Schwinger}
\;\propto\;
\exp\!\left(-\frac{\pi m^{2}c^{3}}{e\hbar E}\right),
\label{eqA1}
\end{equation}
we parameterize the effective conversion coefficient $\bar P(\nu)$ with a similar exponential sigmoidal profile based on photon frequency (or wavelength $\lambda$):
\begin{equation}
\bar P(\nu)
\;\equiv\;
P_{0}\,
\exp\!\left[-\left(\frac{\nu_{c}}{\nu}\right)^{p}\right]
\;=\;
P_{0}\,
\exp\!\left[-\left(\frac{\lambda}{\lambda_{c}}\right)^{p}\right].
\label{eqA2}
\end{equation}
Here, $P_0$ is the asymptotic maximum conversion coefficient at high frequency, $\nu_c$ (equivalently $\lambda_c$) sets the transition scale, and $p$ controls the transition sharpness.

To impose operational consistency with the Planck-anchored CMB baseline, we define the effective WMC channel component such that it is strongly suppressed in the microwave band and approaches a finite maximum in the optical/UV regime:
\begin{equation}
H_{\rm q, eff}(T_{\rm \scriptscriptstyle eff})
\;=\;
H_{\rm q,\,max}\cdot
\exp\!\left[-\left(\frac{T_c}{T_{\rm \scriptscriptstyle eff}}\right)^{p}\right],
\label{eqA3}
\end{equation}
where $H_{\rm q,\,max}$ denotes the maximum WMC channel component.

As an alternative response family with a saturation structure, we also test a Michaelis--Menten/Hill-type form \citep{Michaelis1913, Johnson2011}:
\begin{equation}
    v = v_{\rm max}\, \frac{[S]}{K_m + [S]}.
    \label{eqA4}
\end{equation}
If the quantum vacuum is treated as an operational catalyst that converts radiation into effective mass, this kinetic framework offers a convenient mapping. In this analogy, the substrate concentration $[S]$ may be associated with the incident photon energy (parameterized by frequency $\nu$ or effective temperature $T_{\rm eff}$), and the reaction velocity $v$ with the effective conversion rate. Accordingly, $v_{\max}$ can be interpreted as the finite saturation capacity of the vacuum ($P_{\max}$ or $H_{\rm q,\,max}$), while the Michaelis constant $K_m$ sets a characteristic transition scale ($\nu_c$ or $T_c$) at which the conversion efficiency reaches half of its maximum value. Adapting this concept with a macroscopic power-law dependency yields a Hill/logistic form:
\begin{subequations}
    \begin{align}
       & P(\nu) = P_{\rm max}\, \frac{\nu^p}{\nu_c^p + \nu^p},     \label{eqA5}\\[1.2ex]
        &H_{\rm q, {\rm eff}}(T_{\rm eff}) = H_{\rm q,\,max} \cdot  \frac{[T_{\rm eff}]^p}{[T_c]^p + [T_{\rm eff}]^p}.\label{eqA5b}
    \end{align}
\end{subequations}

\subsection{Asymmetric Standardization and Signal Encoding}
\label{AppC2}

For the operationally assumed WMC mechanism to successfully mimic dust extinction and trigger Dust-mimicking Magnitude Compensation (DMC), the WMC-induced chromatic attenuation must be absorbed by the standard color correction term, $\beta c$ (where $\beta$ is the color-luminosity coefficient and $c$ is the apparent color excess). The Pantheon+SH0ES pipeline empirically applies $\beta \approx 3.1$ \citepalias{Brout2022}. Furthermore, because our framework infers anchor-specific $H_q$ values ($0.4~\mathrm{km\,s^{-1}\,Mpc^{-1}}$) corresponding to the JAGB calibrations by \citetalias{Freedman2025}, the WMC color signature must also align with the $\beta \approx 2.9$ applied in their pipeline. Consequently, we investigate whether the effective WMC color law can approximately satisfy a consensus target of $\beta \approx 3.0$.

Let $A_\lambda$ denote the effective WMC-induced extinction (dimming in magnitudes) at wavelength $\lambda$, and define the normalized attenuation profile $\Pi(\lambda)$ by:
\begin{equation}
A_\lambda \;\propto\; \Pi(\lambda)\equiv \frac{\bar P(\lambda)}{P_0},
\label{eqC5_Apropf}
\end{equation}
so that $\Pi(\lambda)$ specifies the shape of the attenuation independently of its absolute amplitude.

Using the standard $B$ and $V$ bands ($\lambda_B\simeq 440~\mathrm{nm}$, $\lambda_V\simeq 550~\mathrm{nm}$), we define the WMC-implied total-to-selective extinction ratio as:
\begin{equation}
R_{\rm \scriptscriptstyle WMC}
\;\equiv\;
\frac{A_V}{A_B-A_V}
\;\simeq\;
\frac{\Pi(\lambda_V)}{\Pi(\lambda_B)-\Pi(\lambda_V)}.
\label{eqC5b_def}
\end{equation}
Given that the empirical SALT2 color coefficient $\beta$ serves a role analogous to the total-to-selective extinction ratio in standard dust models \citep{Tripp1998, Guy2007}, treating $R_{\rm \scriptscriptstyle WMC}$ as its operational analogue provides a practical test of whether the WMC signature can reproduce the dust-like correction scale required by the standard light-curve calibration. We therefore test both the exponential-cutoff and Hill/logistic response families under the following joint constraints: each model must reproduce the target effective channel amplitudes $H_q^\kappa$ within the adopted tracer-specific effective-temperature ranges, while simultaneously yielding $R_{\rm \scriptscriptstyle WMC}\approx 3.0$ in the optical band.

\subsection{Constrained Fits and Monte Carlo Feasibility }
\label{AppC3}

To evaluate the operational feasibility of the proposed response models, we employ a two-step numerical approach. First, we perform a constrained optimization using the Sequential Least Squares Programming (SLSQP) algorithm. In this step, the optical anchor amplitudes $H_q^\kappa$ are fixed as target values (i.e., $0.4$, $2.4$, $5.0$, and $7.0~\mathrm{km\,s^{-1}\,Mpc^{-1}}$), while the representative effective temperatures for each tracer are treated as latent variables constrained within their respective physically admissible bounds (JAGB: 2000--3000\,K; TRGB: 3000--4500\,K; Cepheid: 5000--7000\,K; SN~Ia: 10000--15000\,K). An explicit condition of the color-law analogue $\beta(R_{\rm \scriptscriptstyle WMC}) = 3.0$ is enforced simultaneously.

Second, to verify that these solutions are not singular or highly fine-tuned, we conduct a forward Monte Carlo sampling ($N = 5 \times 10^6$ draws per response family) where $\beta(R_{\rm \scriptscriptstyle WMC})$ is left unconstrained and free to vary. We uniformly sample the broad parameter space of the response functions (i.e., the maximum amplitude $H_{q,\max}$, the transition scale, and the shape parameter $p$). For each randomly generated curve, we analytically invert the function to determine the exact effective temperatures required to reproduce the fixed anchor amplitudes $H_q^\kappa$. A curve is retained as valid only if all these mathematically derived temperatures simultaneously fall within the physically admissible bounds of their respective tracers.

For the exponential-cutoff response (Table~\ref{tabA1} and Figure~\ref{figA1}), the constrained solution likewise satisfies $\beta(R_{\rm \scriptscriptstyle WMC})=3.0$ by construction, with $(H_{q,\max},\lambda_c,p)=(8.095,\,623.90~\mathrm{nm},\,2.108)$ and derived $T_c=4644.61~\mathrm{K}$. The fitted tracer temperatures again remain within the adopted intervals (SN~Ia: $11593$~K, Cepheid: $6567$~K, TRGB: $4233$~K, JAGB: $2755$~K). At the current microwave background ($T=2.725$~K), the model yields $H_q=0.000~\mathrm{km\,s^{-1}\,Mpc^{-1}}$. The corresponding diagnostic extrapolation at $T(z=1100)=3000.22$~K gives $H_q=0.656~\mathrm{km\,s^{-1}\,Mpc^{-1}}$.

For the Hill/logistic model (Table~\ref{tabA2} and Figure~\ref{figA2}), the constrained solution satisfies the imposed condition with $\beta(R_{\rm \scriptscriptstyle WMC})=3.0$ by construction and yields $(H_{q,\max},\lambda_c,p)=(7.134,\,598.67~\mathrm{nm},\,4.177)$, corresponding to $T_c=4840.35~\mathrm{K}$. The fitted tracer temperatures remain within the adopted intervals (SN~Ia: $12478$~K, Cepheid: $5935$~K, TRGB: $4114$~K, JAGB: $2462$~K). At the current microwave background ($T=2.725$~K), the model yields a negligible $H_q \approx 1.91 \times 10^{-13}~\mathrm{km\,s^{-1}\,Mpc^{-1}}$. The diagnostic extrapolation at $T(z=1100)=3000.22$~K gives $H_q=0.852~\mathrm{km\,s^{-1}\,Mpc^{-1}}$.

We then test how non-fine-tuned these constrained solutions are by sampling broad forward Monte Carlo ensembles ($5\times10^6$ draws per family) and retaining only curves that satisfy all amplitude-intersection and temperature-bound conditions. For the Hill/logistic family, $4484$ valid curves survive ($0.08968\%$), with $\beta$ median $=2.828$, mean $=3.013\pm0.010$, and central ranges [2.448, 3.611] (16--84\%) and [2.315, 4.420] (05--95\%). For the exponential-cutoff family, $6406$ valid curves survive ($0.12812\%$), with $\beta$ median $=3.805$, mean $=3.949\pm0.011$, and central ranges [3.023, 4.929] (16--84\%) and [2.703, 5.726] (05--95\%). In both families, the sampled admissible set contains curves with $\beta_{\rm target}=3.0$.

These results support the operational assumptions adopted in this Appendix. Within both response families, continuous mappings can simultaneously (i) satisfy the tracer-stratified $H_q$ targets within valid temperature bounds, (ii) naturally vanish at the microwave background, and (iii) yield an optical color-law analogue near $\beta \approx 3$. Statistically, the consensus target $\beta=3.0$ resides within the central admissible regions---roughly $0.30\sigma$ and $0.85\sigma$ from the medians of the Hill/logistic and exponential-cutoff families, respectively (using the 16--84 percentile half-width as a dispersion proxy).

\subsection{Operational Consistency}
\label{AppC4}

Given the admissible wavelength-dependent response families identified above, the primary operational consequence is an asymmetry in the standardization pipeline. When a WMC-induced color imprint is processed as ordinary dust, the color-correction term can partially restore the photometric channel (the vertical axis of the Hubble diagram), while the spectroscopic redshift (the horizontal axis) remains unchanged by construction. As a result, the WMC-induced excess redshift remains encoded in the data whereas part of the associated flux attenuation is removed.

\vspace{5mm}

The scale of this processing-induced bias can be expressed as an effective magnitude offset of $\Delta\mu \approx 5\log_{10}(73.0/67.4) \approx 0.17~\mathrm{mag}$. This value is numerically close (in absolute magnitude) to the intercept shifts recovered in the flat $\Lambda$CDM $\Lambda5$ configurations of Table~\ref{tab3} across all covariance modes, namely $M=-0.176\pm0.005$ (D0), $M=-0.173\pm0.004$ (ST), and $M=-0.173\pm0.004$~mag (SS). Within the present operational framework, the $\sim 0.17$~mag discrepancy is therefore interpreted as a modulus--redshift mismatch induced by asymmetric standardization of a WMC imprint.\footnote{We emphasize that this $\sim0.17$~mag offset does not imply an incorrect nearby (anchor-scale) photometric calibration in Pantheon+ or a literal mismeasurement of source luminosity. Rather, it acts as an \emph{equivalent modulus proxy}, translating the difference in inferred effective expansion rates into distance-modulus units.}

The suppression at microwave frequencies is guaranteed by the zero-energy limit ($T_{\rm eff} \to 0$~K) inherent to the model definition, but the result remains nontrivial because the same response families must also satisfy the optical dust-mimicking condition and the tracer-dependent $H_q$ stratification under the joint constraints of Appendix~\ref{AppC}. As shown in Figures~\ref{figA1} and \ref{figA2}, both response families become negligible in the microwave regime, preserving the Planck-anchored CMB baseline and remaining operationally compatible with the FIRAS blackbody constraint~\citep{Fixsen1996}.

In addition, the diagnostic extrapolation to $T(z=1100)=3000.225$~K remains finite and well-behaved in both cases ($H_q=0.656$ and $0.852~\mathrm{km\,s^{-1}\,Mpc^{-1}}$ for the exponential-cutoff and Hill/logistic families, respectively), although this extrapolation is not used as a direct regression constraint. In a Planck-anchored background \citepalias{Planck2020}, the early-universe expansion rate is governed by the standard $\Lambda$CDM Friedmann equation:
\begin{equation}
H(z) = H_{0}^{\scriptscriptstyle \mathrm{CMB}} \sqrt{\Omega_{r}(1+z)^4 + \Omega_{m}(1+z)^3 + \Omega_{\Lambda}}.
\label{eq_friedmann}
\end{equation}
Adopting the ~\citetalias{Planck2020} baseline parameters ($H_{0}^{\scriptscriptstyle \mathrm{CMB}} \approx 67.4~\mathrm{km\,s^{-1}\,Mpc^{-1}}$, $\Omega_{m} \approx 0.315$, $\Omega_{\Lambda} \approx 0.685$, and the radiation density $\Omega_{r} \approx 9.1 \times 10^{-5}$ derived from $T^{\scriptscriptstyle \mathrm{CMB}} = 2.725~\mathrm{K}$ and the effective number of relativistic species $N_{\mathrm{eff}} = 3.046$), evaluating this equation at the recombination epoch ($z \approx 1100$) yields an expansion-rate scale of order $\sim 1.59 \times 10^6~\mathrm{km\,s^{-1}\,Mpc^{-1}}$. Relative to this background scale, these diagnostic $H_q$ values are at the $\sim 4.2\times10^{-7}$ to $5.4\times10^{-7}$ fractional level. Restricted to this negligible level even at $z=1100$, the WMC contribution remains dynamically insignificant throughout the post-recombination epoch ($0 < z < 1100$), allowing the cosmological thermal history to remain materially unperturbed.

\subsection{Observational Signatures of WMC Wavelength Dependence}
\label{C5}

In the hybrid framework, WMC is operationally assumed to be strongly wavelength dependent: its impact grows toward shorter wavelengths (rest-frame UV/blue/optical) while remaining negligible in the microwave regime. Importantly, WMC is not treated as a pure gray-dimming channel. In the present picture, part of the radiative energy is transferred into non-radiative degrees of freedom, so the observable consequences can include both (i) chromatic attenuation and (ii) an effective redward migration of the surviving radiative power. Operationally, this implies that short-wavelength emission can be disproportionately depleted, while a fraction of that radiative power can be redistributed toward longer wavelengths around a characteristic transition scale, resulting in a localized, ``bottleneck''-like flux accumulation near a critical band.

\vspace{5mm}

First, such wavelength-selective suppression and migration can act as a systematic bias pathway for interpreting high-redshift galaxy SEDs in recent JWST analyses \citep[e.g.,][]{Labbe2023}. If rest-frame UV photons are preferentially removed from the radiative channel and/or shifted redward relative to optical/near-IR light, the observed SED would appear artificially redder than the intrinsic source spectrum. Under standard stellar population synthesis assumptions---where reddening is largely attributed to dust and/or older stellar populations---this can bias UV-based star-formation indicators low, while biasing SED-based ages and stellar masses high. In this sense, a WMC imprint could contribute to the appearance of prematurely quenched and over-massive early galaxies, without requiring that interpretation to be uniquely implied by the data.

Second, WMC-induced band-dependent attenuation together with redward spectral migration may provide a qualitative pathway for background-light tensions, including the reported optical excess. Absolute background measurements with \textit{New Horizons} report a cosmic optical background intensity in excess of the integrated galaxy light inferred from resolved counts \citep{Lauer2022}. Within an operational WMC picture, one plausible interpretation is that part of this excess could arise from redward spectral redistribution, in which a fraction of the surviving radiative power is transferred from intrinsically shorter wavelengths into the optical band. 

This interpretation is also consistent with the existence of strong short-wavelength constraints. If UV/blue photons are preferentially depleted and redistributed to longer wavelengths, the residual diffuse UV background is expected to remain low, broadly consistent with the faint far-UV background measurements \citep[e.g.,][]{Murthy1999}. In this sense, WMC provides a qualitative framework that can connect short-wavelength faintness and long-wavelength excess claims.

\section{Rationale for Equal-Count Tomographic Partitioning}
\label{AppD}

In our tomographic analysis (Section~\ref{2.6.4}), we adopt an equal-count partitioning scheme to construct redshift bins rather than employing uniform redshift intervals ($\Delta z$). This methodological choice is driven by the nonuniform redshift distribution of the Pantheon+SH0ES compilation, which is concentrated at low redshifts. Partitioning by a constant $\Delta z$ would yield drastically underpopulated bins at higher redshifts, leading to extreme disparities in the statistical uncertainties of the per-bin $H_{\Lambda}$ estimates.

By enforcing equal sample sizes per bin across varying configurations ($N_{\rm bin}\in\{2,4,6,8\}$), we aim to maintain a consistent statistical power balance. This uniform statistical weighting is advantageous for the reliability of the weighted linear regression used to diagnose the $H_{\Lambda}(z)$ drift (Table~\ref{tab7}). It helps ensure that each bin maintains sufficient sample size for a robust Gaussian approximation, thereby helping to prevent highly populated low-$z$ bins from overpowering the regression or sparse high-$z$ bins from introducing disproportionate noise into the slope ($b$) estimation.

Furthermore, this approach functions as a data-driven adaptive grid. It naturally yields high redshift resolution in the dense low-$z$ regime ($z<0.1$)—which is valuable for resolving localized variations and verifying whether the Hubble-tension signature is already ``baked in'' at local scales ($z\simeq 0.023$; \citetalias{Scolnic2025})—while dynamically widening the bins at higher redshifts to preserve statistical significance. The consistency of the results across the varied $N_{\rm bin}$ configurations suggests that our conclusions are robust against the specific choice of binning resolution and less susceptible to partitioning-induced bias.

\section{Deterministic Pre-fit and MCMC Initialization}
\label{AppE}
To ensure robust convergence of the MCMC ensemble sampler without introducing uninformative pseudo-priors, our numerical pipeline utilizes a deterministic multi-start least-squares pre-fit solely to define a finite sampling window for the walkers.

In this pre-fit stage, initial conditions are not drawn randomly. Instead, they are constructed deterministically from model-specific baseline vectors (e.g., $[H_\Lambda=70\,\text{km s}^{-1} \text{Mpc}^{-1}, \Omega_m=0.315]$ for flat $\Lambda\text{CDM}$, $[H_\Lambda=70\,\text{km s}^{-1} \text{Mpc}^{-1}, H_q=5\,\text{km s}^{-1} \text{Mpc}^{-1}]$ or $[H_q=5\,\text{km s}^{-1} \text{Mpc}^{-1}]$ for the hybrid model) using a predefined grid of multiplicative and additive perturbations. For example, applying scaling factors ranging from 0.5 to 1.5 to the baseline expansion rate ($H_\Lambda = 70\,\text{km s}^{-1} \text{Mpc}^{-1}$) disperses the initial computational seeds across a wide parameter space from 35 to 105 $\text{km s}^{-1} \text{Mpc}^{-1}$. Each starting point is then optimized using the Trust Region Reflective (TRF) algorithm within hard parameter bounds. After all starts converge, the single solution with the minimum whitened $\chi^2$ is selected.

This optimal solution is not used as a Bayesian prior to weight the likelihood, nor do we compute a weighted average of the runs. Rather, it functions as a computational seed to define a bounded, uniform sampling window within which the MCMC walkers are initialized and allowed to explore the non-linear objective surface. The final reported parameter estimates and their associated uncertainties are derived directly from the converged MCMC posterior distributions.

To prevent the deterministic pre-fit from artificially restricting parameter space, the MCMC uniform priors define a broad sampling window spanning 20$\times$ ($\pm 10\sigma$) the least-squares standard error. In cases of numerical instability or singular local curvature, a failsafe defaults this window to 40\% ($\pm 20\%$) of the physically permissible bounds. This ensures the MCMC chains evaluate the global non-linear posterior surface rather than remaining confined near the pre-fit optimum.

The operational validity of our MCMC and GLS regression pipeline is corroborated by the reproduction of key benchmark results from the literature. As detailed in Section~\ref{3.1}, under the standard $\Lambda$CDM configuration, our derived expansion rate ($H_\Lambda = 73.089 \pm 0.119 \text{ km s}^{-1} \text{ Mpc}^{-1}$) is consistent with the late-Universe local measurement by \citet{Riess2022} within a $0.05\sigma$ margin. Furthermore, as discussed in Section~\ref{3.2}, evaluating hybrid configurations conceptually analogous to the dual-redshift framework of \citet{Gupta2023} yields parameter constraints comparable to their findings, showing differences of $\le 0.31\sigma$ for $\Omega_m$ and $\le 0.54\sigma$ for the expansion rate. These cross-validations demonstrate the numerical stability of our methodology, confirming its operational validity across both the standard $\Lambda$CDM model and the hybrid configurations.

\section{Approximate origin of the Q4 and Q6 degeneracies}
\label{AppF}

We emphasize that the purpose of this Appendix is not to reproduce the full regression numerically, but to show, through low-redshift approximations, why Q4 and Q6 are structurally more weakly identifiable than the other hybrid configurations.

The exceptional behaviour of Q4 and Q6 is interpreted here as a consequence of reduced parameter identifiability in two special regression configurations. This interpretation is suggested by the global results (see Section~\ref{3.2}): once Q4 and Q6 are excluded, the hybrid fits recover $(\Omega_m,H_\Lambda)$ values close to the Planck anchors and a stable hybrid-channel scale $H_q \approx 5~\mathrm{km\,s^{-1}\,Mpc^{-1}}$.

\subsection{Q4 Degeneracy}
\label{AppF1}

The Q4 configuration leaves $(\Omega_m,H_\Lambda,H_q)$ all free while fixing $M=0$. In the low-redshift limit, starting from the hybrid decomposition,
\begin{equation}
(1+z_h)=(1+z_\Lambda)(1+z_q)
\;\;\Longrightarrow\;\;
z_h \simeq z_\Lambda + z_q.
\label{eqF1}
\end{equation}
Under the shared-path condition~\citepalias{Gupta2023}, the metric and hybrid contributions satisfy
\begin{equation}
X \simeq \frac{c_{l}}{H_\Lambda}z_\Lambda \simeq \frac{c_{l}}{H_q}z_q,
\label{eqF2}
\end{equation}
which implies
\begin{equation}
z_q \simeq \frac{H_q}{H_\Lambda}z_\Lambda.
\label{eqF3}
\end{equation}
Substituting equation~\eqref{eqF3} into equation~\eqref{eqF1} gives
\begin{equation}
z_h \simeq \left(1+\frac{H_q}{H_\Lambda}\right)z_\Lambda,
\end{equation}
or equivalently,
\begin{equation}
z_\Lambda \simeq \frac{H_\Lambda}{H_\Lambda+H_q}z_h,
\qquad
z_q \simeq \frac{H_q}{H_\Lambda+H_q}z_h.
\label{eqF4}
\end{equation}

Applying the low-redshift approximations $(1+z')^3 \simeq 1+3z'$ and $(1+3\Omega_m z')^{-1/2}\simeq 1-\frac{3}{2}\Omega_m z'$ to the comoving distance integral in equation~\eqref{eq1} yields the expanded form
\begin{equation}
X_\Lambda(z_\Lambda)
\simeq
\frac{c_{l}}{H_\Lambda}
\left[
z_\Lambda
-\frac{3}{4}\Omega_m z_\Lambda^2
\right].
\label{eqF5reason}
\end{equation}
Using the relation $d_\Lambda(z_\Lambda)=(1+z_\Lambda)X_\Lambda(z_\Lambda)$, and neglecting terms beyond second order in $z_\Lambda$, this converts to the luminosity distance:
\begin{equation}
d_\Lambda(z_\Lambda)
\simeq
\frac{c_{l}}{H_\Lambda}
\left[
z_\Lambda
+
\left(1-\frac{3}{4}\Omega_m\right)z_\Lambda^2
\right].
\label{eqF5}
\end{equation}
Together with the hybrid definition
\begin{equation}
d_h=d_\Lambda(z_\Lambda)\sqrt{1+z_q}
\label{eqF6}
\end{equation}
and
\begin{equation}
\sqrt{1+z_q}\simeq 1+\frac{1}{2}z_q,
\label{eqF7}
\end{equation}
while neglecting third-order terms such as $z_\Lambda^2 z_q$, one obtains
\begin{equation}
d_h
\simeq
\frac{c_{l}}{H_\Lambda}
\left[
z_\Lambda
+
\left(1-\frac{3}{4}\Omega_m\right)z_\Lambda^2
+
\frac{1}{2}z_\Lambda z_q
\right].
\label{eqF8}
\end{equation}
Substituting equation~\eqref{eqF4} into equation~\eqref{eqF8} gives
\begin{equation}
d_h(z_h)
\simeq
\frac{c_{l}\,z_h}{H_\Lambda+H_q}
\left[
1+
\frac{1+\frac{1}{2}\frac{H_q}{H_\Lambda}-\frac{3}{4}\Omega_m}
{1+\frac{H_q}{H_\Lambda}}
\,z_h
\right].
\label{eqF9}
\end{equation}

Equation~\eqref{eqF9} shows that, at leading order, the Hubble-diagram slope is controlled primarily by $H_\Lambda+H_q$, rather than by $H_\Lambda$ and $H_q$ separately. At the same time, the first curvature correction is controlled not by $\Omega_m$ alone, but by the coupled combination
\begin{equation}
\frac{1+\frac{1}{2}\frac{H_q}{H_\Lambda}-\frac{3}{4}\Omega_m}
{1+\frac{H_q}{H_\Lambda}}.
\label{eqF10}
\end{equation}
Therefore, in Q4, a decrease in $H_\Lambda$ can be partially compensated by an increase in $H_q$ while preserving nearly the same leading-order Hubble slope, and the remaining subleading curvature differences can be partially absorbed by a coupled shift in $\Omega_m$ and $H_q/H_\Lambda$. The displacement of Q4 from the Planck anchor is therefore more naturally interpreted as a structural degeneracy among $\Omega_m$, $H_\Lambda$, and $H_q$, rather than as evidence against the Planck-recovering hybrid solutions.

\subsection{Q6 Degeneracy}
\label{AppF2}

The Q6 configuration has a different structure. There, $H_\Lambda$ is fixed to the Planck value, while $M$, $\Omega_m$, and $H_q$ remain free. Using the same low-redshift approximation as above,
\begin{equation}
d_h(z_h) \simeq \frac{c_{l}\,z_h}{H_\Lambda+H_q}
\left[
1+
\frac{1+\frac{1}{2}\frac{H_q}{H_\Lambda}-\frac{3}{4}\Omega_m}
{1+\frac{H_q}{H_\Lambda}}
\,z_h
\right].
\label{eqF11}
\end{equation}
The regression modulus is then
\begin{equation}
\mu_h^{\rm reg}
=
5\log_{10}\!\left(\frac{d_h}{\mathrm{Mpc}}\right)+25+M.
\label{eqF12}
\end{equation}
Substituting equation~\eqref{eqF11} into equation~\eqref{eqF12} gives, to first order,
\begin{align}
\mu_h^{\rm reg}
\simeq\;&
25
+
5\log_{10}\!\left(\frac{c_{l}\,z_h}{H_\Lambda\,\mathrm{Mpc}}\right)
+
M
-
5\log_{10}\!\left(1+\frac{H_q}{H_\Lambda}\right)
\nonumber\\[1.0ex]
&+
\frac{5}{\ln 10}
\frac{1+\frac{1}{2}\frac{H_q}{H_\Lambda}-\frac{3}{4}\Omega_m}
{1+\frac{H_q}{H_\Lambda}}
\,z_h.
\label{eqF13}
\end{align}
Since $H_\Lambda$ is fixed in Q6, equation~\eqref{eqF13} shows that the leading-order intercept is controlled not by $M$ alone, but by
\begin{equation}
M - 5\log_{10}\!\left(1+\frac{H_q}{H_\Lambda}\right).
\label{eqF14}
\end{equation}

Thus, even with $H_\Lambda$ fixed, a larger $H_q$ can be partially compensated by a larger fitted $M$ while preserving nearly the same leading-order modulus. At the same time, because $\Omega_m$ remains free, the subleading curvature term can also be partially adjusted through a coupled shift in $\Omega_m$ and $H_q/H_\Lambda$. Therefore, Q6 is prone to a compound degeneracy: at leading order, the intercept freedom is shared between $M$ and $H_q$, while at the next order the residual shape can be partially absorbed by $\Omega_m$. The large magnitude offset returned by Q6 is therefore more naturally interpreted as a regression artifact of reduced identifiability, rather than as evidence that the hybrid framework requires a genuine intrinsic luminosity recalibration. 

Overall, a notable point is that the degeneracies in Q4 and Q6 disappear in Qd and Qf, which highlights the role of the independent additional-redshift parameter in recovering the Planck baseline.

More generally, similar degeneracies may in principle also affect the $w$CDM and CPL fits. However, the present analysis is concerned more specifically with whether a regression both recovers the Planck baseline and attains statistical preference under Planck-anchored priors. Under the configurations examined here, no direct evidence is found that the $w$CDM and CPL families satisfy these two criteria simultaneously (Figure~\ref{fig wCDM CPL Bin} and Table~\ref{tab10}). A fully symmetric comparison may require additional benchmark-specific constrained configurations analogous to Qd and Qf, and this remains for further study.

\section{Non-triviality of the Planck Recovery in the Q2 Configurations}
\label{AppG}

A possible concern is that, in the Q2 configuration with $M=0$ and $\Omega_m=0.315$ fixed, the recovery of $H_\Lambda$ close to the Planck baseline may simply reflect the imposed matter-density prior, rather than the operational WMC interpretation itself.

A fixed $\Omega_m$ can constrain the overall curvature budget, but it does not by itself determine how an additional redshift component enters the luminosity-distance relation, and therefore does not by itself ensure recovery of a Planck-adjacent $H_\Lambda$. To illustrate this point, we consider a counterfactual hybrid formulation in which the extra redshift is attributed to an explicit \textit{time-dilation} channel.

The corresponding WMC hybrid luminosity-distance relation and its low-redshift expansion have already been given in equation~\eqref{eq18} and Appendix~\ref{AppF}, so we derive here only the contrasting time-dilation case. In the counterfactual time-dilation formulation, the hybrid luminosity distance would take the form
\begin{equation}
d_h^{\rm (TD)} = d_\Lambda (1+z_q).
\label{eqG1}
\end{equation}
Using the same low-redshift relations derived in Appendix~\ref{AppF},
\begin{equation}
z_\Lambda \simeq \frac{H_\Lambda}{H_\Lambda+H_q}z_h,
\qquad
z_q \simeq \frac{H_q}{H_\Lambda+H_q}z_h,
\label{eqG2}
\end{equation}
together with
\begin{equation}
d_\Lambda(z_\Lambda)
\simeq
\frac{c_{l}}{H_\Lambda}
\left[
z_\Lambda
+
\left(1-\frac{3}{4}\Omega_m\right)z_\Lambda^2
\right],
\label{eqG3}
\end{equation}
the time-dilation formulation gives, to second order,
\begin{equation}
d_h^{\rm (TD)}(z_h)
\simeq
\frac{c_{l}\,z_h}{H_\Lambda+H_q}
\left[
1+
\frac{1+\frac{H_q}{H_\Lambda}-\frac{3}{4}\Omega_m}
{1+\frac{H_q}{H_\Lambda}}
\,z_h
\right].
\label{eqG4}
\end{equation}

Comparison of equation~\eqref{eqG4} with the operational WMC result in equation~\eqref{eqF9} shows that the two formulations do not share the same second-order structure: in the time-dilation case, the coefficient of $H_q/H_\Lambda$ in the curvature term is $1$, whereas in the WMC case it is $\tfrac{1}{2}$. Therefore, even under the same Q2 constraints, the two formulations do not in general imply the same low-redshift distance--redshift relation, and need not recover the same best-fit $(H_\Lambda,H_q)$ pair.

This comparison shows that the Planck-adjacent recovery of $H_\Lambda$ in Q2 is not a trivial consequence of fixing $\Omega_m=0.315$. If it were, changing the luminosity-distance formulation while keeping the same Q2 constraints would leave the low-redshift distance-redshift relation, and thus the recovered Planck-adjacent $(H_\Lambda,H_q)$ solution, effectively unchanged. Instead, the second-order coefficient changes explicitly, so the recovered $(H_\Lambda,H_q)$ pair is formulation-dependent even under the same Q2 constraints. The Q2 result is therefore better understood as a consequence of the specific operational WMC prescription adopted in this work.  

This non-trivial recovery remains meaningful even though Q2 is statistically less favored than the corresponding w2 configuration in Table~\ref{tab4} (with $\Delta \mathrm{BIC} \approx 0.09$--$0.71$). Since BIC is defined by the balance between in-sample likelihood and parameter-count penalization, it does not necessarily guarantee parameter stability or physical consistency with the Planck baseline.

%%%%%%%%%%%%%%%%%%%%%%%

\section*{Data Availability}

No new observational data were generated in support of this study. %The Pantheon+SH0ES dataset used in this paper is publicly available, and the download link is provided in footnote 2.

% References
\bibliographystyle{mnras}
\bibliography{references}

\bsp	
\label{lastpage}
\end{document}